\pdfoutput=1
\documentclass{aa}
\usepackage[varg]{txfonts}
\usepackage{graphicx}
\usepackage{subfig}
\usepackage{natbib,twoopt}
\usepackage{lscape}
\usepackage{rotating}
\usepackage{mathtools}
\usepackage[usenames,dvipsnames]{color}
\usepackage[normalem]{ulem} 

\bibpunct{(}{)}{;}{a}{}{,}

\newcommand{\HI}{{\rm H\,\scriptstyle I}}
\newcommand{\HII}{{\rm H\,\scriptstyle II}}

\begin{document}

\title{Resolved magnetic structures in the disk-halo interface of NGC\,628}

\author{D.\,D.~Mulcahy\inst{1}\thanks{Corresponding author email: david.mulcahy@manchester.ac.uk
 \newline
 Based on observations with the Karl G. Jansky JVLA of the NRAO at Socorro and the 100-m telescope of the Max-Planck-Institut f\"ur Radioastronomie at Effelsberg}
\and R.~Beck\inst{2}
\and G.\,H.~Heald\inst{3,4,5}}

\institute{Jodrell Bank Centre for Astrophysics, Alan Turing Building, School of Physics and Astronomy, The University of Manchester, Oxford Road, Manchester, M13 9PL, U.K
\and Max-Planck-Institut f\"ur Radioastronomie, Auf dem H\"ugel 69, 53121 Bonn, Germany
\and CSIRO Astronomy and Space Science, 26 Dick Perry Avenue, Kensington, WA 6151, Australia
\and Netherlands Institute for Radio Astronomy (ASTRON), Postbus 2, 7990 AA Dwingeloo, The Netherlands
\and Kapteyn Astronomical Institute, Postbus 800, 9700 AV Groningen, The Netherlands}

\date{Received 14 October 2016 /
Accepted 21 December 2016}

\abstract{Magnetic fields are essential to fully understand the interstellar medium (ISM) and its role in the disk-halo interface of galaxies is still poorly understood. Star formation is known to expel hot gas vertically into the halo and these outflows have important consequences for mean-field dynamo theory in that they can be efficient in removing magnetic helicity.}
{We aim to probe the vertical magnetic field and enhance our understanding of the disk-halo interaction of galaxies. Studying a face-on galaxy is essential so that the magnetic field components can be separated in 3D.}
{We perform new observations of the nearby face-on spiral galaxy NGC\,628 with the Karl G. Jansky Very Large Array (JVLA) at S-band (2.6--3.6\,GHz effective bandwidth) and the Effelsberg 100-m telescope at frequencies of 2.6\,GHz and 8.35\,GHz with a bandwidth of 80\,MHz and 1.1\,GHz, respectively. Owing to the large bandwidth of the JVLA receiving system, we obtain some of the most sensitive radio continuum images in both total and linearly polarised intensity of any external galaxy observed so far.}
{The application of RM Synthesis to the interferometric polarisation data over this large bandwidth provides high-quality images of Faraday depth and polarisation angle from which we obtained evidence for drivers of magnetic turbulence in the disk-halo connection. Such drivers include a superbubble detected via a significant Faraday depth gradient coinciding with a $\HI$ hole. 
We observe an azimuthal periodic pattern in Faraday depth with a pattern wavelength of 3.7\,$\pm$ 0.1 kpc, indicating Parker instabilities. The lack of a significant anti-correlation between Faraday depth and magnetic pitch angle indicates that these loops are vertical in nature with little helical twisting, unlike in IC\,342.
We find that the magnetic pitch angle is systematically larger than the morphological pitch angle of the polarisation arms which gives evidence for the action of a large-scale dynamo where the regular magnetic field is not coupled to the gas flow and obtains a significant radial component. We additionally discover a lone region of ordered magnetic field to the north of the galaxy with a high degree of polarisation and a small pitch angle, a feature that has not been observed in any other galaxy so far
and is possibly caused by an asymmetric $\HI$ hole.}
{Until now NGC\,628 has been relatively unexplored in radio continuum but with its extended $\HI$ disk and lack of active star formation in its central region has produced a wealth of interesting magnetic phenomena. We observe evidence for two drivers of magnetic turbulence in the disk-halo connection of NGC\,628, namely, Parker instabilities and superbubbles.}


\maketitle

\section{Introduction}

Understanding of the disk-halo interaction is vital to explain the evolution of spiral galaxies. Several theories exist that describe how star formation expels hot gas vertically into the halo, namely the
galactic fountain model  \citep{1976ApJ...205..762S, 1980ApJ...236..577B}, the chimney model  \citep{1988LNP...306..155N}, and the galactic wind model \citep{1991A&A...245...79B} relevant to quiescent star formation (SF), clustered SF and starbursts respectively.

These disk-halo outflows have important consequences for mean-field dynamo theory, a process which involves the inductive effect of turbulence and differential rotation to amplify the weak magnetic fields to produce strong large-scale fields \citep{Ruzmaikin1988, 1996ARA&A..34..155B}.
\cite{1995MNRAS.276..651B} studied a simple model of such a galactic fountain flow and found that the horizontal field in the galactic disk can be pumped out into
the halo to a height of several kpcs. The magnetic field strength at a height of several kpcs was found to be comparable to that in the disk.
Mean-field dynamo models including a galactic wind may also explain the X-shaped magnetic fields observed around
edge-on galaxies \citep{2010A&A...512A..61M}. Alternatively, suppression of the $\alpha$-effect and of mean-field
dynamo action
can result from the conservation of magnetic helicity in a medium of high electric conductivity.
\cite{2006A&A...448L..33S} showed that the galactic fountain flow is efficient in removing magnetic helicity from galactic disks.
This allows the mean magnetic field to saturate at a strength comparable to equipartition with the turbulent kinetic energy.
However, fast outflows can advect large-scale magnetic fields and hence suppress mean-field dynamo action. As the outflow is expected to be stronger above and below the spiral arms, large-scale fields are weaker than in the inter-arm regions which may explain the observation of magnetic arms between material arms \citep{2015MNRAS.446L...6C}.

Such outflows driven from stellar winds and supernovae from young massive stars in OB associations and super star clusters (SSCs) create $\HI$ holes \citep{2011AJ....141...23B} and magnetic field loops pushed up by the gas \citep{1980ApJ...236..577B,1988LNP...306..155N}.
As a result, gradients in Faraday rotation measures (RM) and field reversals across these $\HI$ holes are expected. Detecting such RM gradients would constrain the parameters of this outflow model.

Magnetic fields oriented along the disk plane may also bend upwards and become vertical fields as the result of the
magnetic Rayleigh-Taylor instability, commonly known as the Parker instability. Ionised gas can slide down the magnetic loops, thereby reducing the confining weight of the gas and allowing the loops to rise even higher.

Efforts have been made to observe magnetic field loops. \cite{2012ApJ...754L..35H} detected an RM gradient in NGC\,6946 across a 600-pc $\HI$ hole which indicated a vertical magnetic field.
However, this observational result was obtained for an inclined galaxy, so that the observed RM fluctuations could be affected by the variations of field strengths either parallel or vertical
to the disk. Field amplification in the disk can be due to compressive gas motions (density waves) or shearing, while vertical fluctuations can be due to Parker loops, which are of kpc scale, or gas outflows.
A helically twisted Parker loop was recently found in the spiral galaxy IC\,342 \citep{2015A&A...578A..93B}.

Studying a face-on galaxy like NGC\,628 is essential so that the magnetic field components can be separated in 3D. In this case, RMs are only sensitive to vertical fields, while synchrotron emission traces the magnetic field in the disk.

NGC\,628 = M74 is a large, grand-design, isolated spiral galaxy with a low inclination angle and a $\HI$ disk with a diameter of about 30$\arcmin$, about three times the Holmberg diameter \citep{1992A&A...253..335K}. A list of physical parameters is given in Table~\ref{physicalpara}.
The spiral structure is grand-design but with broad spiral arms. The apparent absence of strong density
waves is related to the lack of a strong bar or an interacting companion, leaving the disk largely undisturbed (seen in optical, UV, and in gas kinematics). NGC\,628 is similar to
NGC\,6946 in many respects and hosts many $\HI$ holes \citep{2011AJ....141...23B}. NGC\,628 actually contains nearly twice the number of $\HI$ holes (102 holes) compared to NGC\,6946 (54 holes) found by  \cite{2011AJ....141...23B}, but previous studies performed by \cite{Boomsma2008} found 121 $\HI$ holes for NGC\,6946 with inferior resolution compared to \cite{2011AJ....141...23B}.

The high star-formation rate of NGC\,628 leads to a bright disk in radio continuum (synchrotron + thermal) of about $10\arcmin \times 8\arcmin$
extent \citep{1987ApJS...65..485C}. \cite{2001ApJS..132..129M} observed NGC\,628 extensively in UV. Bright knots embedded in diffuse emission trace the spiral pattern and many of these knots are also bright in H$\alpha$ \citep{Dale2009}.
\cite{2001ApJS..132..129M} suggested that the entire disk of NGC\,628 has undergone active star formation within the past 500\,Myr and that the inner regions have experienced more rapidly declining star formation than the outer regions.

NGC\,628 was part of the Westerbork Synthesis Radio Telescope (WSRT) SINGS (The SIRTF Nearby Galaxies Survey) survey and was also observed in
polarised radio continuum at 1.5\,GHz with moderate resolution \citep{2009A&A...503..409H}. The polarised emission is restricted to the outer disk and the observed Faraday depths are small, due to strong Faraday depolarisation at this
frequency, similar to NGC\,6946 observed at the same frequency. With the exception of this observation, NGC\,628 has not been studied in radio continuum.

\begin{table}
\centering
\caption{Physical parameters of NGC\,628.}
\label{physicalpara}
\begin{tabular}{l c}
\hline\hline
Morphology \tablefootmark{ a} & SAc \\\
Position of the nucleus & RA(J2000) = 01$^\mathrm{h}$ 36$^\mathrm{m}$ 41$^\mathrm{s}.74$\\
 & DEC(J2000) = +15$\degr$ 47$\arcmin$ 01.1$\arcsec$ \\
Position angle of major axis\tablefootmark{ b} &  $20\degr$ \\
D$_{25}$\tablefootmark{ a} & $10.5 \times 9.5 \arcmin$  \\
Inclination of the inner disk\tablefootmark{ b} & $7\degr$ ($0\degr$ is face on)  \\
Inclination of the outer disk\tablefootmark{ b} & $13.5\degr$   \\
Distance\tablefootmark{ c} & 7.3\,Mpc ($1\arcsec \approx$ 35\,pc) \\
Star formation rate\tablefootmark{ d} & 1.21 M$_{\odot}$ yr$^{-1}$ \\
\hline
\end{tabular}
\tablefoottext{a}{\citet{1991rc3..book.....D}} 
\tablefoottext{b}{\citet{1992A&A...253..335K}} 
\tablefoottext{c}{\citet{2004AJ....127.2031K}} 
\tablefoottext{d}{\citet{2008AJ....136.2563W}} 
\end{table}

Due to its small inclination and large angular size, NGC\,628 is one of the best galaxies to study the magnetic field of the disk. However, its low inclination makes it difficult to extract an accurate rotation curve and was not computed in the THINGS $\HI$ Nearby Galaxy Survey \citep{deblok2008}. \cite{1992A&A...253..335K} derived from their $\HI$ observations that the rotation velocity reaches 200\,km\,s$^{-1}$.

In this paper, we present new Karl G. Jansky Very Large Array (JVLA) S-band observations of NGC\,628 along with Effelsberg observations at 2.6\,GHz and 8.35\,GHz including both total power and polarisation. Sect.~\ref{obsanddatareduction} presents details of the observations and data reduction, Sect.~\ref{effresults} shows the Effelsberg maps and comparison between the extracted flux densities with literature. In Sect.~\ref{JVLAdata} we present the new JVLA observations, compare to observations at other wavelengths, and determine magnetic field strengths. In Sect.~\ref{verticalfields} we show results of RM Synthesis in probing the vertical magnetic field of NGC\,628. In Sect.~\ref{discussion} we discuss the implications of our main findings and in Sect.~\ref{conclusions} we summarise our findings.

\section{Observations and data reduction}
\label{obsanddatareduction}

In this section we describe the nature of the observations taken with the JVLA and Effelsberg 100-metre telescope and the steps performed in the data reduction, starting with the Effelsberg observations.
A brief summary of the observational parameters is given in Table~\ref{table:obsparas}.

\begin{table*}
 \centering
  \caption{Radio continuum observational parameters of NGC\,628.}
  \begin{tabular}{@{}llrrrrlclrrrrrrrccrrrr@{}}
  \hline
                         &  &  &  &  &   &  &  JVLA  &  &  & & & & & & & Effelsberg & & & &\\
 \hline
 Frequency (GHz)         &  &  &  &  &   &  & 2-4   &  &  & & & & & & & 2.6 \& 8.35& & & & &\\
 Bandwidth (MHz)         &  &  &  &  &   &  & 2000 (of which 1000 are useful)  &  &  & & & & & & & 80 \& 1100 & & & & &\\
 No. of spectral windows &  &  &  &  &   &  & 16    &  &  & & & & & & & -- & & & & &\\
Total no. of channels    &  &  &  &  &   &  & 1024  &  &  & & & & & & & 8 \& 1 & & & & &\\
  Array configuration    &  &  &  &  &   &  & D; C  &  &  & & & & & & & -- & & & & &\\
 Pointings               &  &  &  &  &   &  & 7     &  &  & & & & & & & -- & & & & &\\
 Observing dates         &  &  &  &  &   &  &  March \& July 2013, respectively  &  &  & & & & & & & October 2011 & & & & &\\
 \hline
\end{tabular}
\label{table:obsparas}
\end{table*}

\subsection{Effelsberg observations at 2.6\,GHz and 8.35\,GHz}

NGC\,628 was observed at 2.6\,GHz and 8.35\,GHz (with bandwidths of 80\,MHz and 1.1\,GHz, respectively) in October 2011 by the authors.
The maps were scanned alternating in RA and DEC with the one-horn secondary-focus systems.
The data was reduced with the NOD2 software \citep{Haslam1974}. As Radio Frequency Interference (RFI) can be a substantial problem at 8.35\,GHz, NGC\,628 was only observed at higher elevations at this frequency. 4 out of 32 coverages had to be discarded due to extremely bad RFI.
NGC\,628 was observed at 2.6\,GHz at lower elevations as the RFI is not as severe at this frequency. However, significant RFI was still present and had to be flagged in Stokes I, Q, \& U.
The final Effelsberg images at 2.6\,GHz and 8.35\,GHz were obtained from 13 and 28 coverages, respectively, and combined using the spatial-frequency weighting method of \cite{1988A&A...190..353E}.

A flux density scaling was applied to the 8.35\,GHz data using maps of two calibrators, 3C286 and 3C48. Using values from the VLA calibrator handbook \footnote{http://www.aoc.nrao.edu/~gtaylor/csource.html}, a common scale factor was found and was applied to the target data. For both calibrators, the linearly polarised intensity and polarisation angle were found by fitting a 2D-Gaussian over the Stokes Q \& U maps and calculated from the following equations:

\begin{equation}
 P_{lin} = \sqrt{U^2 + Q^2 - (1.2 \, \sigma_{QU})^2}
 \label{equation1}
\end{equation}
where $\sigma$ is the noise in Stokes Q and U,

\begin{equation}
 \phi = \frac{1}{2}\, \rm{arctan}(\frac{U}{Q})\, .
  \label{equation2}
\end{equation}

From the seven maps of 3C48, one had a very different value for the polarisation angle and was discarded. The remaining maps gave a degree of polarisation of $5.5\%$ and a polarisation angle of $-65\degr$. This agrees very well
with the values of $5.3\%$ and $-64\degr$ for 8.1\,GHz found by \cite{2013ApJS..206...16P}.
In addition, for two maps of 3C286, the degree of polarisation was $11.65\%$ and the polarisation angle was
$33.8\degr$. Again, this is in excellent agreement to \cite{2013ApJS..206...16P} who found values of $11.7\%$ and $33\degr$. Therefore, we can be confident that the polarisation calibration for our target is accurate. The level of instrumental polarisation is below 1\% at both frequencies.

As the 2.6\,GHz receiver is equipped with a 8 channel polarimeter, an additional bandwidth calibration had to be performed along with the normal flux density scaling. For each of the eight
channels, the flux density of the calibrator 3C295 was compared to its theoretical value, using a power law with a spectral index of -1.04. A scale factor was then determined from the ratio of the two and applied to each channel.
The final images for both frequencies are presented in Figures~\ref{fig:3cmmaps} \&~\ref{fig:11cmmaps}.

\subsection{JVLA observations at S-band}

NGC\,628 was observed by the JVLA
in D and C configurations in February and June 2013 in S-band (2--4\,GHz).
The synthesised beamwidth for a 12 hour observation with uniform weighting scheme is approximately $23\arcsec$ for D configuration at 3\,GHz and $7.0\arcsec$ for C configuration.
The ultimate factor limiting the field of view is the diffraction-limited response of the individual antennas.

An approximate formula for the full width at half power (FWHM) of the primary beam (in arcminutes) is:
FWHM = $45\arcmin/\nu_\mathrm{GHz}$.
For the highest frequency (4\,GHz) in S-band, FWHM is approximately $10\arcmin$.

As NGC\,628 is extended over $10.5\arcmin$ in angular size (Table~\ref{physicalpara}), observations with several pointings were needed.

For this purpose, a hexagonal grid was chosen for both D and C configurations consisting of 7 pointings, separated by $5\arcmin$.
While the pointing grid is oversampled, this is a safe approach as every position in the galaxy is at least covered three times. In addition, this makes the observations more sensitive to NGC\,628's extended disk.

S-band is subject to very strong RFI from a number of satellites in particular those providing satellite radio service. A satellite passing through the initial slew will cause unwanted erroneous attenuator settings. Therefore, in order to prevent this, the observation was started with a quick L-band scan on 3C48 which was in a satellite free zone at this time. After this, the S-band was switched on.

3C48 was observed at the start of the observation so it could be used as the main flux density calibrator to
calibrate bandpass and absolute flux densities.
The source J0240+1848 was observed at least once per hour for use as a phase and polarisation leakage calibrator.
3C138 was observed twice during the observation in order to calibrate the polarisation angle.
Between these calibrator scans, all 7 pointings of the NGC\,628 were performed several times.
The D \& C configuration observations were a total of 5 \& 6 hours long.

\subsection{Calibration of JVLA S-band data}

The data were reduced using the Common Astronomy  Software Applications (CASA) package\footnote{http://casa.nrao.edu} \citep{MCMULLIN2007}. Both D and C configurations were calibrated identically but separately.

Data affected by shadowed antennas were flagged.
As RFI can be extremely strong in S-band and as this RFI can produce sharp edges in the spectrum, this can introduce an oscillation across the frequency channels where no RFI is present. This is called Gibbs' phenomenon and can be brought under control by using Hanning smoothing.
Due to extreme RFI in several spectral windows at the two extremities of S-band, the bandwidths 2.0--2.6\,GHz and 3.6--4\,GHz were completely flagged. The resulting central frequency is 3.1\,GHz.

A table of antenna position corrections was produced and incorporated into our calibration using the task GENCAL.
A preliminary bandpass correction was applied to the data in order to maximise the effectiveness of the automatic flagging.
Visibilities affected by RFI were flagged via the automatic flagging routine RFLAG. The antenna EA21 was flagged due to higher than average amplitudes.

The model of 3C48 was taken from \cite{2013ApJS..204...19P} and was used as our main flux density and bandpass calibrator.
The residual delays of each antenna relative to the reference antenna were found using 3C48. The delays for all antennas were seen to be on average approximately 4\,ns and none were greater than 8\,ns.
Next, the corrections for the complex antenna gains were obtained. The gain amplitudes
were determined by referencing our standard flux density calibrator. To determine the appropriate complex gains for the target source, we used our phase calibrator, J0240+1848.
This was the closest and most appropriate phase calibrator that could be found to our target in order to minimise the differences through the ionosphere and troposphere between the phase calibrator and the target source.

For polarisation calibration, the cross-hand delays due to the residual difference between the R and L correlations on the reference antenna were solved using 3C138 which has a higher polarisation signal than 3C48. Prior to this, the Q and U values were assigned to 3C138 using the model of \cite{2013ApJS..206...16P}.

The phase calibrator (J0240+1848) had sufficient parallactic coverage for both observations and was used to solve for polarisation leakage (usually 5\%).

3C138 was then used to solve for the polarisation angle.
Once this was done, all calibration tables were applied to the calibrators and each pointing for the target.
The C and D configurations were combined in order to increase the signal-to-noise ratio and maximise the combined uv coverage of the observation.

Self-calibration was tested on the target, however the resulting gain solutions were poor. Applying these solutions degraded the image which was most likely caused by the low signal-to-noise ratio and diffuse nature of this particular target. Therefore, self-calibration was not performed.

\subsection{Imaging and mosaicing}

\begin{table}
\caption{Imaging parameters of the NGC\,628 JVLA data.}
\begin{tabular}{l c c}
\hline\hline
& Natural weighting & \\
\hline \hline
Angular resolution & $18\arcsec$ (630\,pc)  &  \\
Cell size & $3\arcsec$ & \\
Min \& max cleaning scales & $3.0\arcsec$ & $180\arcsec$\\
\hline
& Robust weighting (0.0) & \\
\hline \hline
Angular resolution & $7.5\arcsec$ (265\,pc) &  \\
Cell size & $1.5\arcsec$ & \\
Min \& max cleaning scales & $1.5\arcsec$ & $180\arcsec$\\
\hline
\end{tabular}
\label{table:imagingpara}
\end{table}

The Clark CLEAN algorithm \citep{1980A&A....89..377C} was used to deconvolve the dirty beam (point spread function; PSF) from the dirty map.

The best strategy found for cleaning the naturally weighted images was to perform CLEAN using a mask until all residuals are cleaned, followed by 20,000 clean iterations without a mask and without multi-scale CLEAN. In order to model the extended emission in the field of view, multi-scale CLEAN \citep{2008ISTSP...2..793C} was performed with scales ranging from 1.5$\arcsec$ to 180$\arcsec$.  A summary of the imaging parameters is given in Table~\ref{table:imagingpara}.
Multi-scale CLEAN is an extension to CLEAN that models the sky brightness by the summation of components of emission having different size scales.
The noise for each pointing was found on average to be 5.1\,$\mu$Jy/beam for natural weighting and 6.9\,$\mu$Jy/beam for robust weighting
on average for total intensity (beam sizes can be seen in Table~\ref{table:imagingpara}). For Stokes Q \& U, the noise was found to be approximately 4.5\,$\mu$Jy/beam and 6.5\,$\mu$Jy/beam for natural and robust weighting, respectively.

Multi-frequency synthesis \citep{2011A&A...532A..71R} was attempted with the Taylor terms set to 2. However, it was found that the very weak diffuse emission to the north of the galaxy would be degraded. Leaving the Taylor terms set to 1 was found to produce far better results for this weak extended emission.
Unfortunately this means that we are not able to estimate the in-band spectral index, \cite{2015arXiv150205616C} states that the in-band spectral index for JVLA S-band is only reliable for very strong compact sources with a signal-to-noise ratio of $\approx$ 50; in addition, the apparent spectral indices of extended sources would be too steep as the synthesised beam solid angle decreases by a factor of four across the S-band. In our case, NGC\,628 is too weak, diffuse and extended for the in-band spectral index to be reliable.

Each of the seven pointings was imaged separately for each Stokes parameter and then mosaiced together. The primary beam of the final image was constructed from each pointing and applied to the final image.
The final image of NGC\,628 in total intensity at 7.5$\arcsec$ resolution is shown in Fig.~\ref{fig:628_StokesI8arcsec}.

\section{Results from Effelsberg data}
\label{effresults}

\subsection{Effelsberg maps}

\begin{figure*}
	\hspace{0.6cm}
		\subfloat{\includegraphics[width=0.45\textwidth]{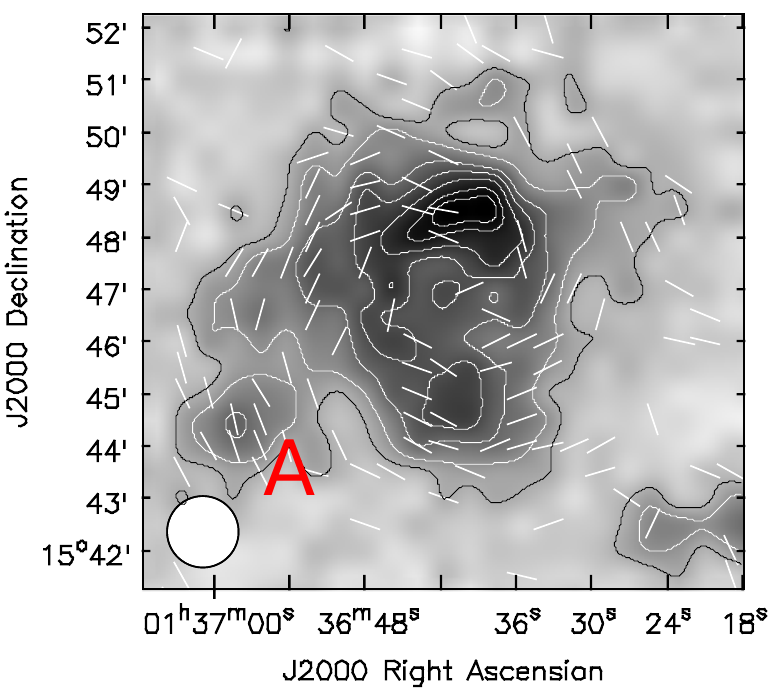}}
		\subfloat{\includegraphics[width=0.45\textwidth]{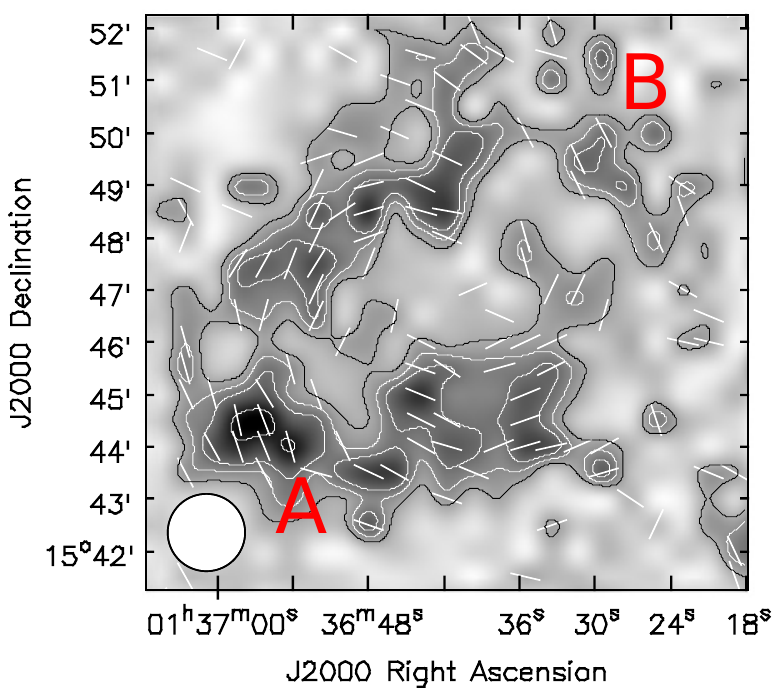}}
	\caption{NGC\,628 observed at 8.35\,GHz with the Effelsberg 100-m telescope at a resolution of 81$\arcsec$. The left image displays the total intensity of NGC\,628 with contours representing 3, 5, 8, 10, 12, 14, 15 $\times$ 30\,$\mu$Jy/beam. The right image displays the linearly polarised intensity of NGC\,628 with contours representing 3, 4, 5, 8 $\times$ 60\,$\mu$Jy/beam.
The vectors show E + 90$\degr$, not corrected for Faraday rotation, which is very small at this frequency.
All vectors are plotted with the same length.
	}
	\label{fig:3cmmaps}
\end{figure*}

\begin{figure*}
	\hspace{0.6cm}
		\subfloat{\includegraphics[width=0.45\textwidth]{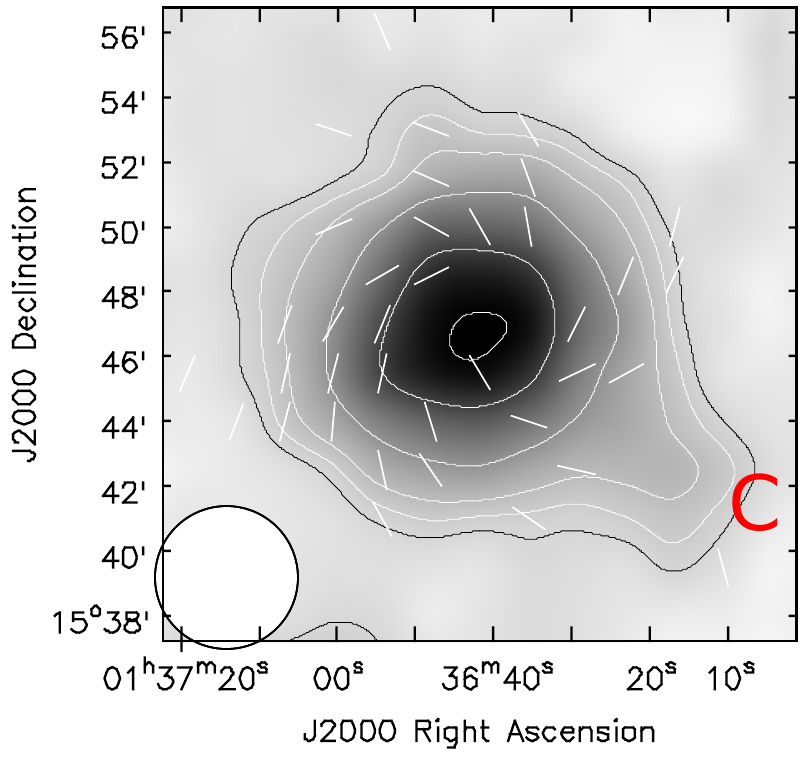}}
		\subfloat{\includegraphics[width=0.45\textwidth]{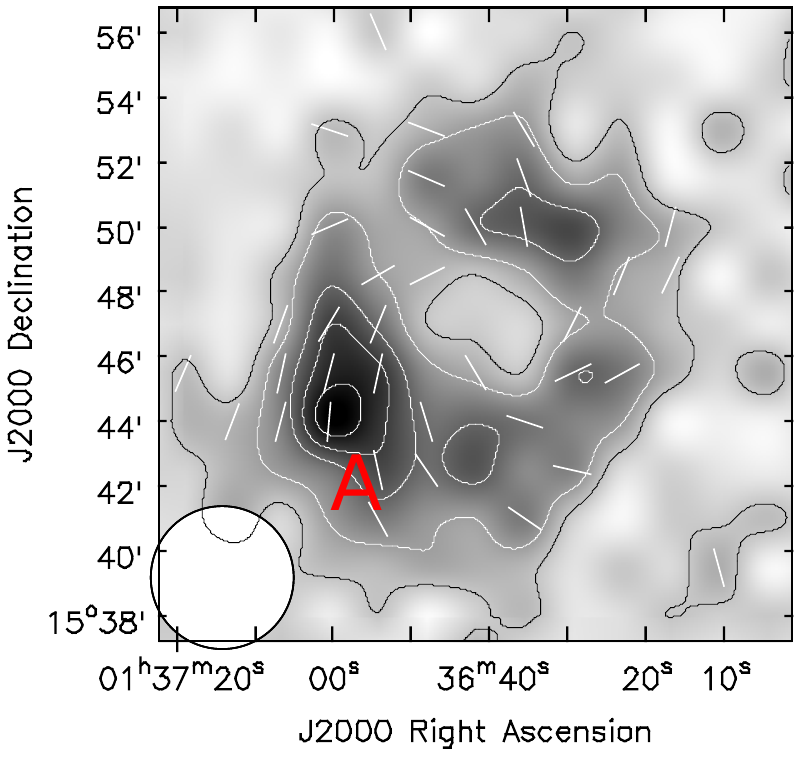}}
	\caption{NGC\,628 observed at 2.65\,GHz with the Effelsberg 100-m telescope at a resolution of 4.6$\arcmin$. The left image displays the total intensity of NGC\,628 with contours representing 3, 5, 8, 16, 32, 44 $\times$ 1\,mJy/beam. The right image displays the linearly polarised intensity of NGC\,628 with contours representing 3, 5, 8, 10, 12 $\times$ 2.5\,mJy/beam. The vectors show E + 90$\degr$, not corrected for Faraday rotation, which is small at this frequency.
All vectors are plotted with the same length.}
	\label{fig:11cmmaps}
\end{figure*}

The final images of NGC\,628 observed with the Effelsberg telescope are shown in Figures~\ref{fig:3cmmaps} \&~\ref{fig:11cmmaps}. The rms noise for total intensity at 8.35\,GHz and 2.6\,GHz were found to be 0.28\,mJy/beam and 1.1\,mJy/beam with their resolutions being 81$\arcsec$ and 4.6$\arcmin$. The average rms noise for Stokes Q and U for 8.35\,GHz and 2.6\,GHz were found to be 0.10\,mJy/beam \& 0.30\,mJy/beam, respectively.

At 8.35\,GHz, NGC\,628 is found to be very asymmetric. The most intense emission is seen to the north of the galaxy, at RA(J2000) = 01$^\mathrm{h}$ 36$^\mathrm{m}$ 41$^\mathrm{s}$, DEC(J2000) = +15$\degr$ 48$\arcmin$ 29$\arcsec$ rather than the centre of the galaxy, suggesting that NGC\,628 does not have an active nucleus like M\,51 \citep{Rampadarath2015}. The brightest region of NGC\,628 is part of the northern spiral arm, consisting of many $\HII$ regions, and is also bright in the infrared spectral range \citep{2011PASP..123.1347K} and in $\HI$ line emission \citep{2008AJ....136.2563W}.
This agrees with the observation of \cite{2001ApJS..132..129M} that the inner regions of the galaxy have experienced more rapidly declining star formation than the outer regions.

In linear polarisation, the 8.35\,GHz map shows diffuse emission
along with the polarisation angles forming a spiral pattern.
A polarised background source (J013657+154422, shown with the letter A in Fig.~\ref{fig:3cmmaps}) is apparent at the south-east of the galaxy. There is no sign of polarisation in the central region of the galaxy, probably due to beam depolarisation.

On the 2.6\,GHz map, no significant features can be made out in total intensity, owing to the poor resolution of the Effelsberg telescope at this frequency.
The total intensity emission is seen to extend to the southwest due to an unresolved background source (seen as the letter C in Fig.~\ref{fig:11cmmaps} and \ref{fig:628_StokesI8arcsec}).
Significant polarisation is detected, with the polarisation angles creating a spiral pattern. No polarisation can be seen in the central region. The strongest polarised emission is seen again at the location of J013657+154422 (letter A in Fig.~\ref{fig:11cmmaps} (right))  but significant diffuse polarised emission is also observed. Most likely this emission originates from the inter-arm regions of NGC\,628 which will be become apparent when inspecting the JVLA data.

\subsection{Integrated radio continuum spectral analysis}

\begin{figure}
	\centering
	\includegraphics[width=0.9\columnwidth]{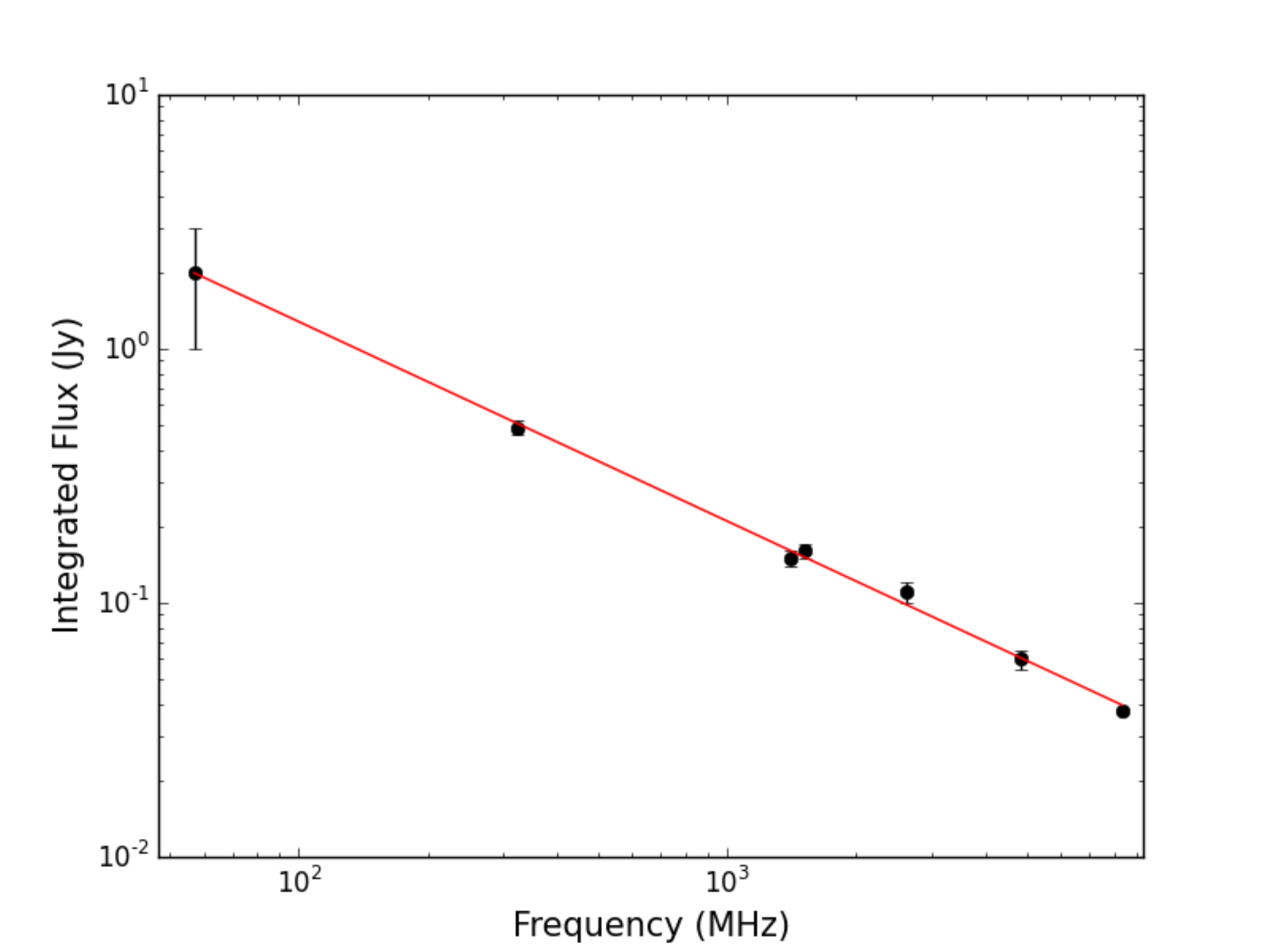}
	\caption{Integrated spectrum of NGC\,628 using Effelsberg and literature integrated flux measurements with the red line representing a single power law with a spectral index of $\alpha = -0.79\pm0.06$.}
	\label{fig:b2007spectrum}
\end{figure}

The integrated flux densities found for NGC\,628 are $38 \pm 3$\,mJy at 8.35\,GHz and $110 \pm 10$\,mJy at 2.6\,GHz. Measurements from the literature (Table~\ref{NGC628integratedflux}) were also used to determine the
spectral index of the galaxy. All these measurements are tied to the flux scale of \cite{1977A&A....61...99B}.
A single power-law fit to the data yields a spectral index of $\alpha = -0.79\pm0.06$. A plot of this data can be seen in Fig.~\ref{fig:b2007spectrum}. This value not only agrees with \cite{2009A&A...503..747P} ($\alpha = -0.78$) but is a normal value for spiral galaxies ($\alpha = -0.74 \pm 0.03$, \cite{1982A&A...116..164G}).

At low frequencies, the spectrum of NGC\,628 shows no indications of flattening all the way down to 57\,MHz. This suggests that neither free-free absorption of the synchrotron emission nor ionisation losses are significant for the galaxy as a whole. The galaxy's integrated spectrum remains a power law, probably due to the clumpy nature of the ISM \citep{Basu2015}.

A straight power-law spectrum was also seen for M\,51 down to 151\,MHz \citep{2014A&A...568A..74M}.
It should be noted that the flux density of NGC\,628 from \cite{1990ApJ...352...30I} is uncertain due to the poor sensitivity and angular resolution of the Clark Lake telescope and should be treated cautiously.
Without this integrated flux, the fitted spectral index is found to be $\alpha = -0.79\pm0.03$. Unfortunately, no other integrated fluxes has been taken at these low frequencies.
Observations with LOFAR \citep{2013A&A...556A...2V} below 100\,MHz are underway and will help to verify whether the integrated spectrum is indeed a straight power law but will additionally be able to resolve this galaxy to arcsecond resolution.

\begin{table}
\caption{Integrated flux densities of NGC\,628.} 
\label{NGC628integratedflux}
\centering 
\begin{tabular}{c c c} 
\hline\hline 
$\nu$ (GHz) & Flux density (Jy) & Reference \\ 
\hline 
8.35 & 0.038 $\pm$ 0.003 & This work  \\
4.85 & 0.06 $\pm$ 0.005 & Paladino et al. (2009) \\
2.614 & 0.11 $\pm$ 0.01 & This work \\
1.515 & 0.16 $\pm$ 0.01 & Paladino et al. (2009) \\
1.4 & 0.15 $\pm$ 0.01 & Condon et al. (1998) \\
0.324 & 0.49 $\pm$ 0.03 & Paladino et al. (2009) \\
0.057 & 2 $\pm$ 1 & Israel \& Mahoney (1990) \\
\hline
\end{tabular}
\end{table}

\section{Results from JVLA S-band data}
\label{JVLAdata}

\subsection{Total intensity}

\begin{figure*}
\hspace{1cm}
\centering
\includegraphics[width=1.0\textwidth]{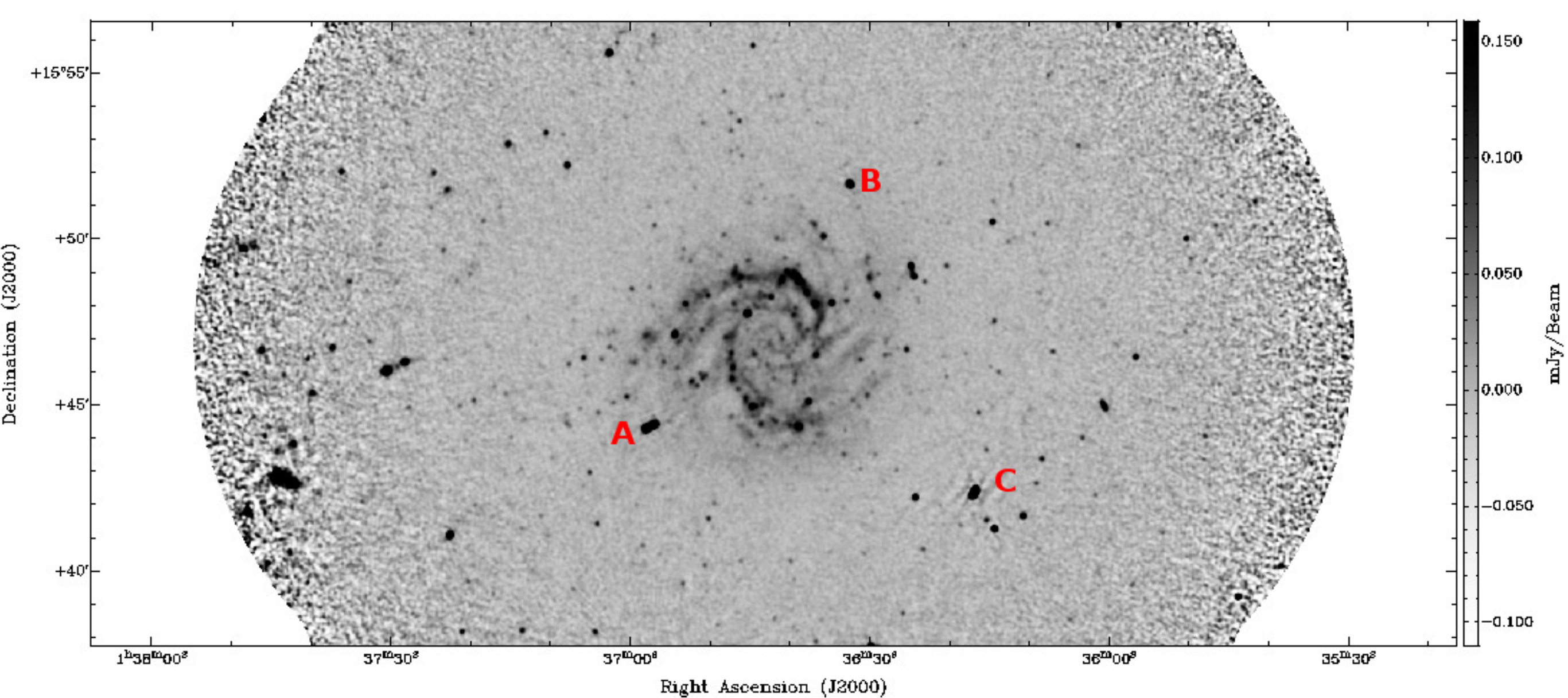}
\caption{Total intensity map of NGC\,628 at 3.1\,GHz at $7.5\arcsec$ resolution, observed with the combined D \& C configurations of the JVLA. The rms noise is approximately 5\,$\mu$Jy/beam. This image has been primary beam corrected. Several background sources are labeled for future reference. }
\label{fig:628_StokesI8arcsec}
\end{figure*}

\begin{figure*}[!ht]
	\vspace{1cm}
    \vspace{0.3cm}
	\centering
		\subfloat{\includegraphics[width=0.6\textwidth]{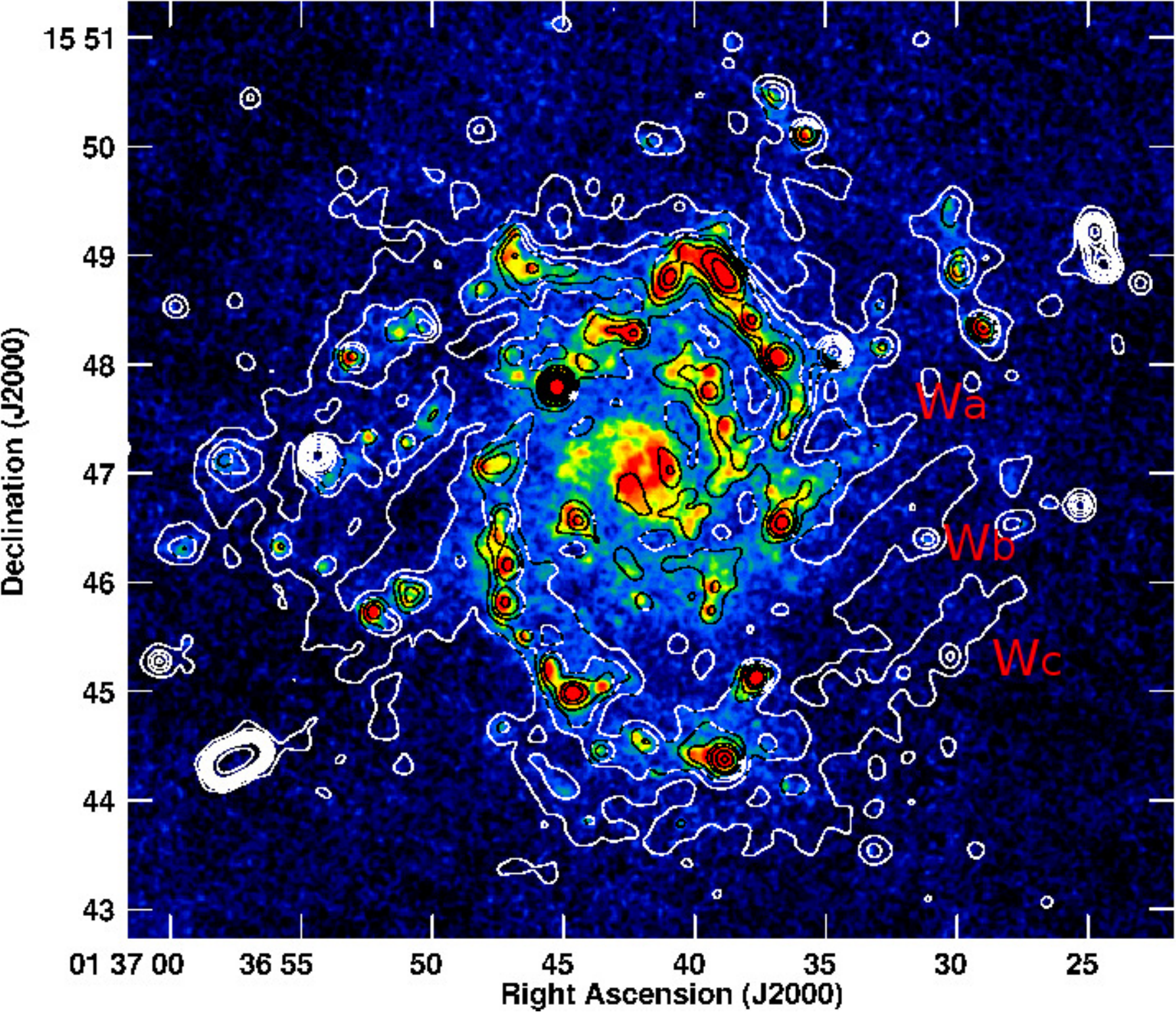}}
    \vspace{0.3cm}
	\centering
		\subfloat{\includegraphics[width=0.6\textwidth]{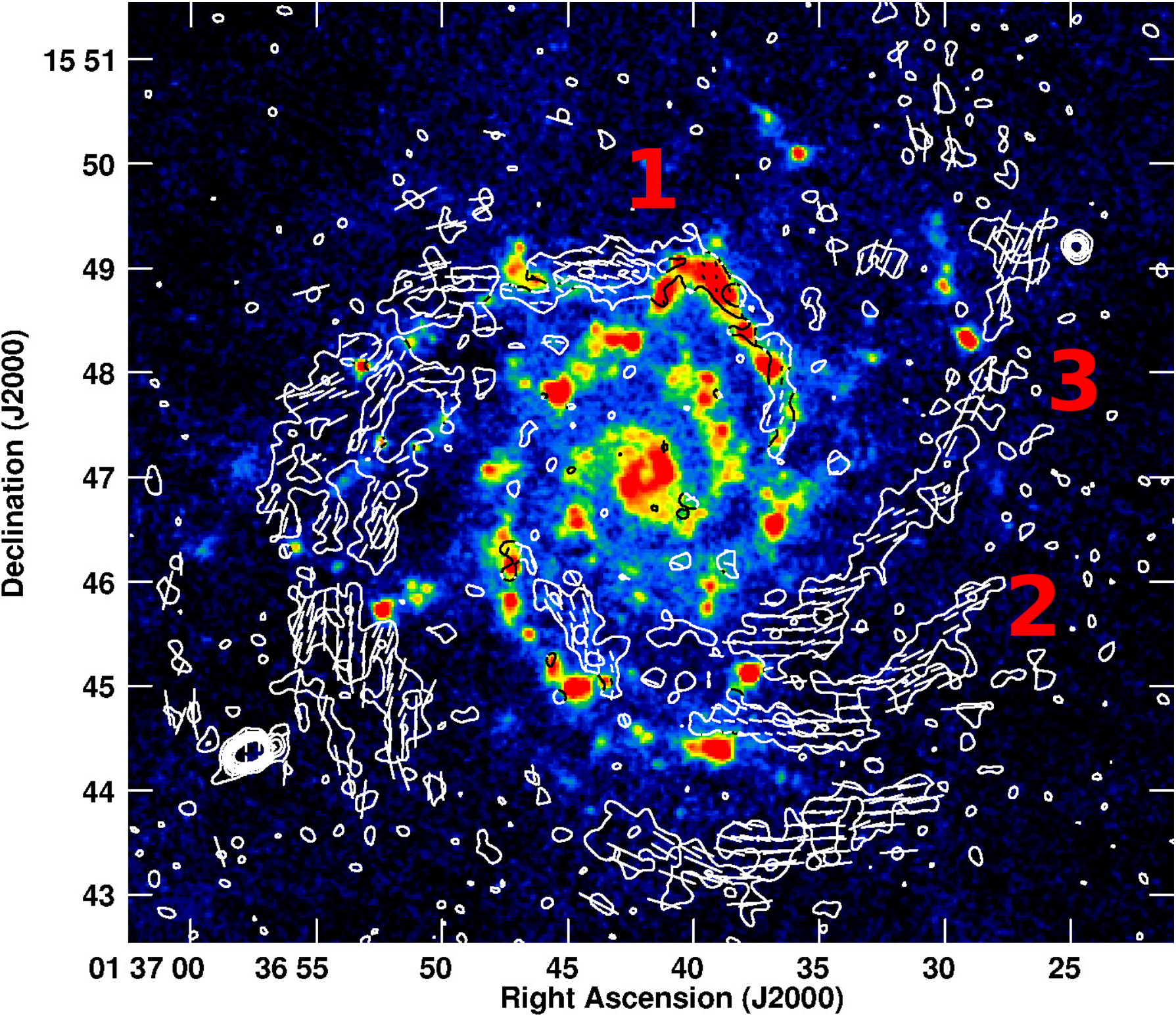}}
	\caption{Robust-weighted JVLA image of NGC\,628 of total intensity (\textsc{top}) and linearly polarised intensity (\textsc{bottom}) at 3.1\,GHz at a resolution of 10$\arcsec$, averaged over all channels,
and overlaid onto an $70\,\mu$m IR image from Herschel \protect{\citep{2011PASP..123.1347K}}.
Contours of total intensity are at 1, 2, 3, 4, 6, 8, 12, 16, 32, 64, 128 $\times$ 40\,$\mu$Jy/beam. The nonthermal arms to the west are marked Wa to Wc for future reference.
Contours of polarised intensity are at 1, 2, 3, 4, 6, 8, 12 $\times$ 12\,$\mu$Jy/beam.
The white lines show the magnetic field orientations (E + 90$\degr$), not corrected for Faraday rotation, with a
length of 10$\arcsec$ representing 30\% degree of polarisation. The main polarised arms are marked from 1 to 3 for future reference.
	\label{fig:robustmaps} }
\end{figure*}

\begin{figure*}[!ht]
	\vspace{1cm}
    \vspace{0.3cm}
	\centering
		\subfloat{\includegraphics[width=0.6\textwidth]{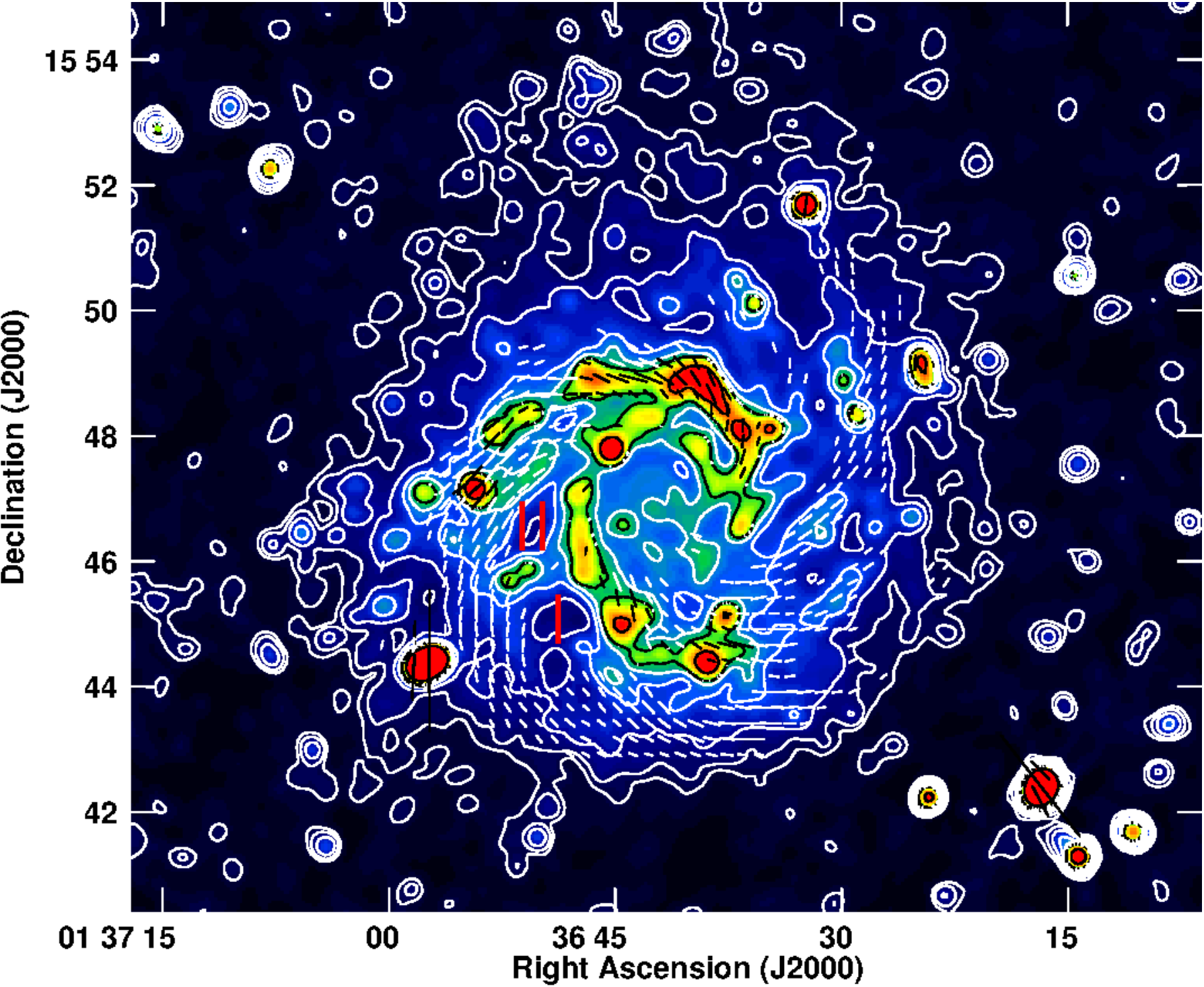}}
    \vspace{0.3cm}
	\centering
		\subfloat{\includegraphics[width=0.6\textwidth]{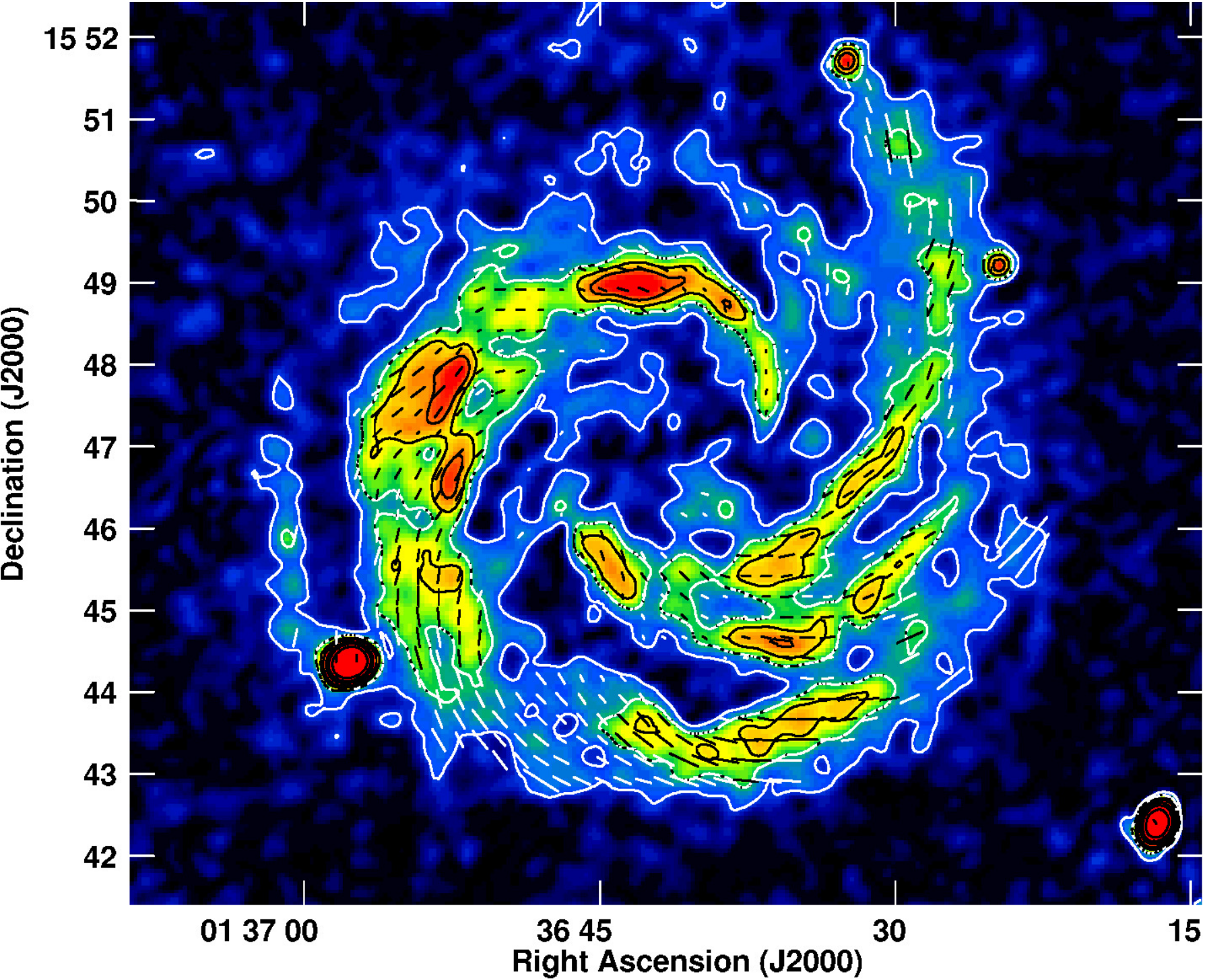}}
	\caption{Natural-weighted JVLA image of NGC\,628 of total intensity (\textsc{top}) and linearly polarised intensity (\textsc{bottom}) at 3.1\,GHz at a resolution of 18$\arcsec$, averaged over all channels. Contours of total intensity are at 1, 2, 4, 8, 12, 16, 32 $\times$20\,$\mu$Jy/beam. Contours of polarised intensity are at
1, 2, 3, 4, 8, 12 $\times$15\,$\mu$Jy/beam. The lines show the magnetic field vectors (E + 90$\degr$), not corrected
for Faraday rotation, with a length of 10$\arcsec$ representing a polarised intensity of $150\,\mu$Jy/beam
(\textsc{top}) and 30\% degree of polarisation (\textsc{bottom}), respectively. The roman numerals  I and II refer to the $\HI$ holes shown in Fig.~\ref{fig:HIradiohole}.
	\label{fig:naturalmaps} }
\end{figure*}

The final total intensity image produced with robust weighting at $7.5\arcsec$ resolution is shown in Fig.~\ref{fig:628_StokesI8arcsec}, with robust weighting and smoothed to 10$\arcsec$ resolution in Fig.~\ref{fig:robustmaps} (upper image), and with natural weighting at 18$\arcsec$ resolution in Fig.~\ref{fig:naturalmaps} (upper image).

The integrated flux density of the JVLA data at 3.1\,GHz is approximately 120\,mJy.
This is comparable to the Effelsberg observation at 2.6\,GHz (Table~\ref{NGC628integratedflux}), demonstrating that the flux density calibration was successful and that there is no significant flux density loss due to the missing small spacings of the JVLA.

Weak radio emission surrounds NGC\,628 and should be mostly nonthermal in nature (Fig.~\ref{fig:naturalmaps}, upper panel). Two main radio spiral arms are present,
both coiling out in a counter-clockwise fashion. One spiral arm dominates the north and east of the galaxy (``outer arm'') with the most intense emission located at RA(J2000) = 01$^\mathrm{h}$ 36$^\mathrm{m}$ 38$^\mathrm{s}$, DEC(J2000) = +15$\degr$ 48$\arcmin$ 47$\arcsec$, corresponding to a cluster of $\HII$ regions which are unresolved in the radio image.
This region shows strong emission in H$\alpha$ \citep{Kennicut2003} and UV \citep{2001ApJS..132..129M}, in addition to $\HI$ \citep{2008AJ....136.2563W}. The outer arm branches out into two components in the east (Fig.~\ref{fig:robustmaps}, upper image).

The second main spiral arm (``inner arm'') travels counter-clockwise from north to the south; several $\HII$ complexes are located along the radio spiral arm (Fig.~\ref{fig:naturalmaps}, upper panel).  These so-called 'beads on a string' are bright knots embedded in diffuse emission tracing the spiral pattern. Many of the UV-bright knots are also bright in H$\alpha$ indicating recent (less than 100\,Myrs) star formation and the radio continuum matches very well to the brightest of these knots. These are especially evident in Fig.~\ref{fig:628_StokesI8arcsec} where the resolution is high enough to resolve these individual knots. The brightest of these knots are located on the inside edges of the spiral arms formed by the diffuse UV continuum.
The youngest and most massive stars are located at the inner edge of the pattern \citep{1992ApJ...395L..41C,Cornett94,2001ApJS..132..129M} which are the knots observed at 3.1\,GHz.
The inner arm continues to the west where the flux density of the arm suddenly decreases and becomes indistinguishable from the radio envelope of the galaxy, with exception of the $\HII$ regions located at RA(J2000) = 01$^\mathrm{h}$ 36$^\mathrm{m}$ 29$^\mathrm{s}$, DEC(J2000) = +15$\degr$ 48$\arcmin$ 50$\arcsec$.

To the west of the galaxy, three narrow radio filaments are seen (marked Wa,Wb and Wc on Fig.~\ref{fig:robustmaps}, upper image), all with a width of approximately 900\,pc. The most southern filament is an extension of the inner arm. No major $\HII$ regions exist in these regions and therefore the emission is most likely nonthermal.

Unlike most nearby galaxies observed in radio continuum, for example M\,51 \citep{2014A&A...568A..74M}, IC\,342 \citep{2015A&A...578A..93B} and NGC\,6946 \citep{2007A&A...470..539B}, NGC\,628 shows no trace of a central source, while this is similar to the flocculent galaxy NGC\,4414 \citep{2002A&A...394...47S}.

\subsection{Comparison to $\HI$ observations \label{STOKESIHIHOLES}}

\begin{figure}
	\centering
	\includegraphics[width=0.9\columnwidth]{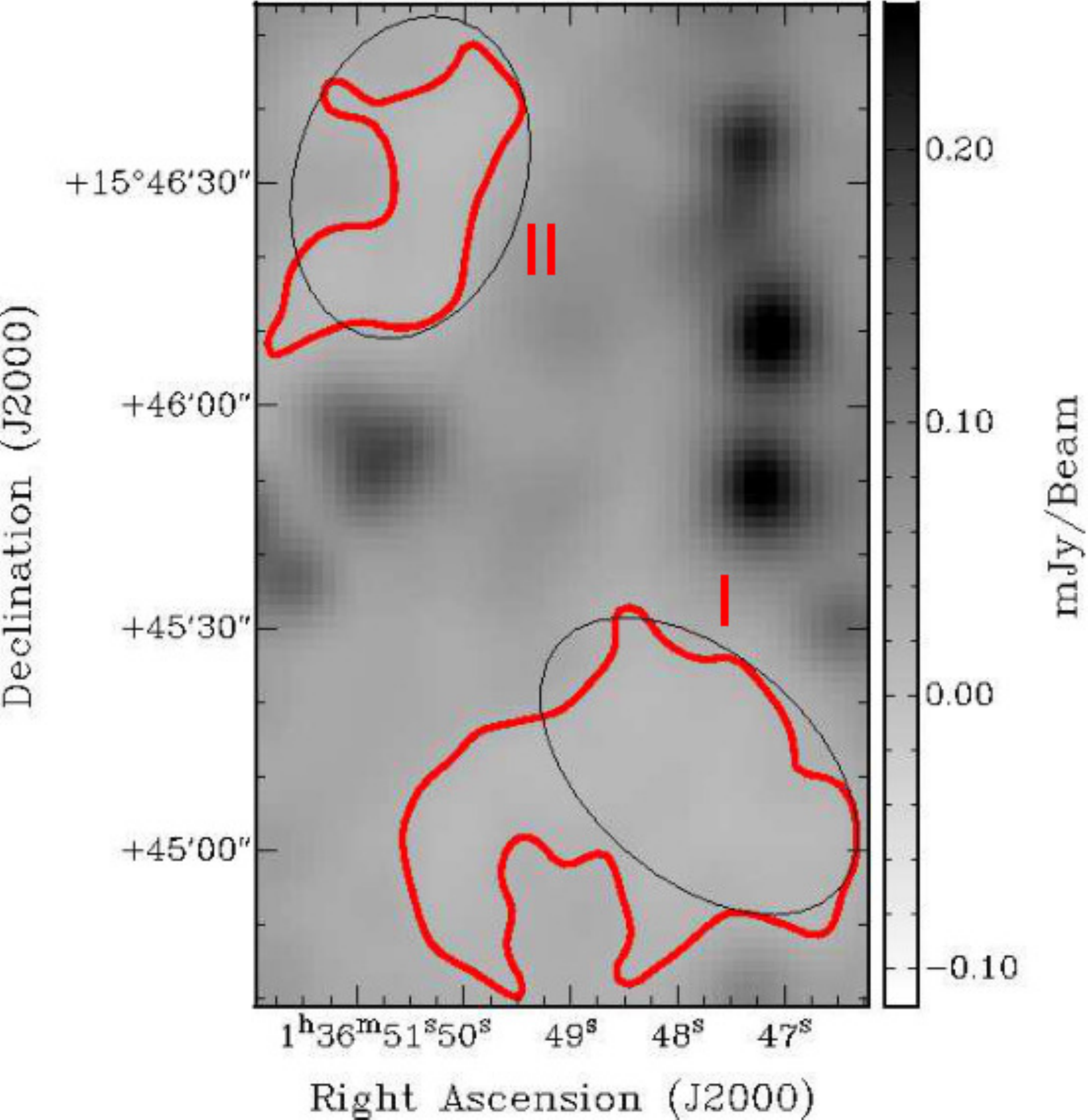}
	\caption{Overlay of two prominent $\HI$ holes on the radio continuum image (Fig.~\ref{fig:robustmaps}) at 10$\arcsec$ resolution. The contour level is at 21\,$\mu$Jy/beam which is approximately 3$\sigma$.}
	\label{fig:HIradiohole}
\end{figure}

Throughout the disk, there are regions where little to no radio emission is seen, most notably the holes located at RA(J2000) = 01$^\mathrm{h}$ 36$^\mathrm{m}$ 48$^\mathrm{s}$, DEC(J2000) = +15$\degr$ 45$\arcmin$ 10$\arcsec$ and RA(J2000) = 01$^\mathrm{h}$ 36$^\mathrm{m}$ 50$^\mathrm{s}$, DEC(J2000) = +15$\degr$ 46$\arcmin$ 27$\arcsec$ (shown in Fig.~\ref{fig:naturalmaps} \& Fig.~\ref{fig:HIradiohole} as Roman numerials). These holes in radio emission correspond to $\HI$ holes detected by \cite{2011AJ....141...23B} (Fig.~\ref{fig:HIradiohole}).
Several more holes can be seen in the west of the galaxy. A list of $\HI$ holes corresponding to holes in radio continuum  and their physical parameters (from \cite{2011AJ....141...23B}) is given in Table~\ref{HIholetable}.
All the holes seen are of type~1 which is a hole where the gas has been completely blown out of the disk of the galaxy. Approximately 75$\%$ of all $\HI$ holes are type~1.

Such holes in total radio continuum corresponding to an $\HI$ hole have been seen previously in NGC\,6946 by \cite{2007A&A...470..539B} who
mentioned that the association between an $\HI$ hole and radio continuum is rare and that the lack of radio emission could be due to two main reasons. The first is that the superbubble creating the $\HI$ holes, which are driven by multiple supernova explosions, are sweeping away the gas and magnetic field and thereby creating a locally weak magnetic field. The second explanation is that the superbubble carries the magnetic field vertically into the halo along with the hot gas \citep{1988LNP...306..155N}. As NGC\,628 is nearly exactly face-on compared to NGC\,6946 ($i=33\degr$), these vertical fields would be along the line of sight, therefore producing no observable synchrotron emission but possibly strong Faraday rotation (see Sect~\ref{subsec:HIholesection}).

The larger number of $\HI$ holes corresponding to holes in the radio continuum in NGC\,628 compared to NGC\,6946 could be an effect of the different inclinations.  When viewing a $\HI$ hole carrying a magnetic field in NGC\,6946 vertical to the disk with its inclination of approximately 33$\degr$, we observe variations of the magnetic field component perpendicular to the line of sight which would give rise to observable synchrotron emission. This would occur less likely for NGC\,628 with its smaller inclination.

The $\HI$ holes seen in NGC\,628 all have a kinetic age greater than 40\,Myrs and should be old enough so that the field configuration has gained a significant vertical offset \citep{2012ApJ...754L..35H}.
However, two holes are significantly older and the vertical shear should have destroyed the vertical offset of the magnetic field. Notably, these holes are the most obvious in our radio continuum image (Fig.~\ref{fig:HIradiohole}). While the vertical shear of NGC\,628 is not known to the authors' knowledge, \cite{heald2007} found that the vertical shear for a sample of three galaxies is approximately 15--25 km s$^{-1}$ per scale height in H$\alpha$, independent of radius. This gives a lower limit of 60\,Myrs for a characteristic shear time. Therefore the first argument is more likely to be true, especially for the two $\HI$ holes in question shown in Fig.~\ref{fig:HIradiohole}, namely that the gas and magnetic field have been radially swept away, leaving a weak local magnetic field.

\begin{table*}
\caption{List of $\HI$ holes corresponding to holes in radio continuum and their physical parameters. t$_\mathrm{kin}$, E$_\mathrm{E}$ \& M$_\mathrm{HI}$ are the kinetic age, energy requirement, and missing $\HI$ mass of the $\HI$ hole
(from \cite{2011AJ....141...23B}).}
\label{HIholetable}
\centering 
\begin{tabular}{c c c c c c} 
\hline\hline 
RA & Dec & Diameter &  t$_\mathrm{kin}$ & log(E$_\mathrm{E})$ & log(M$_\mathrm{HI}$)\\ 
 (h m s)& ($\degr$ \, $\arcmin$ \, $\arcsec$) &  (pc)  &   (Myrs) & (10$^{50}$ ergs) & (10$^{4}$ M$_{\odot}$)\\ 
\hline 
01 36 28.7 &  +15 46 42.0   &   878  &        61   &   3.0   &     2.8 \\
01 36 31.1 &  +15 48 26.9     &  814  &    57   &   2.9    &    2.7 \\
01 36 31.4 &  +15 47 20.9   &  585 &     41    &  2.4   &     2.3 \\
01 36 32.8 &  +15 47 46.4  & 901 &    63   &   3.0    &    2.6 \\
01 36 33.2 &  +15 46 56.8  &  743 &    52   &   2.8    &    2.5 \\
01 36 35.0 &  +15 47 20.8 &  888 &    62   &   3.1   &     2.6 \\
01 36 35.4 &  +15 49 25.3    &   758 &     53  &    2.7   &     2.6 \\
01 36 47.8 &  +15 45 11.4   &  1782  &  125  &    4.1   &     3.4 \\
01 36 50.5 &  +15 46 30.8     &  1573 &   110   &   4.0     &   3.3 \\
\hline
\end{tabular}
\end{table*}

\begin{figure}
	\centering
	\includegraphics[width=0.9\columnwidth]{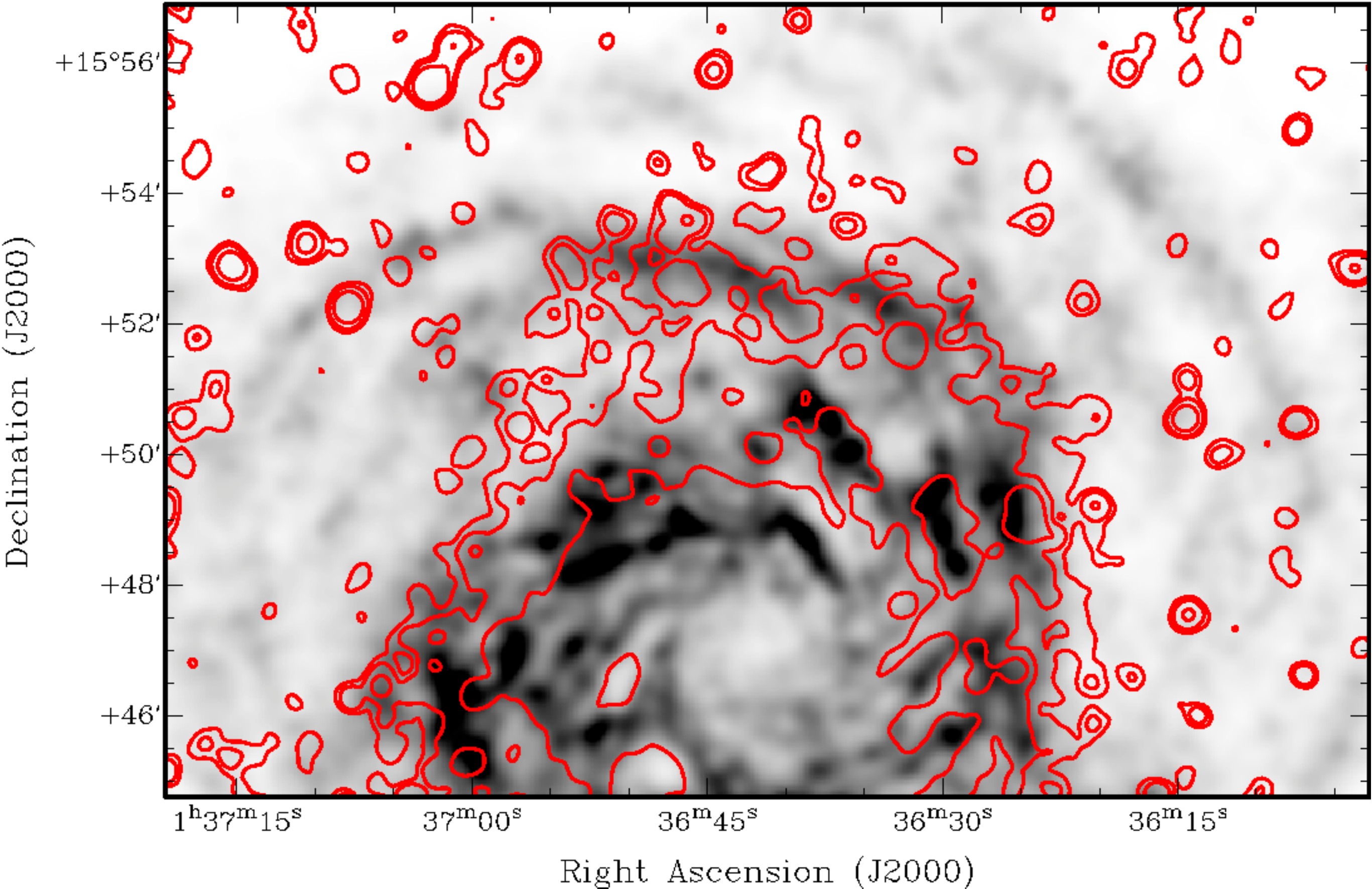}
	\caption{Overlay of the radio continuum emission (red contours) to the north of NGC\,628 onto $\HI$ THINGS data (greyscale), both smoothed to the same resolution of 18$\arcsec$. Contours are at 20, 40, 120\,$\mu$Jy/beam.}
	\label{fig:HIradio}
\end{figure}

To the north of the galaxy, extended, low-level emission is observed (Fig.~\ref{fig:naturalmaps}, top), reaching out to 13\,kpc from the centre of the galaxy. The extension is seen to the north--north-west direction.
Significant $\HI$ emission is present in this region and the radio emission overlays well with the extended northern $\HI$ arm  (Fig.~\ref{fig:HIradio}). Additionally, there are plenty of $\HII$ regions located in the extended disk of NGC\,628. Many of these $\HII$ regions are located on the $\HI$ arm \citep{1998ApJ...506L..19F,2000AJ....120.1306L}. $\HII$ regions in the inner disk usually have a star-formation rate (SFR) of $\approx 10^{3}$ higher than that of the faintest $\HII$ regions in the outer disk. Therefore, it is most likely that we are observing cosmic ray electrons originating from  supernovae occurring at these faint $\HII$ knots from this extreme northern spiral arm.

\subsection{Separation of thermal emission}
\label{separatethermal}

For an accurate determination of the magnetic field strength, separating the two components of continuum emission, namely free-free (thermal) and synchrotron (nonthermal) needs to be performed. There are several methods, the classical approach where a constant nonthermal spectral index is assumed \citep[e.g.][]{1984A&A...135..213K} or using the 24\,$\mu$m infrared emission to directly calculate the thermal emission \citep{2008ApJ...678..828M}. For this work we shall apply the method by \cite{2007A&A...475..133T} where an extinction-corrected H$\alpha$ map is used to estimate the thermal emission.
The continuum-subtracted H$\alpha$ map used was obtained from the ancillary data at the SINGS \citep{Kennicut2003} website. The maps were in units of DN s$^{-1}$ pixel$^{-1}$ which was converted into erg s$^{-1}$ cm$^{-2}$ using the calibration provided in the SINGS Fifth Data Delivery documentation\footnote{\texttt{https://irsa.ipac.caltech.edu/data/SPITZER/\\  SINGS/doc/sings\char`_fifth\char`_delivery\char`_v2.pdf}}.

\begin{figure}
	\centering
	\includegraphics[width=0.9\columnwidth]{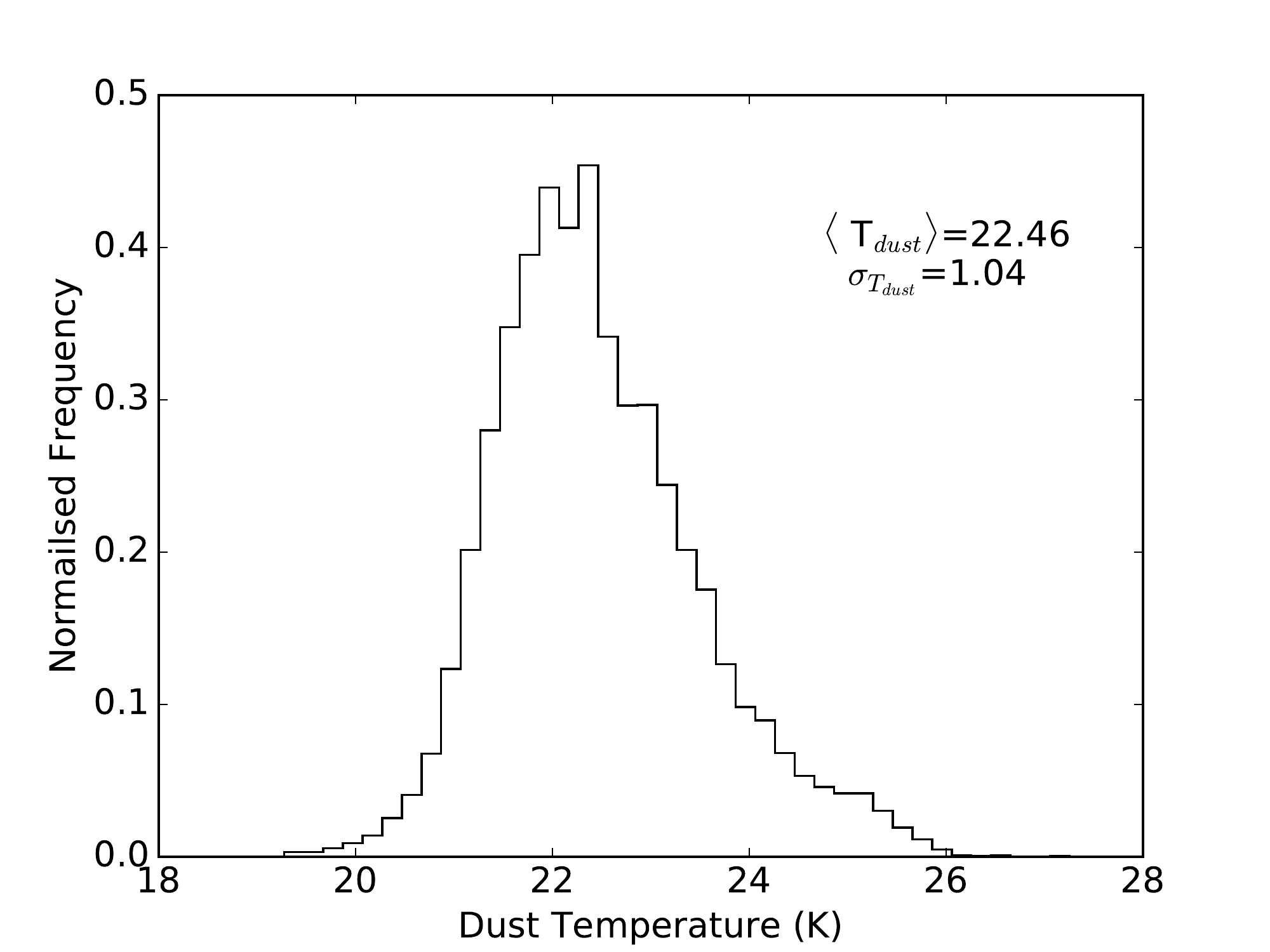}
	\caption{Histogram displaying the pixel-wise distribution of dust temperatures.}
	\label{fig:dusttemp}
\end{figure}

First of all, the dust temperature was calculated using 70 and 160\,$\mu$m Herschel maps \citep{2011PASP..123.1347K}. All maps were smoothed to a resolution of 18$\arcsec$ and normalised to a common grid. Both IR maps were calibrated in surface brightness units of MJy\,sr$^{-1}$. The mean dust temperature across the galaxy was found to be 22.4\,K, very similar to NGC\,6946 with a mean dust temperature of 22.3\,K \citep{2012MNRAS.419.1136B}. A histogram showing the distribution of dust temperatures is shown in Fig.~\ref{fig:dusttemp}.

In the brightest $\HII$ regions, the dust temperature is 25\,K, in the central region and areas of the spiral arms 23\,K, and in other regions approximately 21\,K.
From T$_\mathrm{dust}$, the optical depth at 160\,$\mu$m was derived from Eq.~(2) in \cite{2007A&A...475..133T}. The H$\alpha$ optical depth was calculated using the equation $\tau_\mathrm{H\alpha} \approx f_\mathrm{d} \times 2200 \times \tau_{160\mu m}$ \citep{2003pid..book.....K} assuming an H$\alpha$ filling factor of 0.33 \citep{2003MNRAS.341..369D}.
This optical depth was then used to de-redden the H$\alpha$ flux density, using Eq.~(3) in \cite{2007A&A...475..133T}.

From the de-redded H$\alpha$ map, the emission measure (EM) was found using Eq.~(4) in \cite{2007A&A...475..133T}, assuming an electron temperature of 10$^{4}$\,K \citep{1998PASA...15..111V}.
Eqs.~(5) and (6) from \cite{2007A&A...475..133T} were used to calculate the continuum optical depth and brightness temperature T$_{\mathrm{B}}$. The thermal flux density was then obtained from T$_{\mathrm{B}}$, using the equation in \cite{2012MNRAS.419.1136B}:

\begin{equation}
\frac{S_\mathrm{\nu,th}}{\mathrm{Jy\,beam^{-1}}} = 8.18 \times 10^{-7}\left(\frac{\theta_\mathrm{maj}}{\mathrm{arcsec}}\right)\left(\frac{\theta_\mathrm{min}}{\mathrm{arcsec}}\right)\left(\frac{\nu}{\mathrm{GHz}}\right)\left(\frac{T_\mathrm{B}}{\mathrm{K}}\right),
\end{equation}

where $\theta_\mathrm{maj}$ and  $\theta_\mathrm{min}$ are the major and minor axis of the synthesised beam, respectively.

\begin{figure*}
\centering
    \subfloat{\includegraphics[width=0.45\textwidth]{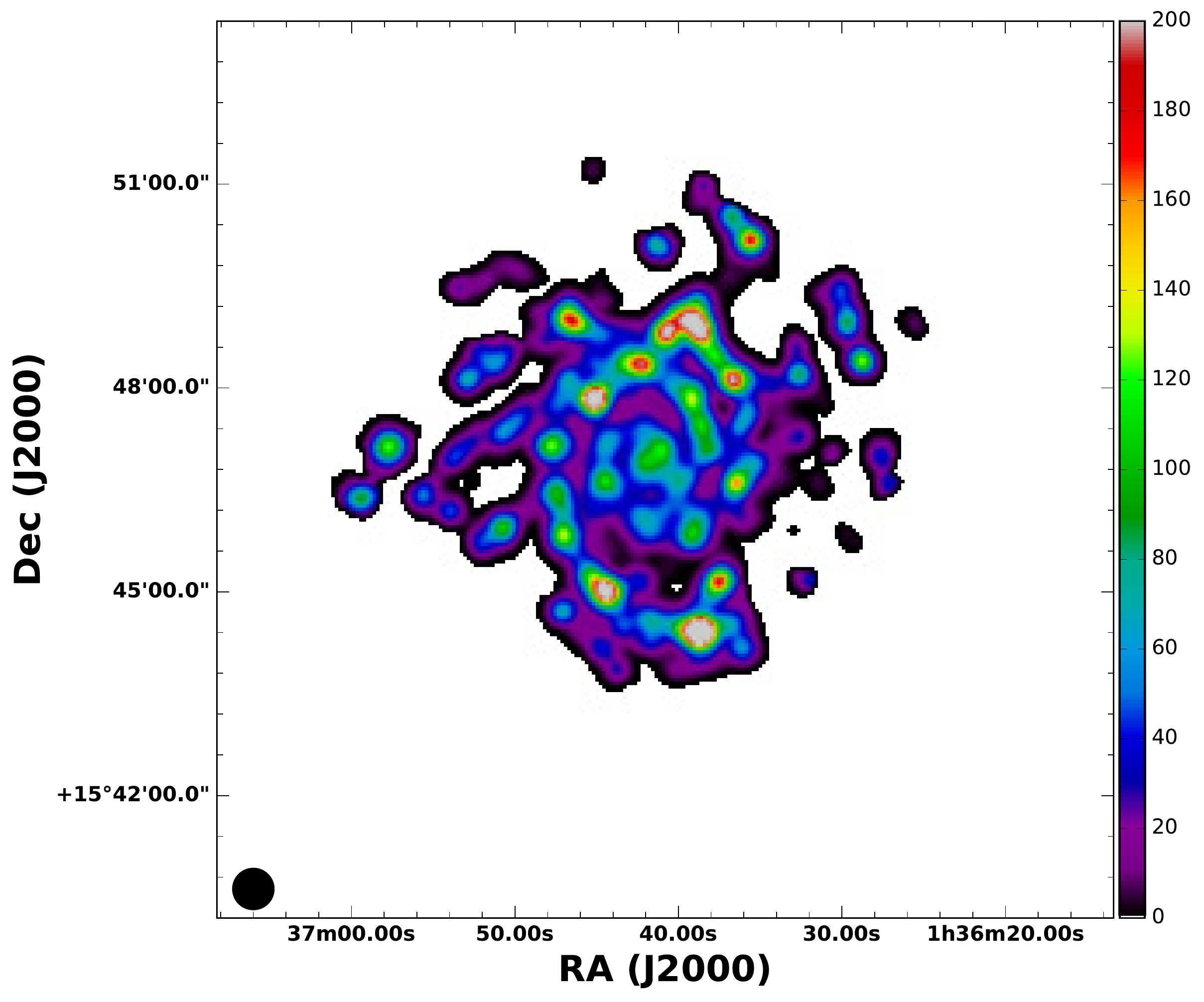}}
    \subfloat{\includegraphics[width=0.45\textwidth]{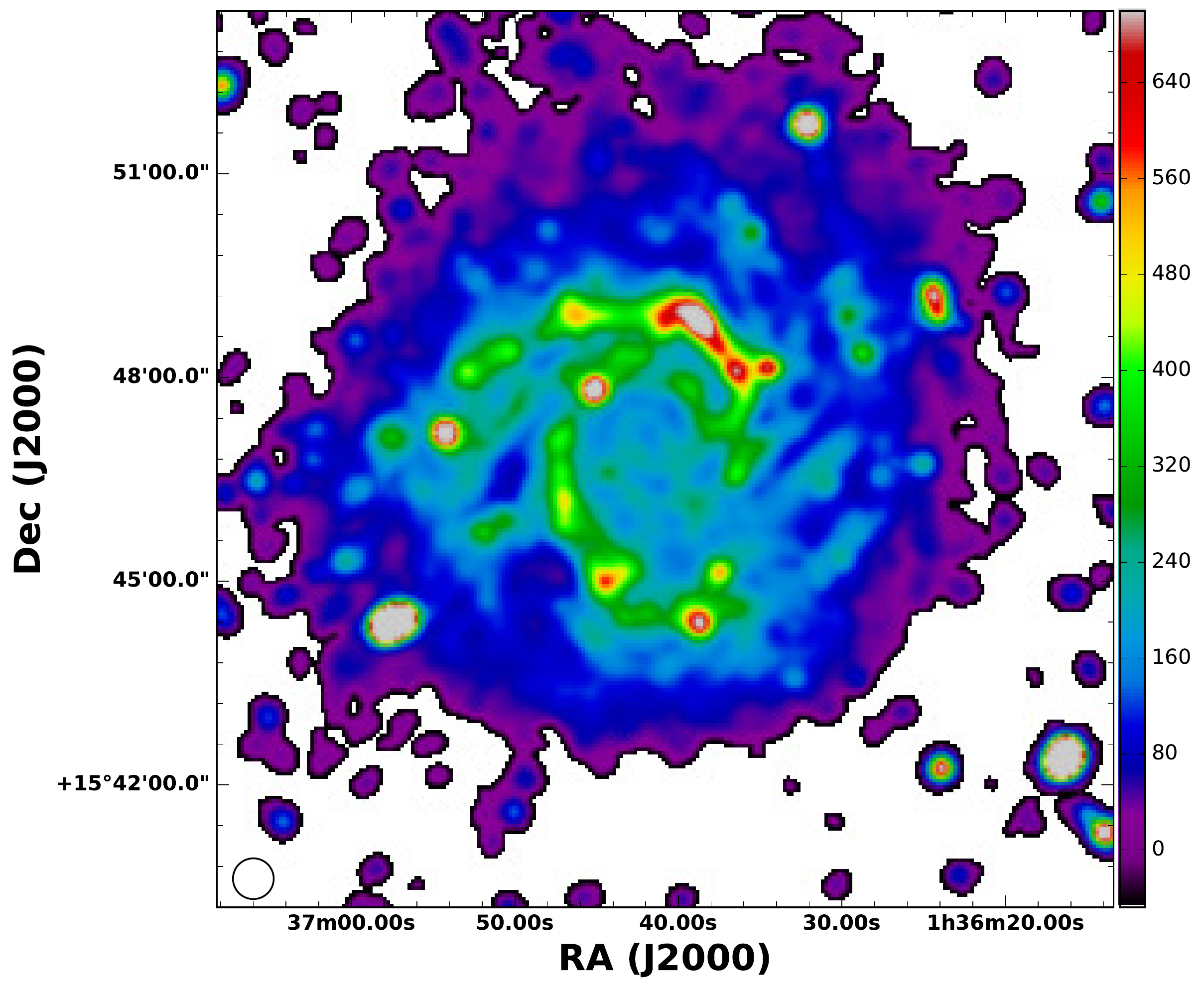}}
  \caption{Images of thermal (left) and nonthermal intensity (right) of NGC\,628 at 3.1\,GHz at 18$\arcsec$ resolution. Colour scale is in $\mu$Jy/beam.}
  \label{fig:thermmaps}
\end{figure*}

Finally, the nonthermal map at 3.1\,GHz is obtained by subtracting the thermal emission map from the JVLA 3.1\,GHz naturally weighted map (Fig.~\ref{fig:naturalmaps}, top). When calculating the nonthermal map for regions where there is no observable dust emission, we set the thermal fraction to zero. This assumption is not entirely true as there could be undetected dust IR emission present in the Hershel map.
Both the thermal and nonthermal maps of NGC\,628 at 3.1\,GHz are shown in Fig.~\ref{fig:thermmaps}.

\begin{figure}
	\centering
	\includegraphics[width=0.45\textwidth]{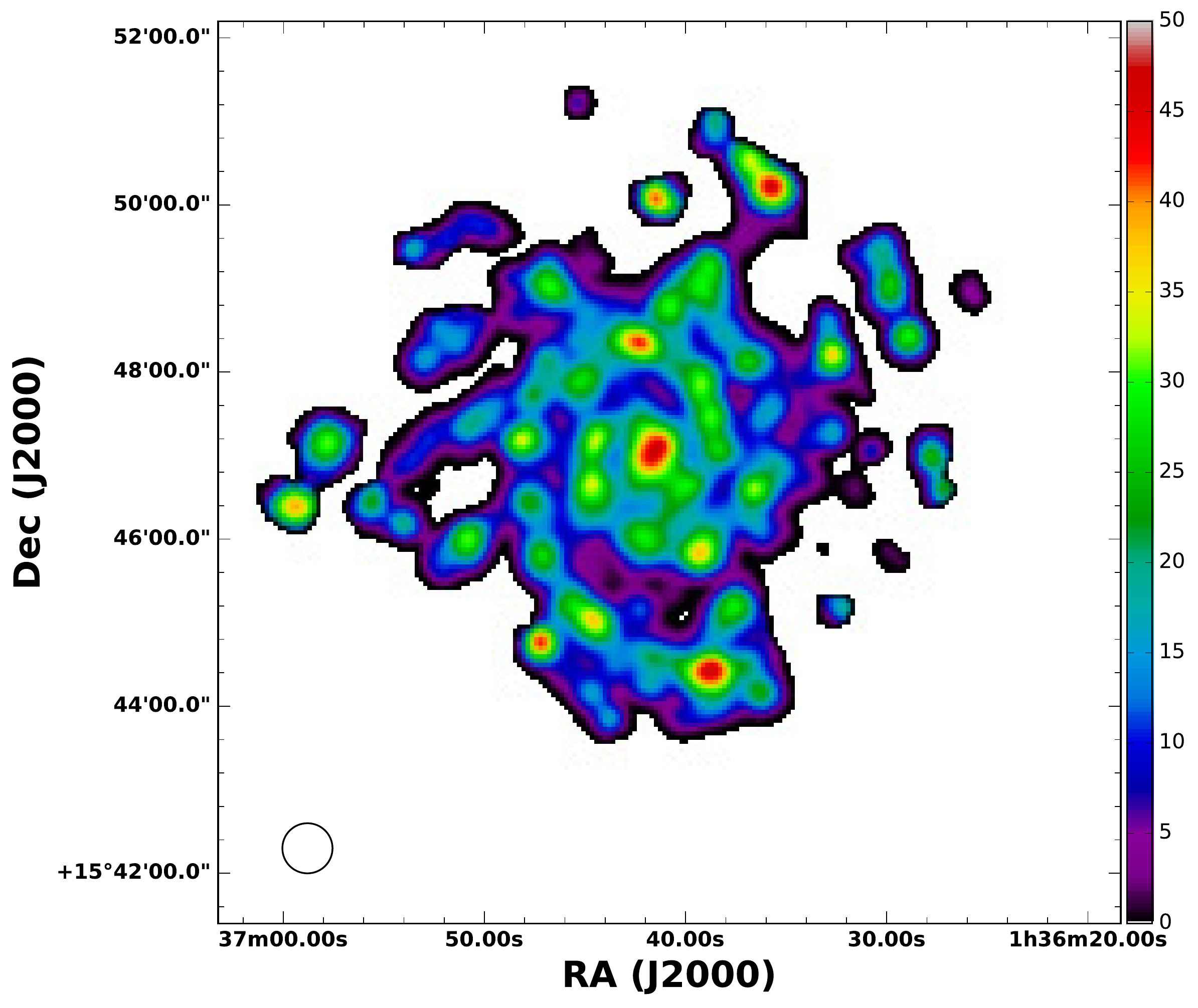}
	\caption{Map of the thermal fraction at 3.1\,GHz, computed from the total intensity and thermal images shown in Fig.~\ref{fig:naturalmaps} \& Fig.~\ref{fig:thermmaps} (left). Colour scale is in percent.}
	\label{fig:thermalfraction}
\end{figure}

The map of the thermal fraction created from the nonthermal and thermal maps of NGC\,628 is shown in Fig.~\ref{fig:thermalfraction}. The spiral arms show a thermal fraction of 10--20$\%$, with typical $\HII$ regions showing 20--30$\%$ and the largest $\HII$ regions having thermal fractions greater than 40$\%$.

The very centre of the galaxy is observed to have thermal fractions that reach up to 47$\%$, indicating that a significant fraction of the emission in this region is of thermal origin. This could be explained by a lack of cosmic ray electrons originating from supernovae in this region, comparable to the central region of M\,31 where \cite{Tabatabaei2013} found the thermal fraction to be approximately 20$\%$ at $\lambda$20cm due to weak synchrotron emission caused by a lack of cosmic ray electrons (CREs).

\cite{Cornett94}, utilising both far and near UV photometry images, found that NGC\,628's central region shows no significant population of OB stars. They also observed that the azimuthally averaged scale lengths for the UV continuum emission decreases with increasing wavelength. However, the UV profiles are clearly non-exponential, despite the approximately exponential behaviour of the R-band profile. Analysing these colour gradients, \cite{Cornett94} concluded that they reflect the star formation history rather than metallicity or internal extinction.
Together with findings from \cite{2001ApJS..132..129M}, they conclude that the entire disk has undergone active star formation within the past 500\,Myr but the inner regions have experienced a more rapidly declining star formation than the outer regions.

The Initial Mass Function (IMF) seems to be universal \citep{2001MNRAS.322..231K,2002Sci...295...82K}, while its form has only been determined directly on star cluster scales.
This canonical IMF has traditionally been applied on galaxy-wide scales and has to be constructed by adding all young stars of all young star clusters \citep{2003ApJ...598.1076K,2006MNRAS.365.1333W}. This integrated galactic initial mass function (IGIMF) is steeper than the usual IMF in star clusters and steepens with decreasing total SFR \citep{2005ApJ...625..754W,2007ApJ...671.1550P}. This is due to the combination of two effects. The first is that the most massive star in a star cluster is a function of the total stellar mass of the young embedded star cluster \citep{2005ApJ...625..754W}. The second is that the most massive young embedded star cluster is a function of the total SFR of a galaxy \citep{2004MNRAS.350.1503W}. Similar to the IMF in star clusters, the embedded cluster mass function (ECMF), which describes the mass spectrum of newly formed star clusters, follows a power-law distribution function in galaxies \citep{2003ARA&A..41...57L}.
Therefore low-mass clusters do not contain massive stars, and this yields an IGIMF that depends on the SFR, since the most massive cluster that can form depends on the SFR.

For NGC\,628, we see UV emission from the central region from the B stars which are still there and can ionise the gas and give rise to H$\alpha$ emission.This is also observed in M31 as high mass stars appear to be ruled out as the primary source for H$\alpha$ emission in the inner region \citep{Devereux1994}. However, the current IGIMF lacks O stars if the SFR is very low, such that Type II supernova
explosions will not be occurring. No radio synchrotron emission is visible because only O stars produce Type II supernovae which in turn generate CREs emitting radio synchrotron emission.

Star formation continues in the spiral arms where we have both O and B stars, both contributing to the UV emission and giving rise to radio synchrotron emission. In the outer regions of the galaxy, we observe the usual H$\alpha$ cut-off but extended UV emission still exists. There the SFR is so low that the small clusters do not contain massive stars and thus there is neither H$\alpha$ nor radio emission \citep{2008Natur.455..641P}.  The extended northern
spiral arm is seen in both H$\alpha$ and UV emission but is much weaker compared to the rest of the galaxy.

Our observations indicate that the IGIMF in the central region is steeper compared to the rest of the galaxy. NGC\,628 is an ideal galaxy to apply an IGIMF model with radial variation.

If there is no significant population of O stars, one cannot expect supernovae to inject fresh CREs. Given the nature of diffusion without fresh injection, a region of smooth radio continuum emission would be expected after sufficient time.
The remaining cosmic ray electrons from the previous star formation period would have aged and diffused  around the central region, creating a flat gradient and a steep spectrum in these regions. 
This explanation would require some CREs present to emit synchrotron emission and therefore that the CRE lifetime to be greater than the lifetime of O stars. Using the total magnetic field strengths found in Sec.~\ref{subsec:magfield} of 8 - 10\,$\mu$G for the central region we find the CRE lifetime to be between 20-30\,Myrs. This is found to be several times greater compared to stellar evolution models of O stars e.g. found in \cite{Weidner2010} where a 120\,M$_{\odot}$ O type star can evolve to a carbon-rich Wolf-Rayet star in 3\,Myrs, one evolutionary stage before going supernovae. For a lower mass of 20\,M$_{\odot}$ it would 9\,Myrs to reach the red super giant phase, again one evolutionary stage before going supernovae.

Low-frequency observations will greatly help along with CRE modelling with time-dependent injection profiles. This is a topic of further investigation.

\subsection{Magnetic field strength of NGC\,628 \label{subsec:magfield}}

\begin{figure}
	\centering
	\includegraphics[width=0.45\textwidth]{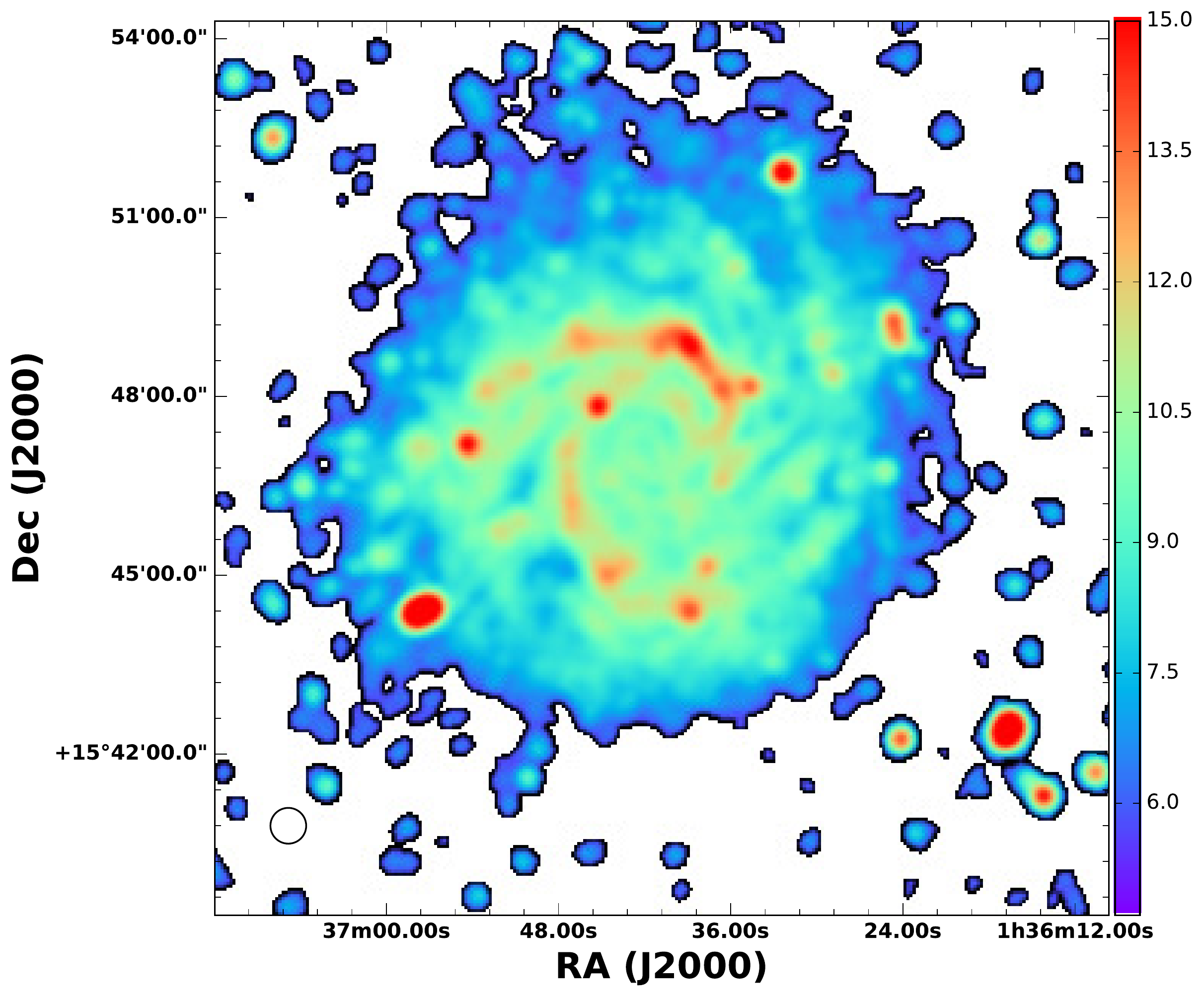}
	\caption{Strength of the total magnetic field at 18$\arcsec$ resolution, in units of $\mu$G, determined by assuming energy equipartition.}
	\label{fig:bfieldmap}
\end{figure}

The total magnetic field strength of NGC\,628 can be determined from the nonthermal emission by assuming equipartition between the energy densities of cosmic rays and magnetic field, using the revised formula of \cite{2005AN....326..414B}. The total magnetic field strength scales with the synchrotron intensity I$_\mathrm{syn}$ as:
\begin{equation}
B_\mathrm{tot,\perp} =I_\mathrm{syn}/((K_0+1)\,L)^{\,\,\,1/(3-\alpha_\mathrm{n})}
\end{equation}
where B$_\mathrm{tot,\perp}$ is the strength of the total field perpendicular to the line of sight. Further assumptions required are the synchrotron spectral index of $\alpha_\mathrm{n}$ = -1.0 and the effective path length through the source of $L = 1000$\,pc$/\cos{i} \simeq 1007$\,pc where $i$ is the inclination of the galaxy. We also assumed that the polarised emission emerges from ordered fields in the galaxy plane.
The adopted ratio of CR proton to electron number densities of $K_0 = 100$ is a reasonable assumption in the star-forming regions in the disk \citep{1978MNRAS.182..443B}. Realistic uncertainties in $L$ and $K_0$ of a factor of about two would effect the result only by about 20\%. The effect of adjusting $\alpha$ to between -0.7 and -0.9 produces an error of less than 5\% in magnetic field strength. Using these assumptions, we created an image of the total magnetic field in NGC\,628 from the nonthermal map (Fig.~\ref{fig:thermmaps}, right) which is shown in Fig.~\ref{fig:bfieldmap}.

The mean total magnetic field strength is around 9\,$\mu$G and 11--12\,$\mu$G in the spiral arm regions (Fig.~\ref{fig:bfieldmap}).
The largest and brightest star-forming complexes seen across the spiral arms have a total magnetic field strength around 14\,$\mu$G, with a maximum strength greater than 15\,$\mu$G in the star-forming complex in the north. In the extended disk of the galaxy (a galactic centric radius of $\sim$ 8.2\,kpc) we observe a magnetic field strength of approximately 8\,$\mu$G.

NGC\,628 has a similar magnetic field strength as NGC\,6946 but a somewhat smaller field strength compared to IC\,342 \citep{2015A&A...578A..93B}, where the mean magnetic field strength is 14--15\,$\mu$G in the spiral arms. The main difference between NGC\,628 and other galaxies in this respect is the central region, with a lower total magnetic field strength of 9-10\,$\mu$G compared to the central regions of NGC\,6946 (about 25\,$\mu$G) \citep{2007A&A...470..539B} and of IC\,342 (about 30\,$\mu$G) \citep{2015A&A...578A..93B}.

This method of calculating the total magnetic field strength most likely causes an underestimation of the magnetic field strength especially in the centre of the galaxy and extended disk due to the uncertainty of $K_0$ caused by ageing of CREs propagating into regions of low star formation.
Ideally, low-frequency observations would be far more useful in determining the magnetic field strengths, not being contaminated by thermal emission \citep{2014A&A...568A..74M}.

\subsection{Polarised intensity}

Images of linearly polarised intensity with overlaid B-vectors at $10\arcsec$ and $18\arcsec$ resolutions are shown in the lower panels of Figs.~\ref{fig:robustmaps} and \ref{fig:naturalmaps}. These maps were obtained by averaging Q and U over all channels, without correction for the effects of Faraday rotation between the channels.

Ordered magnetic fields are traced by polarised synchrotron emission and form spiral patterns in nearly every galaxy \citep{2016A&ARv..24....4B},
even in flocculent \citep{2002A&A...394...47S} and ring galaxies \citep{2008ApJ...677L..17C}.
In NGC\,628, three polarisation arms are prominent and are labeled in Fig~\ref{fig:robustmaps}.
These arms resemble the arms observed in IC\,342 \citep{1993IAUS..157..305K,2015A&A...578A..93B} and NGC\,6946 \citep{2007A&A...470..539B}.

\begin{figure}
	\centering
	\includegraphics[width=0.8\columnwidth,trim={5.0cm 0 5.0cm 0},clip]{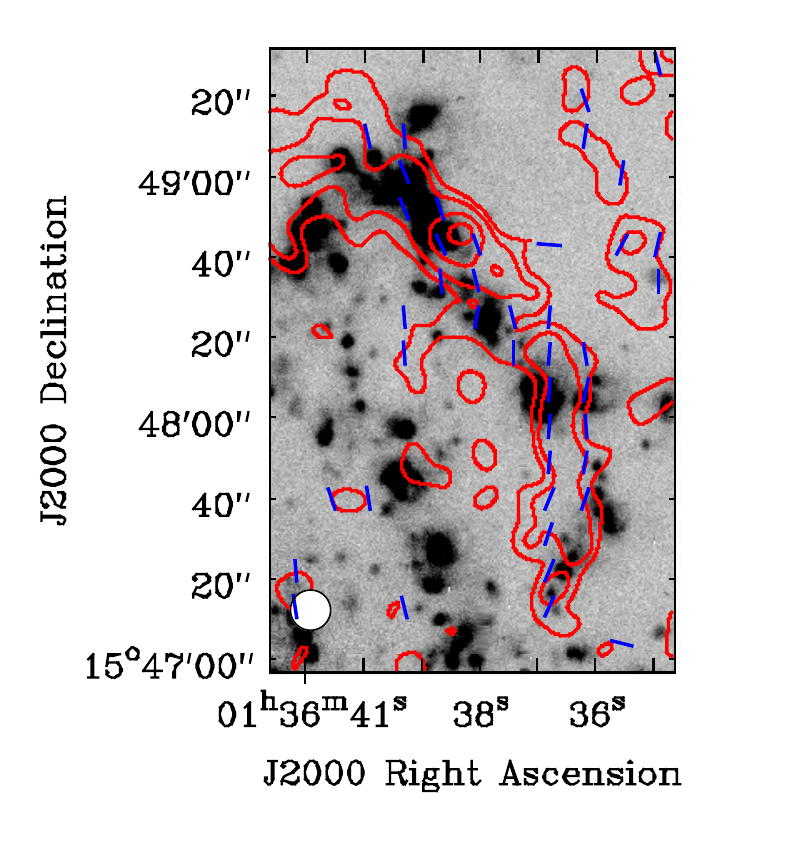}
	\caption{Polarised emission following the largest $\HII$ complexes in the galaxy. Polarised intensity and magnetic vectors are overlaid onto an H$\alpha$ image \citep{Dale2009}. Contours of polarised intensity are at 9, 15, 24, and 30\,$\mu$Jy/beam. The resolution of the JVLA image at 10$\arcsec$ is illustrated by the ellipse located in the bottom left corner.}
	\label{fig:HIIpol}
\end{figure}

The first and most prominent polarisation spiral arm (arm~1) runs counter-clockwise, starting along the large $\HII$ regions in the north-west, which are seen in total intensity (Fig.~\ref{fig:628_StokesI8arcsec}).
This indicates that a fraction of the isotropic turbulent field is compressed or sheared and has become anisotropic turbulent. The region at the most northern point has the brightest polarised flux density and several interesting features makes it stand out from the rest of the galaxy. This region will be discussed in more detail in Sect.~\ref{subsec:faradaywindow}.
Arm~1 spreads out into the inter-arm region of the galaxy in the south-east and continues to the south. The pitch angle of the polarisation vectors decreases along arm~1, from about 50\degr in the north-west to about 30\degr in the east and south (see Table~\ref{tab:pitchangletable}).

The spiral arm features in polarised intensity are located mostly between the optical arms,
except in the inner (north-western and northern) parts of the polarisation arm~1
(at approximately RA(J2000) = 01$^\mathrm{h}$ 36$^\mathrm{m}$ 37--41$^\mathrm{s}$; DEC(J2000) = +15$\degr$ 47$\arcmin$--49$\arcmin$). Here,
the polarised emission closely follows the $\HII$ complexes (Fig.~\ref{fig:HIIpol}).
At small radii the polarised emission is located at the inner edge of the optical spiral arm and then crosses the ridge line delineated by $\HII$ regions to the outer edge of the arm. Finally, the polarisation coincides with a bright $\HII$ complex.

The second and less prominent polarisation arm (arm~2) begins south-east of the centre of the galaxy and travels to the south-west. Arm~2 is narrower than arm~1. Another narrow arm (arm~3) begins in the south and continues into the extended disk in the north-west. The pitch angle of the polarisation vectors in arms~2 and 3 are large (about 45\degr) and do not vary significantly with increasing distance from the galaxy's centre (see Table~\ref{tab:pitchangletable}). Arms~2 and 3 resemble magnetic arms \footnote{For polarised arms to be considered magnetic arms they must follow the required criteria as defined by \citet{2015A&A...578A..93B}: \begin{itemize} \item exist entirely in the interarm region of the galaxy, i.e between the optical arms \item be narrow ($\approx$ 1\,kpc) and filamentary with an almost constant pitch angle \item have a high degree of polarisation \end{itemize}}. 

\begin{figure}
	\centering
	\includegraphics[width=0.75\columnwidth,trim={6cm 0 6cm 0},clip]{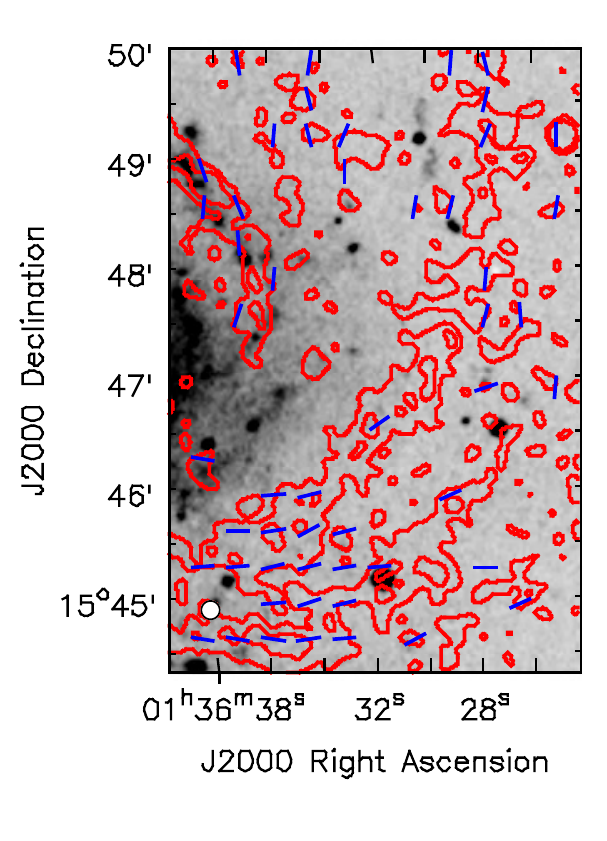}
	\caption{Close-up of the magnetic arms located in an interarm region shown with polarised intensity and magnetic vectors overlaid onto an optical DSS image. Contours of polarised intensity are at 10, 20 $\mu$Jy/beam. The resolution of the JVLA image at 10$\arcsec$ is illustrated by the ellipse located in the bottom left corner.}
	\label{fig:magneticarms}
\end{figure}

They are clearly offset from the optical arm (Figs.~\ref{fig:robustmaps} and \ref{fig:magneticarms}), narrow (about 1.1\,kpc), and show a high degree of polarisation to the total intensity (on average 25\% and up to 40\% locally).

\begin{figure}
	\centering
	\includegraphics[width=1.0\columnwidth]{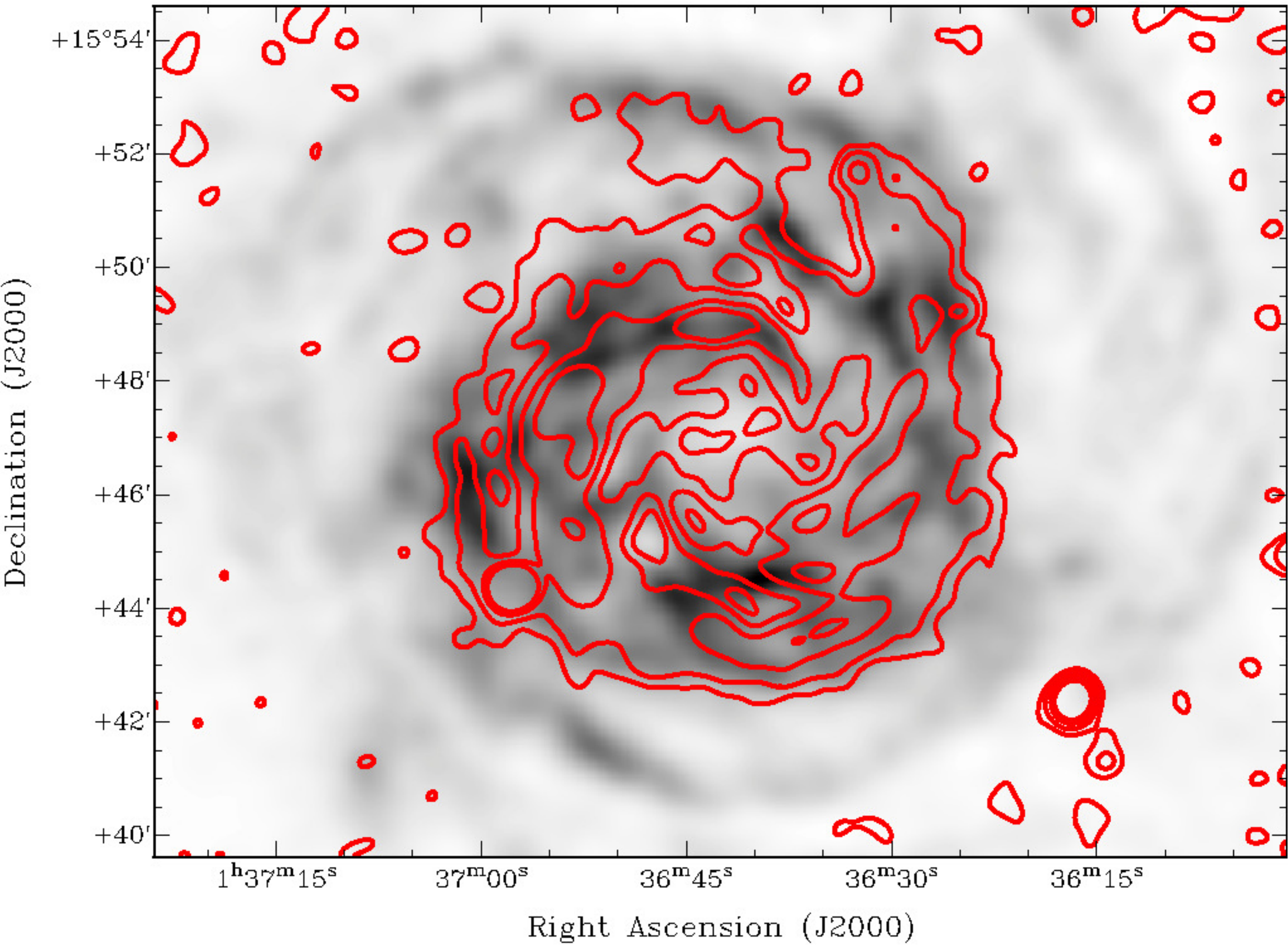}
	\caption{Overlay of the polarised radio continuum emission onto the $\HI$ THINGS data, both smoothed to the same resolution of 30$\arcsec$. Contours are shown at 40, 80, 120\,$\mu$Jy/beam.}
	\label{fig:HIradiopi}
\end{figure}

Both polarisation spiral arms are partly coincident with the $\HI$ gas (THINGS data, \citet{2008AJ....136.2563W}), as shown in Fig.~\ref{fig:HIradiopi}. Given that the Faraday depolarisation at this frequency is small (Fig.~\ref{fig:628depolarisationmap}) and the inclination is low, we do not expect to detect any more polarisation arms. The polarisation arms may extend even further into the outer disk, especially the polarisation (magnetic) arm~3 in the north-west.

We find degrees of polarisation for the inner polarisation arms between 10--25\% at small radii, increasing to 50 \% at larger radii.
This indicates an exceptionally ordered field. These degrees of polarisation are comparable to those at 1.5\,GHz measured by \cite{2009A&A...503..409H} while we are able to detect significantly more polarisation, especially in the northern region of the galaxy, due to the smaller amount of depolarisation at 3.1\,GHz.

A number of background sources are seen in polarisation, those closest to the phase centre are shown in Table~\ref{tab:discretesources}. These sources are used to estimate the Faraday depth for the Galactic foreground in Sect.~\ref{subsec:Milkywayforeground}.

\subsection{Radial scale lengths of the radio continuum emission}

The observed extent of disk emission is sensitivity limited. To characterise the emission along the disk we use the exponential scale length $l$, that is $I_{\nu}  \propto \mathrm{exp}(-r/l)$ where $r$ is the galactocentric radius.

The radial profile of NGC\,628 at 3.1\,GHz was taken from the total, nonthermal, thermal and polarised intensities averaged in concentric rings with the position angle of the major axis and the
inclination of the galaxy taken into account, using the values from Table~\ref{physicalpara}. Surrounding background
point sources were removed by fitting and subtracting Gaussians before measuring the radial profile.
Several background point sources located behind the inner disk were blanked out.

\begin{figure}
\resizebox{\hsize}{!}{\includegraphics{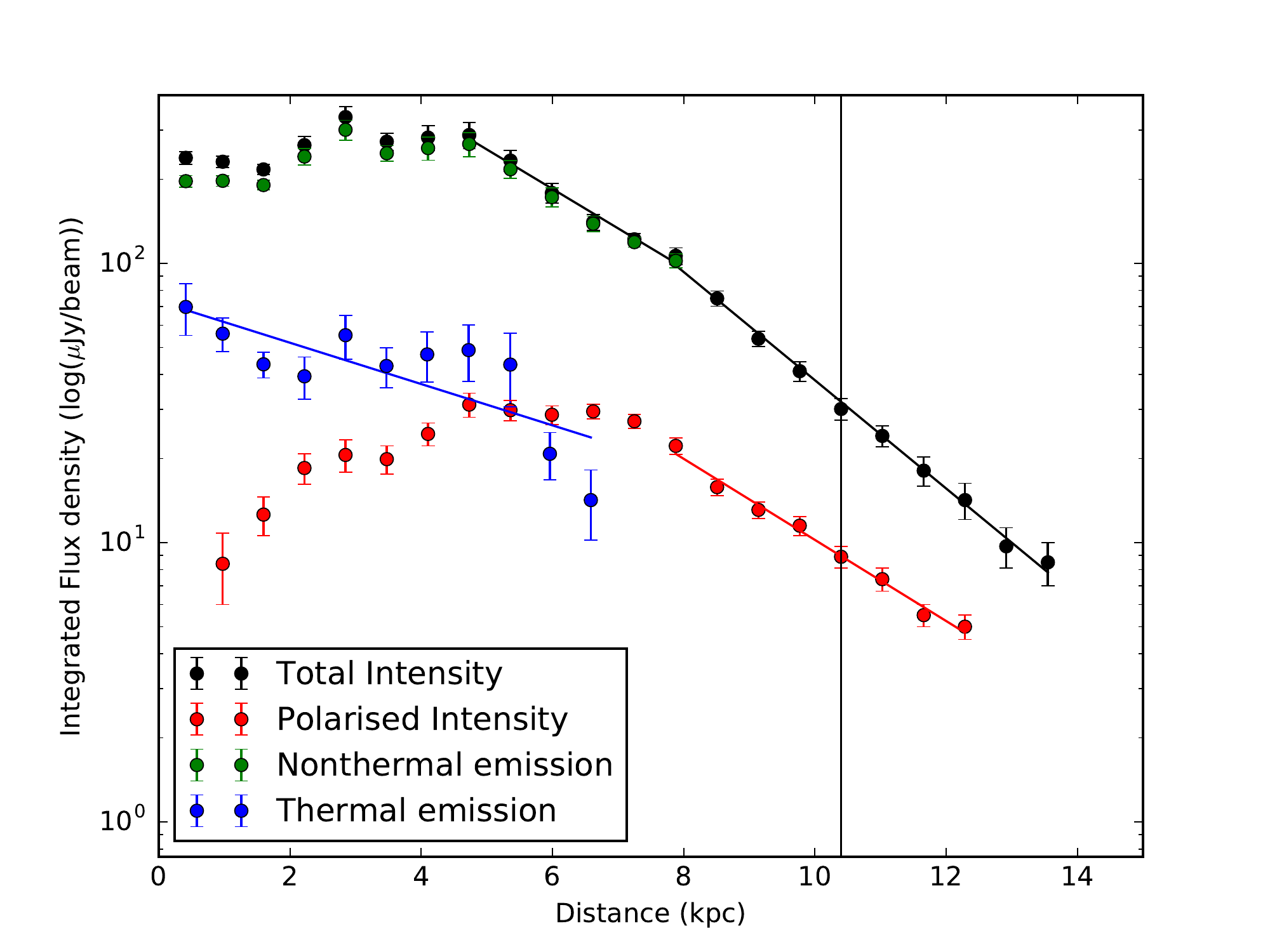}}
\caption{Radial profile of NGC\,628 for the total, nonthermal, thermal, and polarised radio intensities.
The vertical line marks R$_{25}$.}
\label{stokesiradial}
\end{figure}

Fitting a single exponential profile is not possible, as a change in the slope of the total, nonthermal and polarised emission occurs at several points in the radial profile (Fig.~\ref{stokesiradial}). One such break in the slope occurs at approximately 6.5\,kpc which is approximately the radius where the star-formation rate declines. This is the location where thermal emission and the injection of fresh CREs ends. Such a change of slope was a result in the cosmic ray electron propagation model of \cite{Mulcahy2016} for the galaxy M\,51, using a realistic distribution of the injection of CREs. Beyond this radius, the total, nonthermal and polarised emission decrease exponentially.

It is unlikely that the lack of short spacings can cause the break in the radial profile of NGC\,628. The largest angular scale of the JVLA at D configuration at S-band is 490$\arcsec$, comparable to the angular size of NGC\,628 (Table~\ref{physicalpara}). To double check this, the JVLA map was smoothed and regridded to the same grid size as the Effelsberg 2.6\,GHz map and subtracted. No significant residual flux density was observed meaning that the JVLA observation was able to detect the same structures as the single-dish observation, resulting in no missing flux density.

As there is a lack of emission at the centre of the galaxy in the total intensity, nonthermal, and polarised intensity maps, we were not able to fit an exponential function from the centre of the galaxy.
Two separate exponential functions were fitted to the total intensity radial profile, inner (4--6\,kpc) and outer disk ($\ge$ 6.5\,kpc) for both images:

\begin{equation}
  I(R) = \left\{
  \begin{array}{l l}
    I_\mathrm{4}  \, \mathrm{exp}(-r/l_\mathrm{inner}) & \quad \text{4\,kpc $\le$ r $\le$6.5\,kpc}\\
    I_\mathrm{6.5} \, \mathrm{exp}(-r/l_\mathrm{outer}) & \quad \text{r $\ge$ 6.5\,kpc} \, .
  \end{array}
  \right.
\end{equation}

As l$_\mathrm{inner}$ cannot be fit to the polarised intensity radial profile, only l$_\mathrm{outer}$ was fitted. 
The thermal emission profile extracted from $\HII$ displays an overall exponential decrease,with an arm region enhancement observed between 3 and 5\,kpc. Here we fitted a single exponential function.

The radial profiles of the continuum emission for the total, nonthermal, thermal, and polarised intensity with the fitted functions are shown in Fig.~\ref{stokesiradial}.
The obtained scale lengths for the inner and outer parts of the galaxy are given in Table~\ref{scalelengthtable}.

Many galaxies studied possess a radial profile where the maximum flux density is observed at the centre of the galaxy and decreases exponentially usually displaying arm and inter-arm features.
Such galaxies include M\,33 \citep{2007A&A...472..785T}, M\,51 \citep{2014A&A...568A..74M}, IC\,342 \citep{2015A&A...578A..93B}, and M\,101 \citep{berkhuijsen2016}.

In contrast, NGC\,628 is much more complicated, indicating that a continuous injection of CREs, especially in the central region of the galaxy, is not valid.
The total, nonthermal, and polarised emission show a minimum in the centre of the galaxy, peaking at around 3-4\,kpc, the polarised emission peaking furthest out at 4\,kpc.
NGC\,628 is unique with this absence of a bright central region and, as explained in Subsect.~\ref{separatethermal}, the CRE population is not sufficient to produce significant synchrotron emission.

\begin{table}
\caption{Scale lengths of the inner and outer disk of NGC\,628.}
\centering
\begin{tabular}{l c c}
\hline\hline
 & l$_\mathrm{inner}$ (kpc) & l$_\mathrm{outer}$ (kpc) \\
\hline
Total intensity & 3.1 $\pm$ 0.2 & 2.24 $\pm$ 0.1  \\
Polarised intensity & - & 3.0 $\pm$ 0.1 \\
Thermal intensity & 5.8 $\pm$ 1.2 & - \\
\hline
\end{tabular}
\label{scalelengthtable}
\end{table}

\section{Observing the vertical magnetic field of NGC\,628}
\label{verticalfields}

RM Synthesis \citep{2005A&A...441.1217B, 2009IAUS..259..591H} allows us to measure Faraday rotation and investigate vertical magnetic fields. The resulting Faraday map sheds light on phenomena such as Parker loops and $\HI$ holes which shall now be described.

\subsection{RM Synthesis}

When linearly polarised electromagnetic radiation passes through a magnetic-ionic medium, the plane of polarisation will rotate in a process known as Faraday rotation.
When the polarised emission from a background source passes through a medium that is non-emitting (this type of medium is called a Faraday screen) or if Faraday rotation within the emitting medium is small (a Faraday thin medium), then the plane of polarisation will rotate by the following amount in radians:
\begin{equation}
\Delta \chi = \mathrm{RM} \, \lambda^2
\end{equation}
where $\lambda$ is the wavelength of the polarised emission and RM is the rotation measure whose unit is rad\,m$^{-2}$ and is defined as the slope of the polarisation angle versus $\lambda^{2}$.  For  more general situations and in situations when regions with more than one rotation measure within the telescope beam are present, rotation measure is replaced with the quantity ``Faraday depth'' $\phi$.
Faraday depth is defined as:

\begin{equation}
\phi = 0.812 \int^{\mathrm{telescope}}_{\mathrm{source}} n_\mathrm{e}(l) \, B_{\parallel}(l) \, dl \, .
\end{equation}
Here n$_\mathrm{e}(l)$ is the thermal electron density in cm$^{-3}$,  B$_{\parallel}$ is magnetic field strength along the line of sight in $\mu$G, and $dl$ is the path length in parsecs.

RM Synthesis is a novel approach \citep{2005A&A...441.1217B} to extract the distribution of Faraday depths (``Faraday spectrum'') of a source through polarimetric data whose effectiveness depends on the $\lambda^{2}$ coverage of the observation. With the JVLA's large bandwidth, RM Synthesis is an ideal tool to investigate the nature of the vertical magnetic field of NGC\,628.

\begin{figure}
	\centering
	\includegraphics[width=1.0\columnwidth]{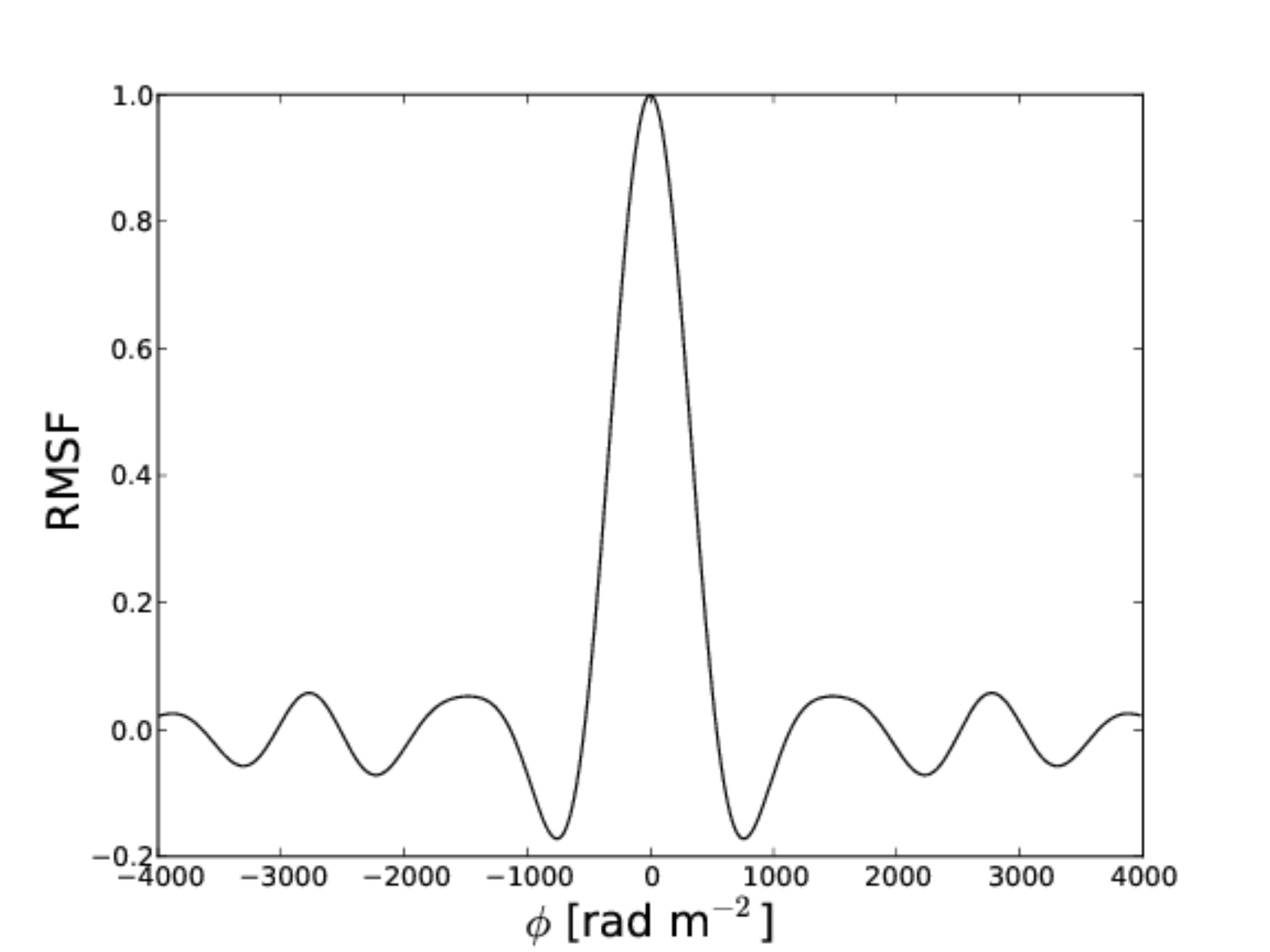}
	\caption{The RMSF of the remaining unflagged data channels (2.6--3.6\,GHz). Note due to the continuous bandwidth, the sidelobes of the RMSF are relatively low. The first sidelobe is approximately 20\% whereas it was 78\% for WSRT SINGS \citep{2009A&A...503..409H}.}
	\label{fig:rmsf}
\end{figure}

The data were averaged in frequency channels in bins of 8, resulting in 64 channels with a bandwidth of 16\,MHz each. The maximum Faraday depth for this channel width at 2.6\,GHz is 1477\,rad\,m$^{-2}$, much larger than what is expected for NGC\,628. With the coverage in $\lambda^2$, the maximum theoretical Faraday resolution is 570\,rad\,m$^{-2}$, while the maximum observable scale in the Faraday spectrum is 458\,rad\,m$^{-2}$.
The resulting RM Spread Function (RMSF) of this $\lambda^2$ coverage is shown in Fig.~\ref{fig:rmsf}.

We created channel images with natural weighting at a resolution of 18$\arcsec$, using the mosaic option of the CASA clean task rather than mosaicing all the fields manually after cleaning. These images then had the primary beam correction applied.
Each channel was visually inspected for RFI and one channel was deemed unusable. Therefore the final number of channel images used for RM Synthesis was 63. 

RM Synthesis and RM Clean \citep{2009A&A...503..409H} were both applied to the data using the {\em pyrmsynth}\ software \footnote{https://mrbell.github.io/pyrmsynth/}.
A search for large Faraday depths in NGC\,628 was performed over the range of -2$\times$10$^{5}$\,rad\,m$^{-2}$ to +2$\times$10$^{5}$\,rad\,m$^{-2}$ with a coarse $\phi$ sampling of 1000\,rad\,m$^{-2}$.
No such polarised emission was detected at very large Faraday depths. Hence, RM Synthesis was performed only
in the range -3000\,rad\,m$^{-2}$ to +3000\,rad\,m$^{-2}$ with a sampling of 25\,rad\,m$^{-2}$.
The rms noise measured in the Stokes Q and U cubes is 3.6 and 4.2\,$\mu$Jy/beam, respectively.

The maximum Faraday depth ($\phi_\mathrm{max}$) in the Faraday spectrum at each pixel of the map was measured by fitting a parabola using the three maximum points at the observed peak in the Faraday spectrum. Only polarised flux densities above a 6$\sigma$ level of 21.6\,$\mu$Jy/beam were used.
The Milky Way contribution (see Sect.~\ref{subsec:Milkywayforeground}) of -34\,rad\,m$^{-2}$ was subtracted from this Faraday depth.

The corresponding Q and U values were found from the nearest pixel to that of the measured maximum Faraday depth and were used to find the intrinsic polarisation angle (corrected for Faraday rotation via RM Synthesis) by:

\begin{equation}
\chi_0 = \frac{1}{2}\, \mathrm{arctan}\left(\frac{U(\phi_{max})}{Q(\phi_{max})}\right) \, .
\end{equation}

The magnetic field orientation is obtained by rotating the polarisation angle by 90$\degr$.

The polarised intensity image was created using the AIPS task `COMB' with the `POLC' options which takes into account the bias polarised intensity caused by noise in Stokes Q and U.
We used the average rms noise of the Q and U maps to correct for this bias (see Equation~\ref{equation1}).
The polarised intensity map is very similar to the channel averaged image (Fig.~\ref{fig:naturalmaps}). This is due to the fact that Faraday rotation between the frequency channels at this frequency is small.
The only significant difference is the slightly more extended emission seen in the south east of the galaxy.

The errors in Faraday depth and intrinsic polarisation angle were determined by:

\begin{equation}
\Delta \phi= \frac{\phi_{rmsf}}{2 \, SR}
\end{equation}
and
\begin{equation}
\Delta \chi_0 = \frac{1\,\mathrm{rad}}{2 \, SR}
\end{equation}
where $\phi_{rmsf}$ is the FWHM of the RMSF and SR is the signal-to-noise ratio of the polarised intensity.

\begin{figure*}
\centering
    \subfloat{\includegraphics[width=0.45\textwidth]{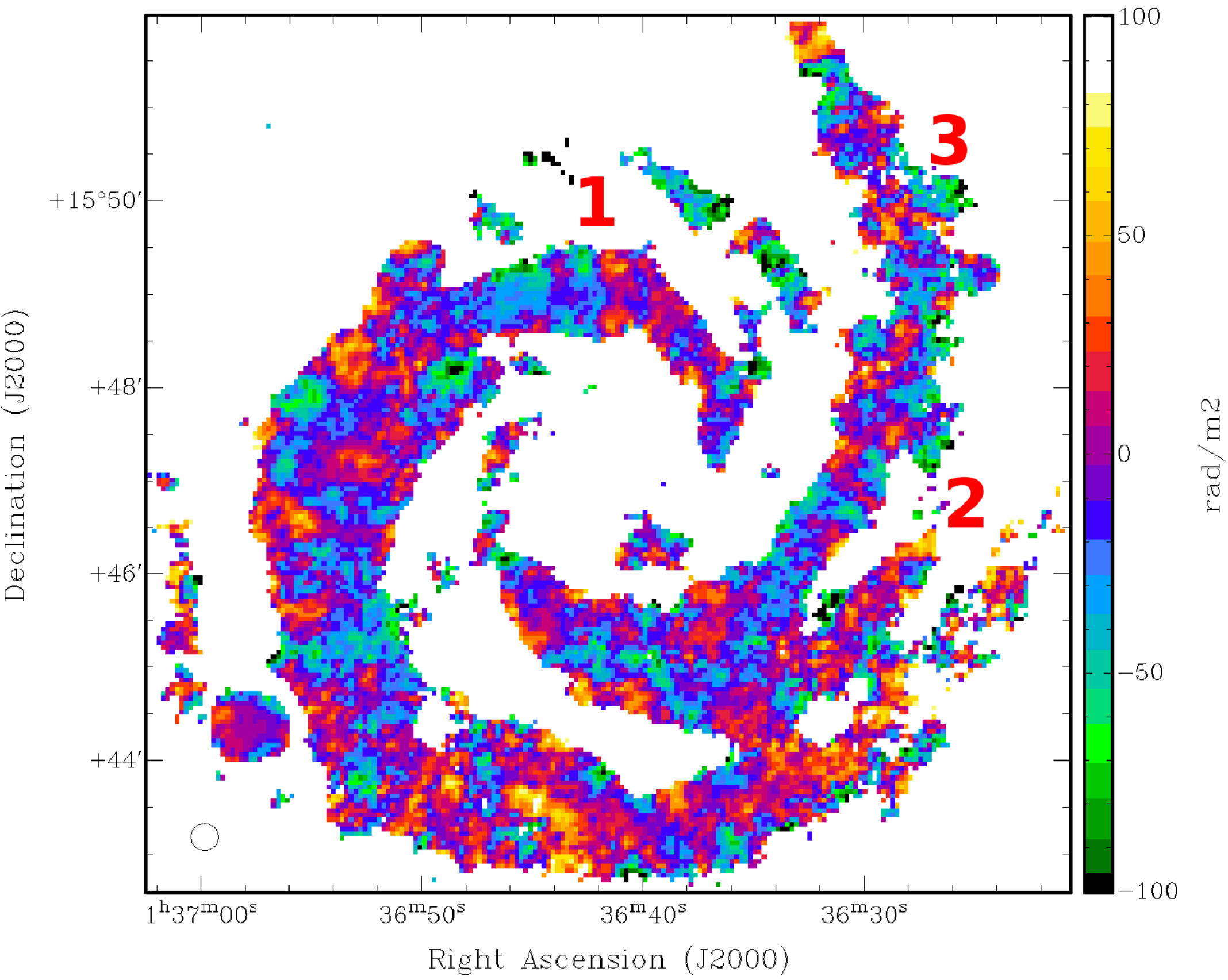}}
   \subfloat{ \includegraphics[width=0.45\textwidth]{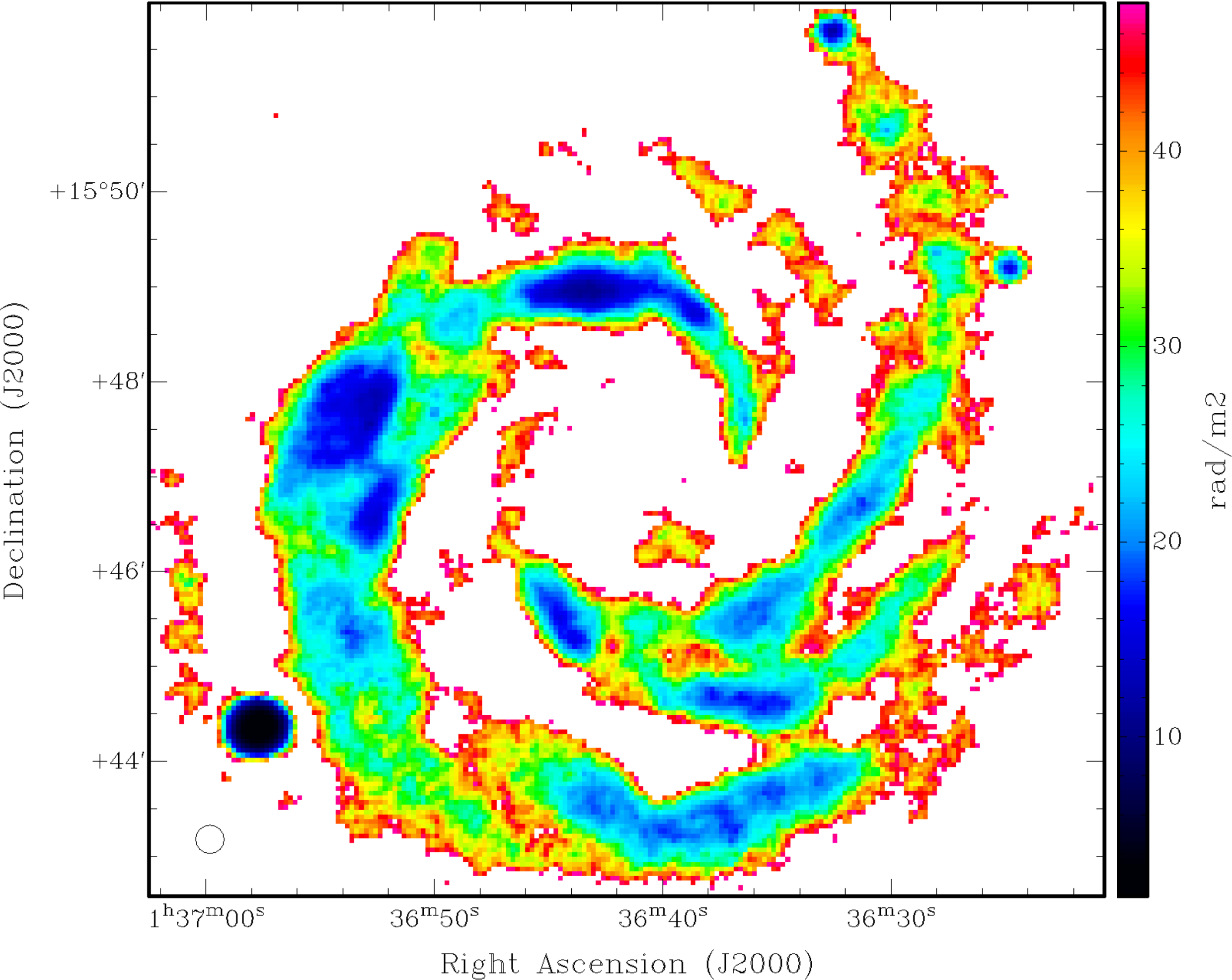}}
  \caption{Maps of the maximum fitted Faraday depth $\phi_\mathrm{max}$ (left) and the corresponding Faraday depth error $\sigma_{\phi}$ (right) at a resolution of 18$\arcsec$ marked by an ellipse in the bottom left. Both quantities are measured in units of rad m$^{-2}$. The polarised arms defined in Fig.~\ref{fig:robustmaps} are shown here marked in red.}
  \label{fig:Maxfaradaydepth}
\end{figure*}

The maps of maximum Faraday depth and associated error are shown in Fig.~\ref{fig:Maxfaradaydepth}.

RM Synthesis has been performed on NGC\,628 previously at 18 and 22\,cm wavelengths with the WSRT by \cite{2009A&A...503..409H} with a Faraday depth resolution of 144\,rad\,m$^{-2}$, with the dominant Faraday depth component centred on -30 \,rad\,m$^{-2}$.
While their Faraday resolution was superior compared to the present work, the sidelobe level was much higher due to the gap between the 18 and 22\,cm bands. Additionally, this work has much better angular resolution and sensitivity and therefore we are able to resolve different features in our Faraday depth map (Fig.~\ref{fig:Maxfaradaydepth} left).

The main Faraday depth component observed in NGC\,628 varies between +100\,rad\,m$^{-2}$ and -110\,rad\,m$^{-2}$. The mean Faraday depth is -8 \,rad\,m$^{-2}$ with a standard deviation of 30 \,rad\,m$^{-2}$. This agrees with the dominant Faraday depth component at -30 \,rad\,m$^{-2}$ observed by \cite{2009A&A...503..409H}. The standard deviation of 30\,rad\,m$^{-2}$ contains real structure that is likely to be complex. The presence of a vertical magnetic field and the large scale height of cosmic-ray electrons tend to make the standard deviation of the Faraday dispersion function larger \citep{Ideguchi2014}.

 We see an interesting and striking periodic pattern alternating between negative and positive Faraday depth values in the northern and eastern parts of arm~1. In the southern part of arm~1 no such pattern is seen, and the Faraday depths are mostly positive. Arms~2 and 3 neither show such a periodic pattern nor any obvious large-scale pattern in Faraday depth but has more negative Faraday depths than the southern part of arm~1 (see Sections~\ref{subsec:parkerloops} and \ref{subsec:loops} for details).

\cite{2010A&A...514A..42B} detected secondary components for NGC\,628 at $\phi$ = -213 and +145 rad m$^{-2}$ in addition to other mildly inclined galaxies such as NGC\,6946 and M\,51.
These secondary components may originate from polarisation at the far side of the midplane, becoming more Faraday rotated when passing through the midplane, resulting in larger values of Faraday depths, approximately $\pm$ 200 rad m$^{-2}$.
\cite{2015ApJ...800...92M} searched for these secondary components in M\,51 with the JVLA at L-band and did not detect any significant polarisation coinciding with these secondary components.
We have searched the Faraday cube at 18$\arcsec$ and could not find indications in the Faraday spectra of these secondary components. It should be noted, however, that the Faraday depth resolution for this observation is not enough to fully resolve the main component from the secondary components.

\subsection{Magnetic field order}
\label{subsec:regularity}

The nonthermal polarisation degree $p_n$ is a measure of the ratio $q$ of the field strength of the ordered field in the sky plane and the random field, named the degree of order of the field, $q=B_\mathrm{reg}/B_\mathrm{ran}$ \citep{2007A&A...470..539B}.
The maximum possible degree of polarisation is given by $p_0=(3-3\alpha_\mathrm{n})/(5-3\alpha_\mathrm{n})$ where $\alpha_\mathrm{n}$ is the nonthermal spectral index taken here as -1.0.

In case of equipartition between the energy densities of magnetic field and cosmic rays \citep{2007A&A...470..539B}:

\begin{equation}
\frac{p_n}{p_0} = \frac{q^2}{(q^2+\frac{1}{3})}
\end{equation}

\begin{equation}
q \simeq \left(\frac{(p_n/p_0)}{[2(1-(p_n/p_0))]}\right)^{0.5}
\end{equation}

\begin{figure*}
	\hspace{0.6cm}
		\subfloat{\includegraphics[width=0.45\textwidth]{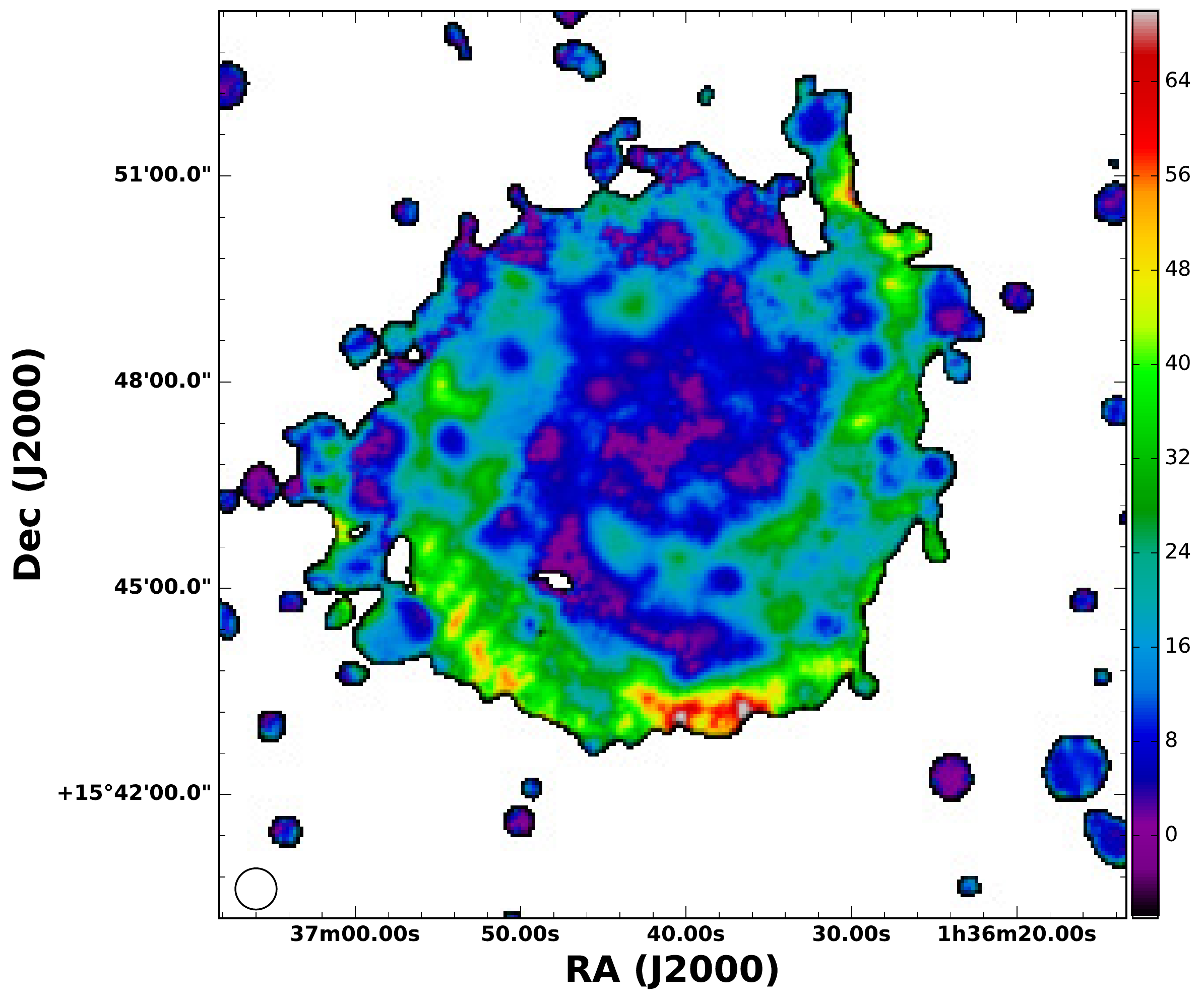}}
		\subfloat{\includegraphics[width=0.45\textwidth]{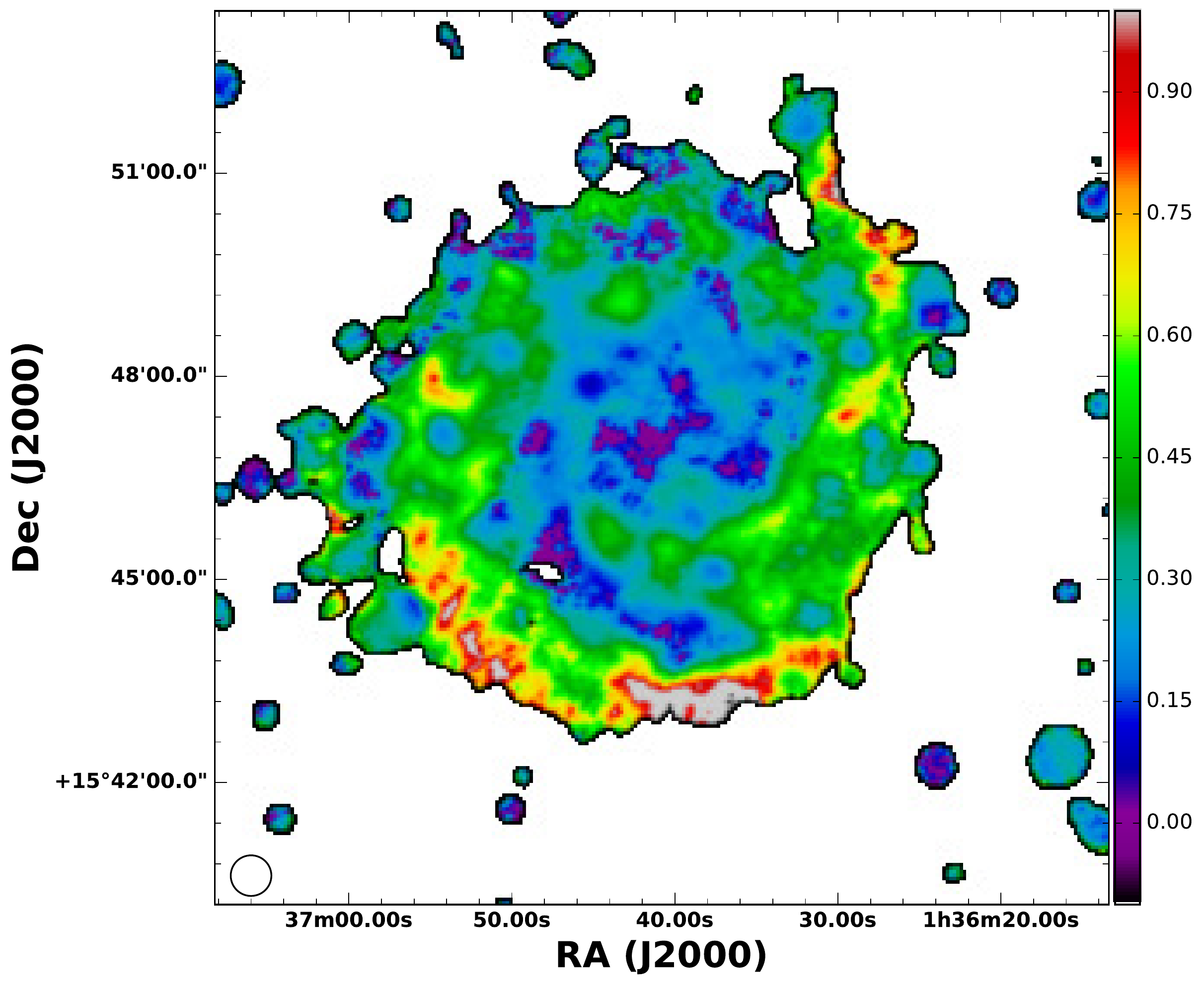}}
	\caption{Nonthermal polarisation degree (left) and the degree of field order $q$ (right), the ratio of the ordered field strength in the sky plane to the isotropic random field strength, derived from the fractional polarisation of the nonthermal intensity (Fig.~\ref{fig:thermmaps}, right) The colour scale is in percent (left) and degree of field order (right).}
	\label{fig:fieldreg}
\end{figure*}

Fig.~\ref{fig:fieldreg} shows the nonthermal polarisation degree (left) and the degree of magnetic field order (right) for NGC\,628 in the case of equipartition which was derived from the polarised intensity map found from RM Synthesis (Fig.~\ref{fig:Maxfaradaydepth}, left) and the nonthermal map (Fig.~\ref{fig:thermmaps}, right). $q$ is almost zero in the optical spiral arms, meaning that depolarisation is enhanced in the spiral arms. $q$ increases toward the outer radii, specifically in the inter-arm regions where $q$ reaches 0.8 where depolarisation is low. An isolated region with a maximum $q$ of 0.56 exists in the northern part of polarisation arm~1 and coincides with the region of the most intense polarised emission, surrounded by $q$ values of around 0.2. This region will be discussed in further detail in Sect.~\ref{subsec:faradaywindow}.

\subsection{Estimate of the Galactic foreground in the direction of NGC\,628}
\label{subsec:Milkywayforeground}

Based on three polarised sources detected in the NGC\,628 field, \cite{2009A&A...503..409H} estimated that the likely Galactic foreground Faraday depth is about -34$\pm$2\,rad\,m$^{-2}$.
In our NGC\,628 field we detect seven polarised sources, four of which are not directly located behind NGC\,628. Table~\ref{tab:discretesources} shows the locations, Faraday depths, and polarised intensities of these discrete sources. The average of the Faraday depth of these four sources gives us a likely value of -20$\pm$14\,rad\,m$^{-2}$. It is possible that the Faraday depth of these sources used is still affected by NGC\,628. These sources are found in the range of 16--23\,kpc from the centre of NGC\,628. \cite{Han1998} found through RM data of background sources that in M\,31 the regular magnetic field probably extends to a radius of 25\,kpc.

A Galactic foreground Faraday depth of -20$\pm$14\,rad\,m$^{-2}$ is comparable to that of \cite{2009A&A...503..409H}. As the Faraday depth resolution in \cite{2009A&A...503..409H} was superior to that of our work, -34 rad\,m$^{-2}$ was taken as the likely value when subtracting the Galactic foreground contamination. This also highlights the importance of L-band observations to achieve better Faraday depth resolution and thus ensure a more accurate removal of the Faraday rotation caused by the Galactic foreground.

\begin{table*}
\centering
\caption{Discrete polarised sources detected in the NGC\,628 field, uncorrected for Faraday rotation caused by the Galactic foreground. Sources used to calculate the Galactic foreground are marked in bold.}
\label{tab:discretesources}
\begin{tabular}{cccccccc}
  \hline
  RA&DEC& Galactocentric radius & Peak flux density&$\phi$&$\bigtriangleup\phi$&PA&$\bigtriangleup$PA\\
 (h m s)&($\degr$ \, $\arcmin$ \, $\arcsec$)& (kpc) & ($\mu$Jy/beam)&(rad/m$^{2}$)&(rad/m$^{2}$)&($\degr$)&($\degr$)\\
  \hline
\textbf{01 36 00} & \textbf{+15 44 58} & \textbf{21.42} & \textbf{87} & \textbf{-13.7} & \textbf{11.6} & \textbf{12.7}  & \textbf{1.1} \\
\textbf{01 36 14} & \textbf{+15 41 16} & \textbf{18.37} & \textbf{69} & \textbf{-6.5}  & \textbf{14.7} & \textbf{-68.9} & \textbf{1.4} \\
\textbf{01 36 16} & \textbf{+15 42 25} & \textbf{16.14} & \textbf{405} & \textbf{-39.1} & \textbf{2.5}  & \textbf{-52.0} & \textbf{0.3} \\
01 36 24 & +15 49 13 & 9.80 & 67 & -49.5 & 15.5 & -86.8 & 1.5 \\
01 36 32 & +15 51 43 & 10.80 & 85 & +6.1  & 11.49& 76.0  & 1.2 \\
01 36 57 & +15 44 22 & 9.87 & 532 & -34.5 & 1.9  & -88.9 & 0.2 \\
\textbf{01 37 28} & \textbf{+15 46 18} & \textbf{23.34} &\textbf{140} & \textbf{-21.8} & \textbf{7.3}  & \textbf{40.2}  & \textbf{0.7} \\
   \hline
\end{tabular}
\end{table*}

\subsection{Depolarisation within S-Band}
\label{subsec:depolarisation}

Faraday depolarisation is an important tool in retrieving information on the density of ionised gas, the strength of the ordered and turbulent field components, and the typical length scale (or integral scale) of turbulent magnetic fields.
Significant depolarisation is to be expected for NGC\,628, as \cite{2009A&A...503..409H} observed little to no polarised emission in the northern regions of the galaxy.

\begin{figure}
	\includegraphics[width=1.0\columnwidth,trim={3.4cm 0.4cm 3.4cm 0.4cm},clip]{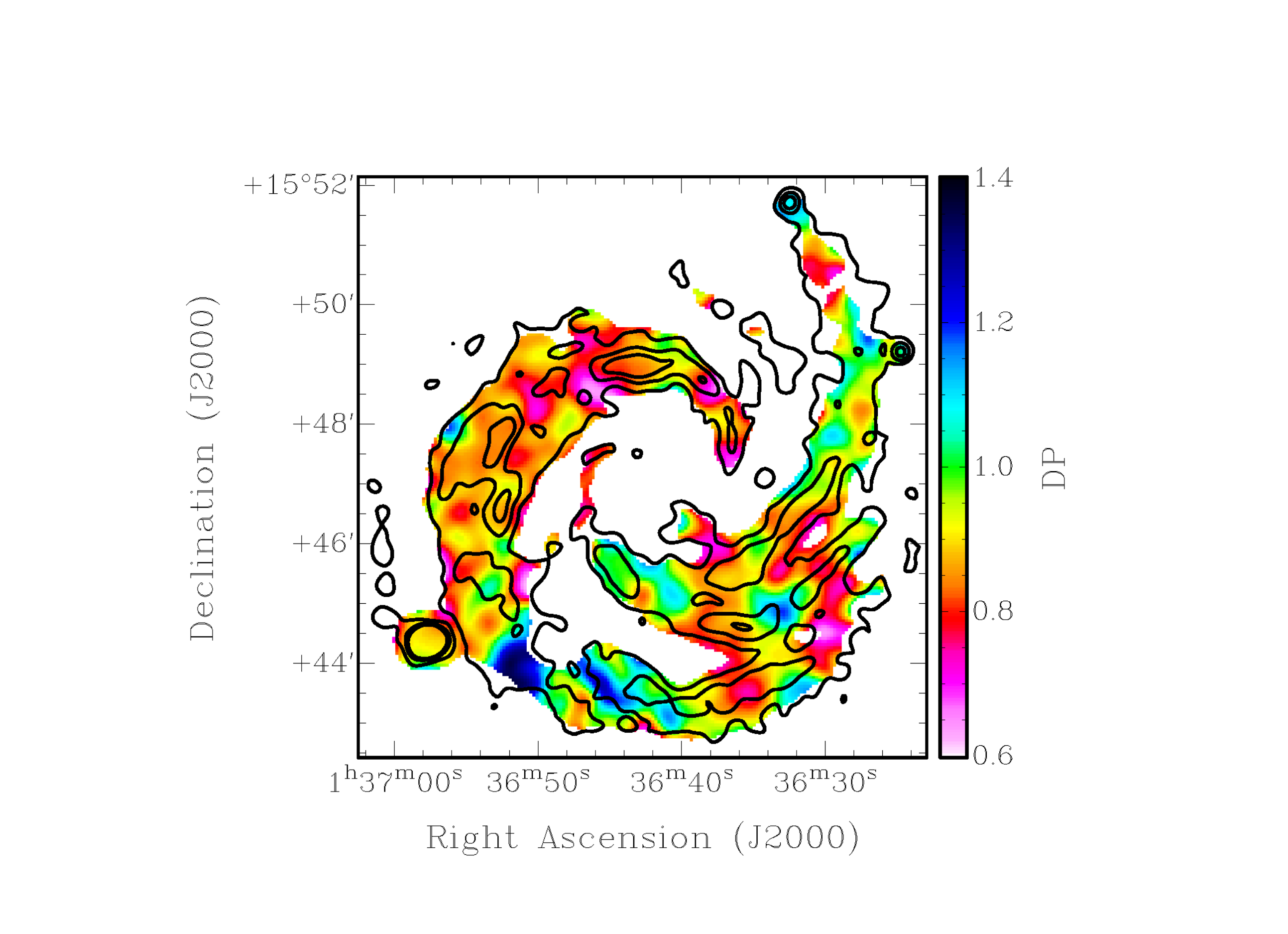}
	\caption{Map of the Faraday depolarisation ratio of NGC\,628 across the S-band at $31\arcsec$ resolution, with contours of polarised intensity at 20, 40, 60\,$\mu$Jy/beam.}
	\label{fig:628depolarisationmap}
\end{figure}

In order to get an estimation of the Faraday depolarisation we define the ratio DP \citep{2007A&A...470..539B} by:

\begin{equation}
DP = \left(\frac{PI_\mathrm{2.87\,GHz}}{PI_\mathrm{3.43\,GHz}}\right) \times \left(\frac{\mathrm{3.43\,GHz}}{\mathrm{2.87\,GHz}}\right)^{\alpha_\mathrm{n}}
\end{equation}
where $\alpha_\mathrm{n}$ = -1.0 is the nonthermal spectral index and is assumed to be constant across the entire galaxy. DP = 1 signifies no depolarisation and DP = 0 means total depolarisation.
We image two sections of the band which have the same $\lambda^2$ coverage. These two bands have central frequencies of 2.87 \& 3.43\,GHz.
The computed depolarisation map is shown in Fig.~\ref{fig:628depolarisationmap}. We observe depolarisation ratios between 0.6 and 1.4 but the largest ratios are uncertain due to low signal-to-noise ratios.
DP is about 0.9 on average and 0.7 in star-forming regions.
Varying $\alpha_\mathrm{n}$ by 0.2 changes the DP ratio by 0.03, likewise a 10\% difference in polarised intensity changes DP by 0.1.

Generally, lower DP ratios, i.e. stronger depolarisation, are seen in the northern parts of the of the galaxy, located around the major axis (PA = 20$\degr$).
This is consistent with the findings of \cite{2009A&A...503..409H} and \cite{2010A&A...514A..42B} that for many galaxies Faraday depolarisation at $\lambda$20\,cm is asymmetric along the major axis of the projected galaxy disk in the sky plane, caused by the large-scale halo field. This results in less polarised emission on the kinematically receding side of the major axis. Other regions with low DP are located where star formation and thus B$_\mathrm{ran}$ and thermal electron density are greatest.

There are two main mechanisms that cause depolarisation. The first, differential Faraday rotation, arises when relativistic and thermal electrons occupy the same region of the magnetised medium. The degree of polarisation $p$ depends on the observable Faraday rotation measure RM and $\lambda$ as \citep{burn1966,1998MNRAS.299..189S}:

\begin{equation}
p = p_0\frac{\mathrm{sin} |2 \, \mathrm{RM} \, \lambda^2 |}{ |2 \, \mathrm{RM} \, \lambda^2 |}
\end{equation}
where $p_0$ is the intrinsic degree of polarisation. The polarisation vanishes when
$\left| \mathrm{RM} \right|$ = RM$_0$ where

\begin{equation}
2 \, \mathrm{RM}_0 \, \lambda^2 = \pi \, n
\end{equation}
with n = 1, 2, ... . This causes depolarisation canals in the ISM as seen extensively in M\,31 at $\lambda$\,20.5cm \citep{2003MNRAS.342..496S}.
For NGC\,628 at $\lambda$9.6\,cm, we find RM$_0$ = 170\,rad m$^{-2}$ which is larger than the average Faraday depth observed in Fig.~\ref{fig:Maxfaradaydepth}. Thus we expect little depolarisation due to differential Faraday rotation at this frequency. Indeed, no depolarisation canals can be seen in any of the polarisation maps.

The second mechanism that causes depolarisation, internal Faraday dispersion by turbulence in the magnetised ISM, is the most probable source of depolarisation in NGC\,628 \citep{1998MNRAS.299..189S}:

\begin{equation}
p = p_0\frac{1-\mathrm{exp}(-2 \, \mathrm{S})}{2 \, \mathrm{S}}
\end{equation}
where $S=\sigma_{\mathrm{RM}}^2\,\lambda^4$. $\sigma_{\mathrm{RM}}$ is the dispersion in intrinsic rotation measure $\mathrm{RM}_i$.
$\sigma_{\mathrm{RM}}$ of 43\,rad m$^{-2}$ is needed to produce DP = 0.9 between $\lambda 0.087$\,m and $\lambda 0.104$\,m.

Internal Faraday dispersion is a consequence of the turbulent ISM and can be written as $\sigma_{\mathrm{RM}}=0.812 \, n_\mathrm{e}\, \sqrt{1/3} \,\, B_\mathrm{ran}\,\sqrt{L\,d / f}$ \citep{2016A&ARv..24....4B}
where $n_\mathrm{e}$ is the average thermal electron density of the diffuse ionised gas along the line of sight (in cm$^{-3}$), $B_\mathrm{ran}$ the random field strength (in $\mu$G), $L$ the path length through the thermal gas (in pc), $d$ is the turbulent scale (in pc), and $f$ the filling factor of the Faraday-rotating gas. Standard values found in NGC\,6946 \citep{2007A&A...470..539B} of $n_\mathrm{e}$ = 0.03\,cm$^{-3}$, $B_\mathrm{ran}$ = 10\,$\mu$G, $L$ = 1000\,pc, $d$ = 50\,pc, and $f$ = 0.5 yield $\sigma_{\mathrm{RM}} \simeq44$\,rad m$^{-2}$, as required for NGC\,628. For the star-forming regions, increasing the thermal electron density or the random field strength by a factor of two yields the required value of DP = 0.7.

\subsection{A lone ordered magnetic field in NGC\,628?}
\label{subsec:faradaywindow}

The most intense polarised emission is seen in the northern part of polarisation arm~1 at RA(J2000) = 01$^\mathrm{h}$ 36$^\mathrm{m}$ 44$^\mathrm{s}$, DEC(J2000) = +15$\degr$ 48$\arcmin$ 58$\arcsec$.
This region could be a Faraday window with low Faraday depolarisation (i.e. a high DP ratio). However, the S-band depolarisation map (Fig.~\ref{fig:628depolarisationmap}) shows normal DP ratios of around 0.8, similar to other regions in the galaxy and must then originate from truly ordered fields.

The region has a degree of magnetic field order $q$ of approximately 0.56 (Fig.~\ref{fig:fieldreg}).
The degree of polarisation in this region has a mean of 19$\%$ and a maximum of 29$\%$. The immediate area around this particular region has degrees of polarisation of less than 10$\%$.
While this is not especially high compared to the outer regions of the galaxy, it is quite high with respect to the immediate vicinity ($q$ of 0.2). This region is also apparent and out of place with its small pitch angle of 25$\degr$ compared to its surroundings with a pitch angle of approximately 45$\degr$ (Fig.~\ref{fig:pitchanglemap}).
This feature is unique as nothing similar has been detected so far in any other galaxy.

A compressed magnetic field due to substantial ram pressure caused by the motion of the galaxy through the intergalactic medium (IGM) (like in the Virgo cluster, see \cite{Vollmer2007}) does not seem to be a valid explanation as NGC\,628 shows no signs of gas stripping in the outer parts; instead, it has a large intact $\HI$ disk, especially in the north where the polarised region is located.

The magnetic field in this region may be compressed due to an expanding superbubble caused by several supernovae. We consider this to be a plausible explanation,  since tentative signs of gas expansion are present in the $\HI$ data from the THINGS survey. Although there are no previously catalogued $\HI$ holes in this region \citep{2011AJ....141...23B}, a position velocity diagram (Fig.~\ref{fig:PVdiagram}) of the region shows tentative kinematic signs of a bubble. This double-peaked kinematic signature corresponds to a depression in the local $\HI$ column density that has an elliptical morphology overlapping the extent of the polarised region. In fact, this region was identified as a potential hole during the preparation of the catalog presented by \citet{2011AJ....141...23B}, but was excluded from the final list due to the relatively low apparent expansion velocity $(\sim20~\mathrm{km\,s}^{-1}$ as compared to the resolution of the data, $\sim8~\mathrm{km\,s}^{-1})$ combined with the fact that the expansion is not seen along the minor axis of the elliptical feature. The $\HI$ features are suggestive, but not conclusive, of an expanding bubble (Y. Bagetakos, priv. comm.). On the other hand, bright H$\alpha$ features are seen around the periphery of the $\HI$ feature, which could indicate recent star formation induced by the compression of ISM material at the edges of an expanding feature.  Taken all together, these features could be consistent with a barrel-shaped expansion taking place preferentially along the direction of the ordered magnetic field.

\begin{figure}
\centering
\includegraphics[width=0.75\columnwidth,angle=-90]{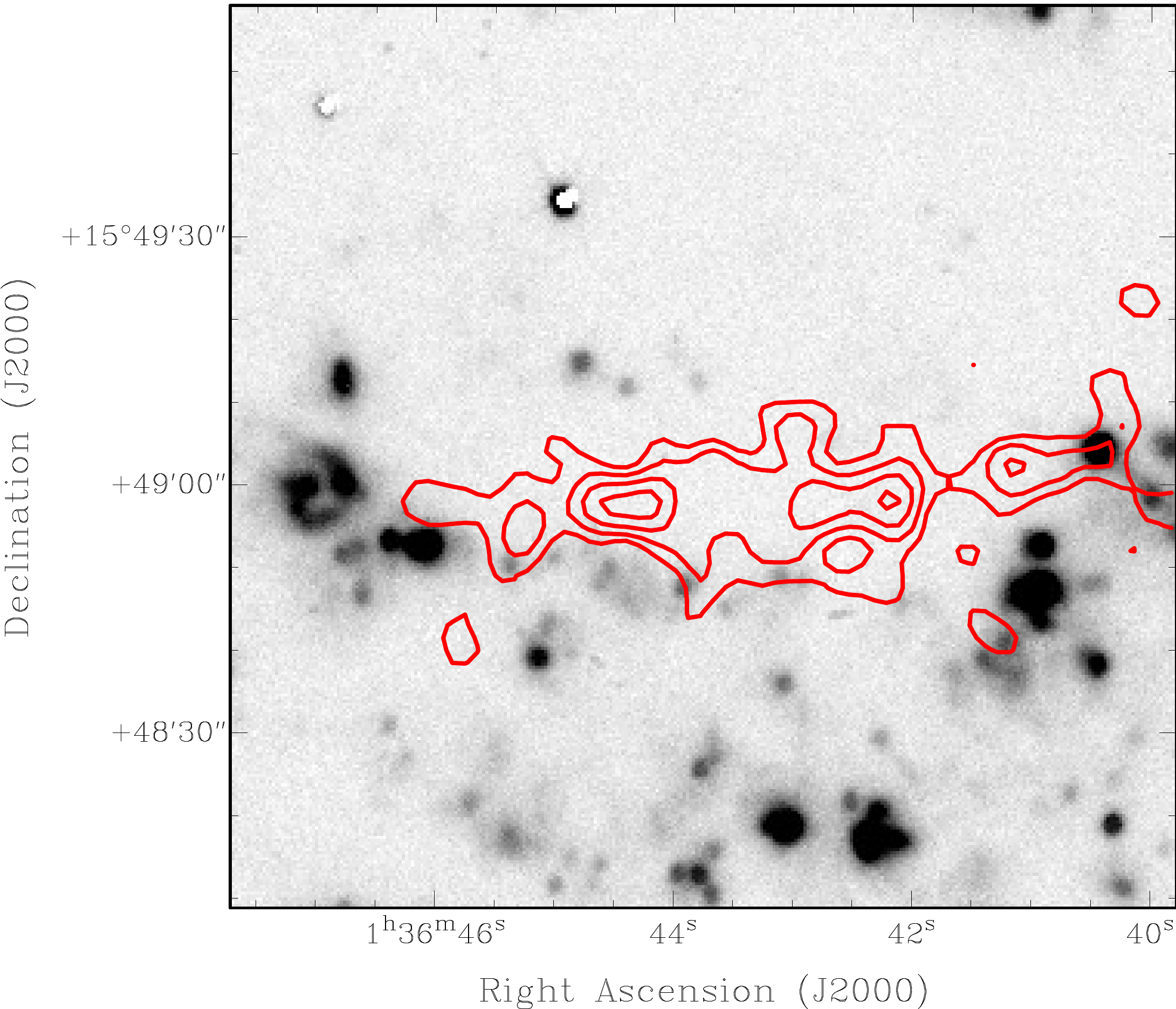}
\caption{Northern region of NGC\,628 hosting an exceptionally strong and ordered magnetic field discussed in Sec.~\ref{subsec:faradaywindow} shown with polarised intensity overlaid onto a H$\alpha$ image \citep{Dale2009}. Contours are at 10, 15, 20, 25\,$\mu$Jy/beam. Resolution is 7.5 \arcsec.}
\label{fig:faradaywindow}
\end{figure}

\begin{figure}
\centering
\includegraphics[width=1.0\columnwidth]{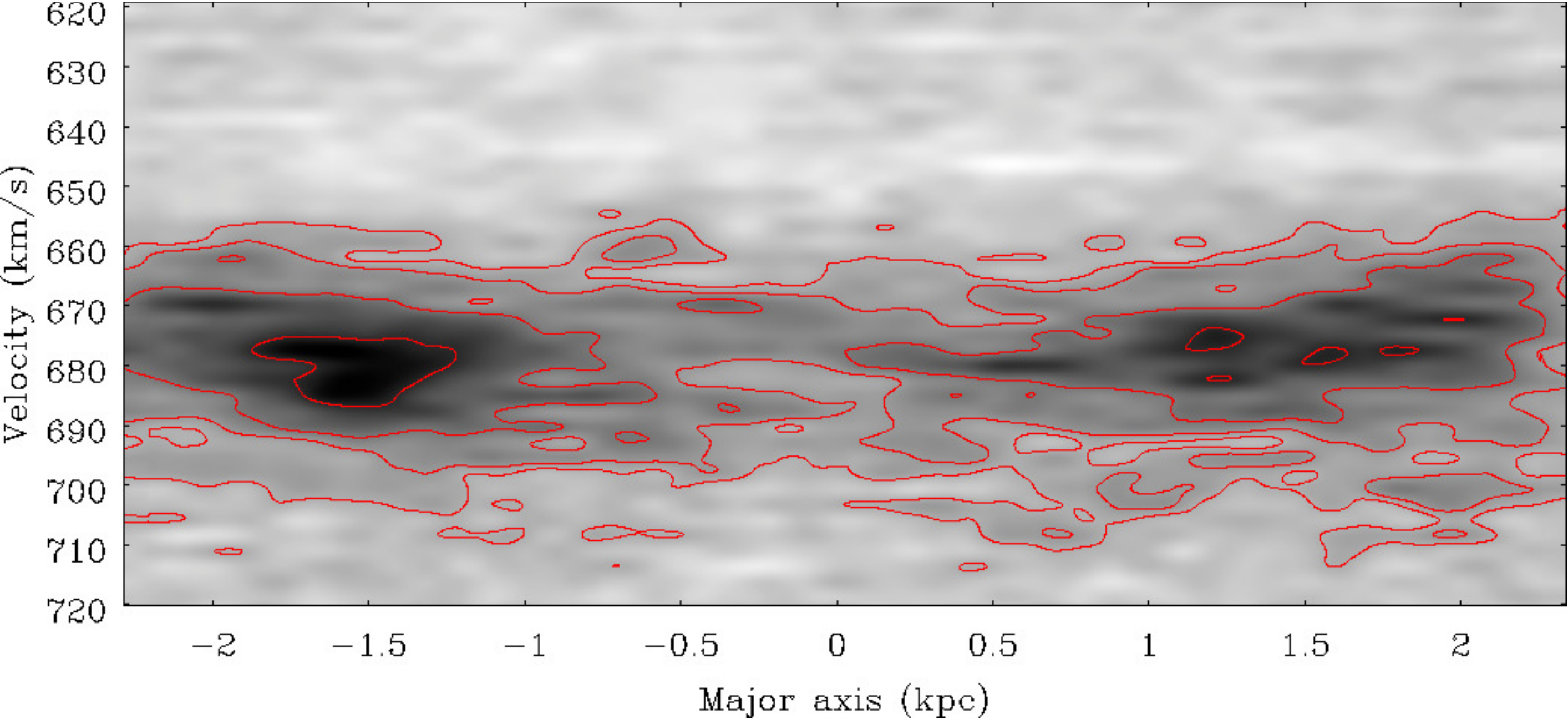}
\caption{$\HI$ position velocity diagram of a 1-D slice through the region of a lone ordered magnetic field, as described in subsect.~\ref{subsec:faradaywindow}. The centre of this region is located at 0\,kpc and extends approximately from -1 to 1\,kpc. Contours are at 3, 5, 8, 14 $\times \,\sigma$ ($\sigma \simeq 5\,\mu$Jy/beam).}
\label{fig:PVdiagram}
\end{figure}

We conclude that the northern region hosts an exceptionally strong and ordered field. This region has a small Faraday depth of -5 to -10 rad/m$^{2}$, very similar to other regions of the galaxy and thus signifies that, while the field in the sky plane is highly ordered, the magnetic field in the line of sight is not particularly strong.
We propose that the polarisation peak is caused by a strongly ordered magnetic field driven by an asymmetrically expanding $\HI$ bubble, and possibly accentuated by low turbulence due to the absence of recent star formation in the centre of the feature, evident from the lack of H$\alpha$ and UV emission \citep{gildepaz2007} as illustrated in Fig.~\ref{fig:faradaywindow}.
Observations at other frequencies could help to further constrain the cause of such an isolated, highly ordered magnetic field.

\subsection{$\HI$ holes}
\label{subsec:HIholesection}

\begin{figure}
	\centering
	\includegraphics[width=1.0\columnwidth]{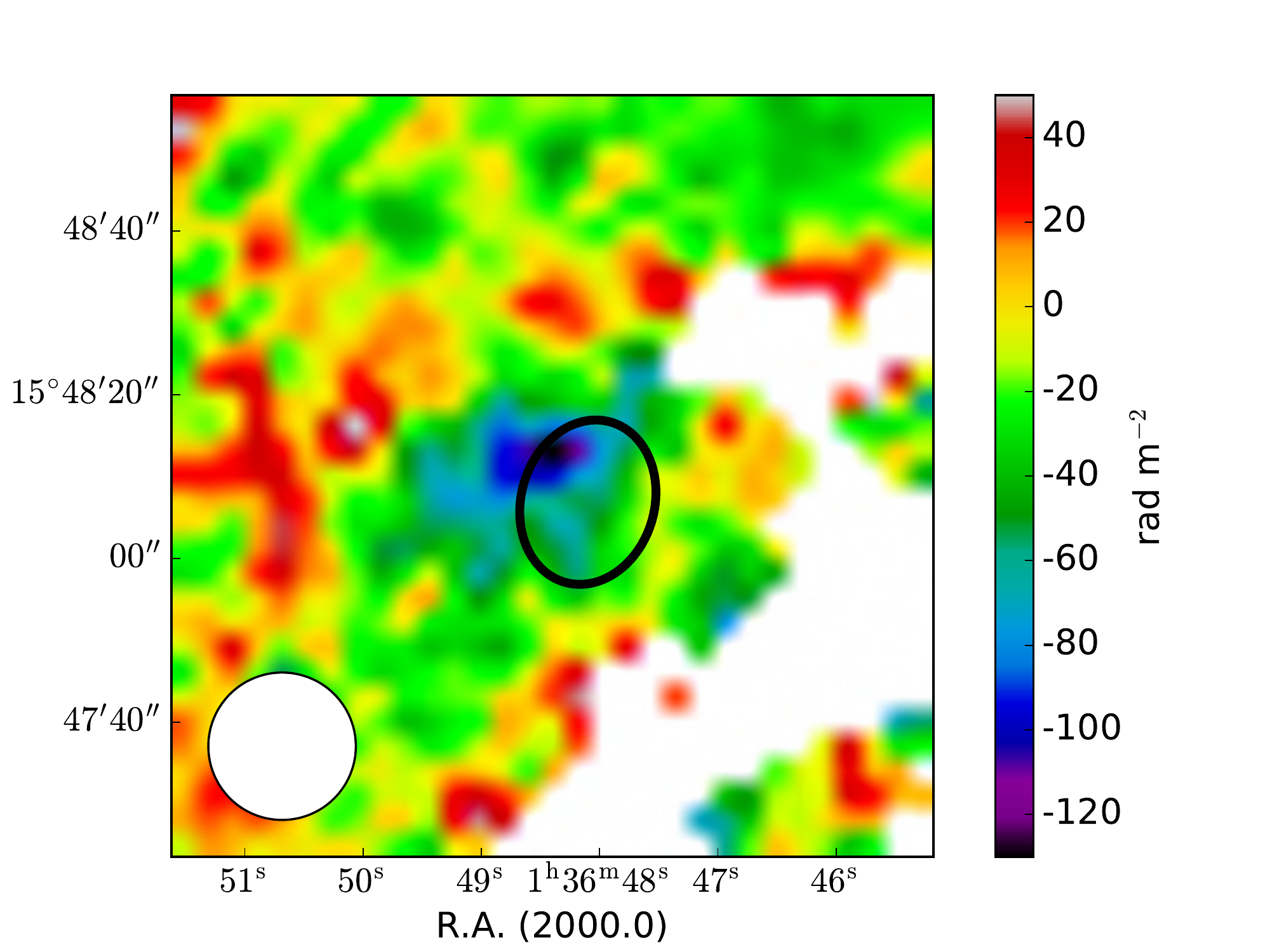}
	\caption{Region of NGC\,628 with a large Faraday depth coinciding with a known $\HI$ hole marked by an ellipse. The 18\arcsec\ beam is shown in the bottom left, filled in white.}
	\label{fig:HIfdhole}
\end{figure}

The Faraday depth (FD) map (Fig~\ref{fig:Maxfaradaydepth}) was compared by eye to the $\HI$ hole catalogue of \cite{2011AJ....141...23B}. A Faraday depth gradient like the one seen in NGC\,6946 by \cite{2012ApJ...754L..35H} is detected, as we observe a region with a large Faraday depth feature, the largest one in NGC\,628 coinciding with a $\HI$ hole (Fig.~\ref{fig:HIfdhole}) with a Faraday depth gradient, with FD decreasing from -20 rad m$^{-2}$ at the western edge of the $\HI$ hole to -130 rad m$^{-2}$ at the eastern edge. This $\HI$ hole is located at
RA(J2000) = 01$^\mathrm{h}$ 36$^\mathrm{m}$ 48$^\mathrm{s}$, DEC(J2000) = +15$\degr$ 48$\arcmin$ 06$\arcsec$ with an expansion velocity of 7\,km/s and a kinetic age of 50\,Myrs.

The age of an $\HI$ hole is crucial in order to observe any Faraday depth gradient, as it must be old enough to have gained a certain vertical offset but young enough so that vertical shear
has not destroyed the Faraday signature.
For a vertical shear of $\approx$ 15 km s$^{-1}$ kpc$^{-1}$, similar to NGC\,891 \citep{2006ApJ...647.1018H}, the characteristic age is approximately 60\,Myrs. Therefore, this $\HI$ hole has an ideal age to observe such a Faraday depth gradient caused by magnetic fields carried by hot gas motions originating from star formation episodes in the disk \citep{1988LNP...306..155N}.

No other Faraday depth feature coincides with $\HI$ holes across the galaxy. The reason for this is deferred until the discussion later in the paper (Sect.~\ref{subsec:holes}).

\subsection{Pitch angles of the magnetic field}
\label{subsec:pitchangles}

\begin{figure}
	\centering
	\includegraphics[width=0.8\columnwidth,angle=-90]{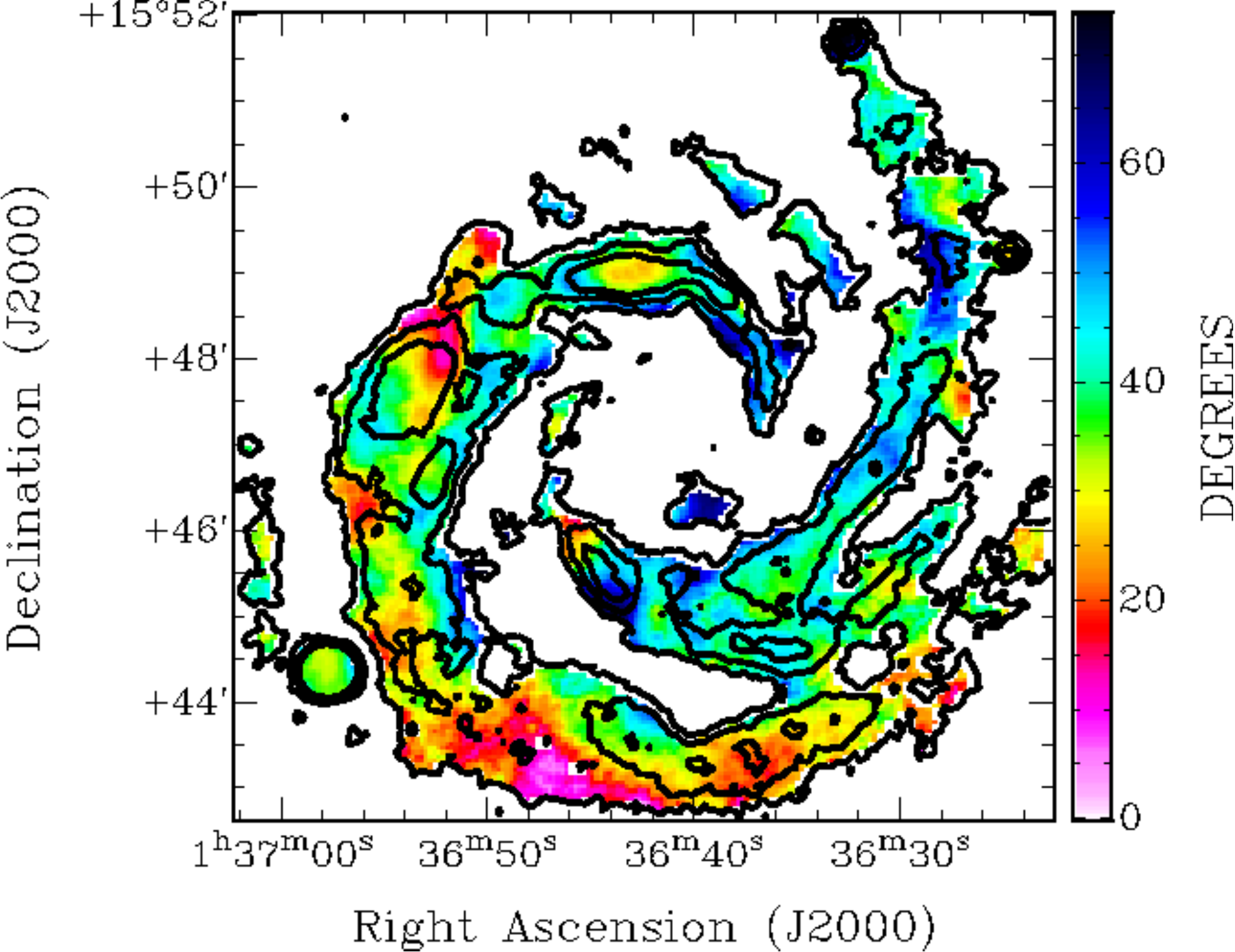}
	\caption{Intrinsic pitch angles of the magnetic field in the plane of NGC\,628 at 18$\arcsec$ resolution. Contours show the polarised intensity at 5, 10, 15 $\times \, \sigma_{rms}$.}
	\label{fig:pitchanglemap}
\end{figure}

To determine the intrinsic magnetic pitch angle, the observed B-vector map taken from the RM Synthesis cube and thus corrected for Faraday rotation is transformed into the galaxy's plane and the position angle of the local circumferential orientation is subtracted.
The resulting map (Fig.~\ref{fig:pitchanglemap}) shows that the polarisation arms~2 and 3 extending from the central region in the south-east to the north-west have a relatively constant and large magnetic pitch angle, while the (broader) polarisation arm~1 extending from the central region in the north-west over east to the south shows strongly fluctuating pitch angles. In the southern part of arm~1, the pitch angle is smaller at the outer edge compared to the inner edge.

\begin{figure*}
\hspace{1cm}
\centering
\includegraphics[width=0.9\textwidth]{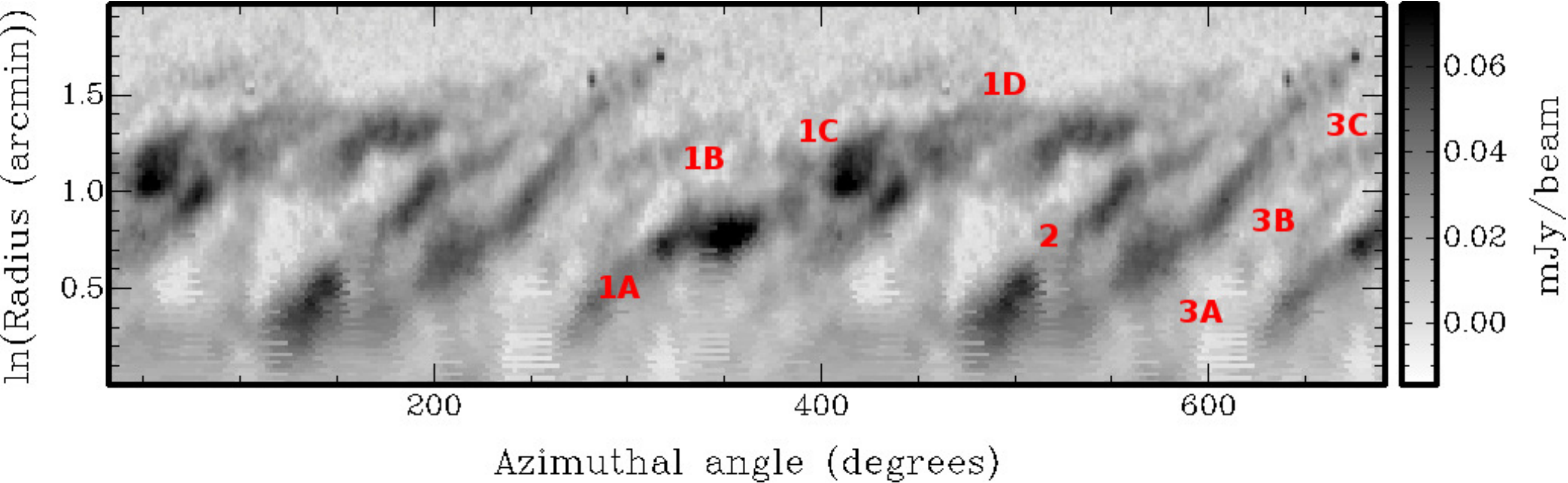}
\caption{Polarised intensity of NGC\,628 at 18$\arcsec$ in polar coordinates (azimuthal angle in degrees, measured counterclockwise from the north-eastern major axis in the galaxy plane, and ln of radius in arcminutes). The range of azimuthal angles is plotted twice for better visibility of the spiral arms. The recognised structures in the polarisation arms for which the structure and magnetic pitch angles are found (Table~\ref{tab:pitchangletable}) are labeled in red.}
\label{fig:628_polarpi}
\end{figure*}

The magnetic field lines seem to generally follow the orientations of the spiral pattern seen in polarised intensity. A quantitative analysis needs measurement of the pitch angle of the spiral arm structures observed in polarised intensity. Spiral arms with constant pitch angles are linear structures in polar coordinates [$ln$(r), $\phi$] where r is the radius and $\phi$ the azimuthal angle. Fig.~\ref{fig:628_polarpi} shows the polar transform of the PI map retrieved from RM Synthesis (Fig.~\ref{fig:Maxfaradaydepth}).

Table~\ref{tab:pitchangletable} lists the spiral features that are marked in Fig.~\ref{fig:628_polarpi}. The average morphological pitch angles follow from the slopes in the polar plot, while the average magnetic pitch angles are measured from Fig.~\ref{fig:pitchanglemap}. The remarkable result is that the magnetic pitch angles are {\em systematically larger}\ than the morphological pitch angles of the spiral arms. The average difference is about $23\degr$ for arm~1, about $12\degr$ for arm~2, and about $15\degr$ for arm~3. The morphological pitch angle varies much more along a spiral arm than the magnetic pitch angle; the magnetic pattern is {\em smoother}\ than the spiral arm pattern.
A magnetic pitch angle that is systematically larger than the morphological pitch angle of the spiral arms was also found for the galaxies M\,83 \citep{2016A&A...585A..21F} and M\,101 \citep{berkhuijsen2016}.
This finding is further discussed in Sect.~\ref{subsec:pitch}.

\begin{table*}
 \centering
  \caption{Average pitch angles of spiral arm structures of polarised intensity and magnetic pitch angles of the ordered field of NGC\,628. Note that the pitch angle is positive for a spiral that is winding outwards in the counter-clockwise sense.}
  \label{tab:pitchangletable}
  \begin{tabular}{@{}llrrrrlrlrrrrrrrrrrrrr@{}}
  \hline
       & Spiral feature & Azimuthal range ($\degr$) & Radial range ($\arcmin$) & Morphological pitch angle ($\degr$) & Magnetic pitch angle ($\degr$)\\
 \hline
& 1A & 264-344  & 0.22-1.04 & +25 $\pm$ 2 & +50 $\pm$ 8 \\
& 1B & 310-402  & 0.72-0.98 & +9 $\pm$ 1  & +40 $\pm$ 7 \\
& 1C & 336-76  & 0.66-1.38 & +20 $\pm$ 2  & +32 $\pm$ 10 \\
& 1D & 114-210 & 1.34-1.36 & +2 $\pm$ 2 & +27 $\pm$ 8\\
& 2 & 118-206 & 0.26-1.18 &  +30 $\pm$ 3& +42 $\pm$ 8 \\
& 3A & 162-244 & 0.26-0.86 & +23 $\pm$ 2 & +43 $\pm$ 7 \\
& 3B & 244-284 & 0.86-1.42 & +40 $\pm$ 2 & +43 $\pm$ 7 \\
& 3C & 284-324  & 1.42-1.76 & +24 $\pm$ 2& +46 $\pm$ 9\\

 \hline
\end{tabular}
\end{table*}

\subsection{Faraday depth signatures of Parker loops and helical fields}
\label{subsec:parkerloops}

\begin{figure}
	\centering
	\includegraphics[width=0.9\columnwidth]{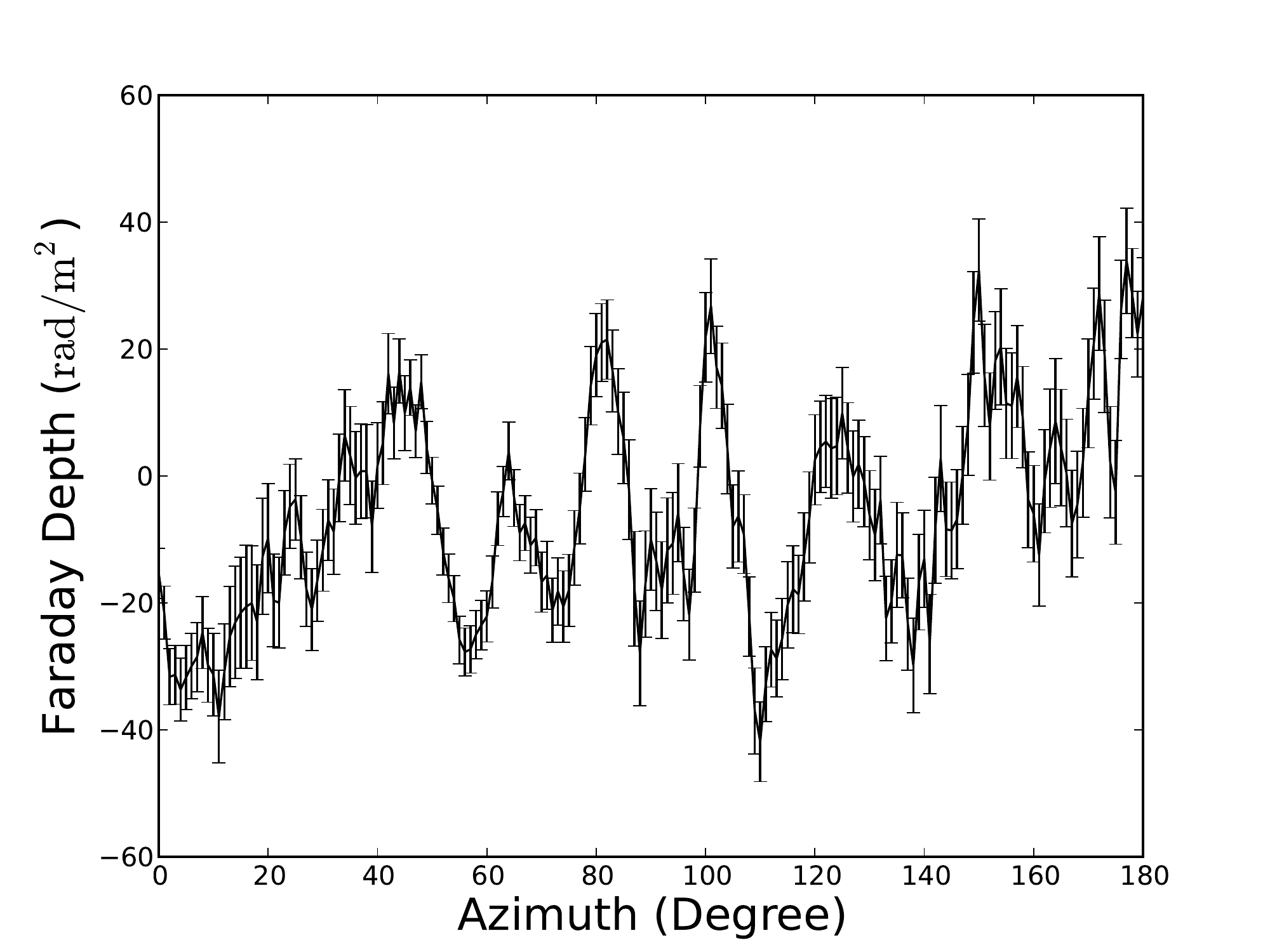}
	\caption{Azimuthal profile of Faraday depth measured in sectors of a ring along the eastern part of polarisation arm~1. North is at $0\degr$, east at $90\degr$ and south at $180\degr$. The error bars give the error of the mean value in each sector, using the error map of Fig.~\ref{fig:Maxfaradaydepth} (right).}
	\label{fig:FDprofile}
\end{figure}

An azimuthal profile of Faraday depths was created by averaging the map along an annulus from a central point at RA(J2000) = 01$^\mathrm{h}$ 36$^\mathrm{m}$ 42$^\mathrm{s}$.4, DEC(J2000) = +15$\degr$ 46$\arcmin$ 10$\arcsec$ with a radius of $2.85\arcmin$ and a width of $0.9\arcmin$ which traces the eastern polarisation arm (polarised arm~1) from north to south (Fig.~\ref{fig:FDprofile}). Peaks are located at azimuthal angles of 31, 49, 67, 83, 101, 118, 143, 164, 168, and $175\degr$. If we restrict the analysis to the range of azimuthal angles from $30\degr$ to $130\degr$, we are left with five regularly spaced peaks with an average separation of $17.5\pm0.5\degr$, corresponding to a pattern wavelength of $35\pm1\degr$ or $3.7\pm0.1$\,kpc.

The numerical simulations by \cite{2002ApJ...581.1080K} predict a wavelength of the most unstable symmetric mode of the Parker instability of between 4$\pi h$ and 17$h$ where $h$ is the $\HI$ gas scale height.  The wavelength measured from our observations corresponds to a $\HI$ gas scale height of 108--147\,pc. The average scale height for $\HI$ gas in NGC\,628 is 490\,pc \citep{2011AJ....141...23B}.
This discrepancy is likely due to the fact that \cite{2002ApJ...581.1080K} assumed $\beta$ = 1, where $\beta$ is the ratio of thermal to magnetic energy densities, whereas $\beta \ll 1$ was found in NGC\,6946 and IC\,342.
Additionally, the calculated scale height could also be inaccurate due to assumptions made in \cite{2011AJ....141...23B} such as adopting a constant velocity dispersion for the gas.

Field loops are observed only in the eastern part of polarisation arm~1. \cite{Rodrigues2016} performed numerical simulations of Parker instabilities including cosmic rays that can exert a pressure comparable to thermal gas, turbulence and magnetic fields. They showed that the observed Faraday depth signature is a periodic pattern, with its peak magnitude and wavelength depending on the ratio of cosmic ray pressure to gas pressure (see Fig.~12 of that paper). When the cosmic ray pressure is dominant this Faraday depth signature can develop within 200\,Myrs, albeit, the peak magnitude and separation are small. When the gas pressure is dominant, the Faraday depth signature develops more slowly but the peak magnitude and wavelength are large. In the eastern part of polarisation arm~1, where the field loops are observed, significant amounts of gas and star formation are present. In other regions, such as the magnetic arms~2 and 3 where we do not observe an alternating Faraday depth signature, the cosmic ray pressure will be more comparable to the gas pressure which in turn produces smaller Faraday depth peak magnitudes and pattern wavelengths, so much smaller in fact that we may not possess the Faraday depth and angular resolution needed to resolve these features.

Faraday depth signatures of Parker loops could exist throughout the galaxy but better Faraday depth and angular resolution as well as sensitivity are needed to reveal them. Further analysis and observations will be performed to further explore this issue.

Models by \cite{2002A&A...386..347H} indicate that the Parker instability develops a wave-like deformation with a small amount of twisting at the base of the instability that occurs after 400\,Myrs before magnetic reconnection begins and becomes a helical field at large heights (see Fig.~9 of their paper).
Helical fields on scales of the typical turbulence length of 50--100\,pc can also be generated by the large-scale dynamo \citep[e.g.][]{1996ARA&A..34..155B} from which larger loops may be formed by reconnection.
Our resolution of 18$\arcsec$ corresponds to about 640\,pc.

\begin{figure}
	\centering
	\includegraphics[width=1.2\columnwidth,trim={2.5cm 0.4cm 0.4cm 0.4cm},clip]{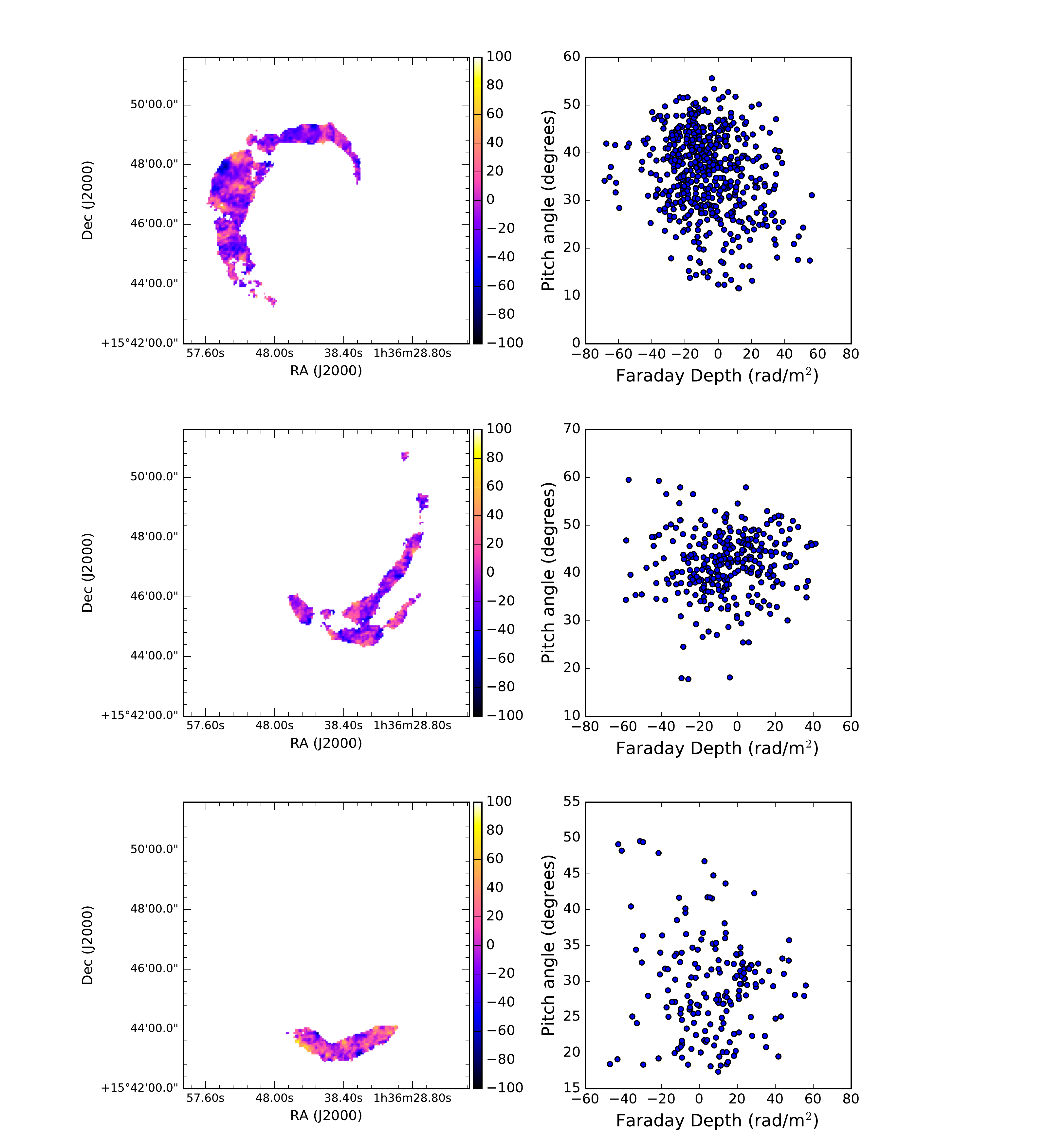}
	\caption{A pixel to pixel correlation between Faraday depth and pitch angle (right) corresponding to each polarised arm shown in the colour maps of Faraday depth to the left, with the colour scale in rad m$^{-2}$.}
	\label{fig:FDpitchanglecorr}
\end{figure}

A signature of helical fields on such scales would be a correlation between Faraday depth (Fig.~\ref{fig:Maxfaradaydepth}, left) and intrinsic pitch angle (Fig.~\ref{fig:pitchanglemap}). We computed a pixel to pixel correlation for every fifth pixel between these two quantities at pixels where the polarisation signal-to-noise ratio is larger than 8, and for each spiral arm individually. 
We found a weak anti-correlation ($\rho$=-0.2 with a p-value of 2.3$\times10^{-6}$) between the Faraday depth and the pitch angle in the eastern part of the polarisation arm~1 (Fig.~\ref{fig:FDpitchanglecorr}), which is in conflict with the helical field interpretation (see Sect.~\ref{subsec:loops} for a discussion).

No correlation whatsoever is seen for the southern part of the polarisation arm~1 ($\rho$=-0.1) or for the polarisation arms~2 and 3 ($\rho$=0.06) that do not show any regularly spaced peaks, unlike the eastern part of arm~1. Correlations between Faraday depth and total \& polarised intensity were also checked but no meaningful correlation could be seen in any of the arms.

\section{Discussion}
\label{discussion}

\subsection{Why does only one Faraday depth feature coincide with an $\HI$ hole?}
\label{subsec:holes}

The co-location of a Faraday depth gradient with an $\HI$ hole found in this work suggests that the discovery in NGC\,6946 \citep{2012ApJ...754L..35H} is not artificial. This is consistent with superbubbles being the main energy driver in the disk-halo interface.
The Faraday depth observed in the Faraday depth gradient in NGC\,6946 \citep{2012ApJ...754L..35H} ranges from 18.8 to 57.0\,rad m$^{-2}$, a range considerably smaller than the Faraday resolution of our observation. Possibly we do not possess the Faraday depth resolution to detect similar Faraday depth gradients.
We may be observing the most extreme Faraday depth gradient in NGC\,628 of about 110\,rad m$^{-2}$ (Fig.~\ref{fig:HIradiohole}), more than twice compared to what is seen in NGC\,6946.

There are several possible reasons why only one Faraday depth gradient coinciding with an $\HI$ hole is observed.
Firstly, more than half (58 out of 102) of the $\HI$ holes in the disk of NGC\,628 are located in regions devoid of S-band polarised emission due to more intense star formation, accompanied by an increased level of depolarisation.
The age of the $\HI$ hole is also important and must be within a certain range, in that it must be old enough so that the magnetic field configuration is able to gain a significant vertical offset but young enough that the vertical
shear has not destroyed the observational evidence of the Faraday depth gradient.
Additionally, the azimuthal location is important in order to create an ideal geometric situation and thus produce an observable gradient. The small Faraday depolarisation effect (Fig.~\ref{fig:628depolarisationmap}) indicates that polarimetry in S-band is probing the disk close to the midplane where strengths of the turbulent magnetic field and thermal electron densities are large. The magnetic field may not have had enough time to stabilise its configuration due to the turbulence near the midplane.

JVLA polarimetric observations in L-band (1--2\,GHz) with its better Faraday depth resolution (about 90\,rad m$^{-2}$) are superior for finding smaller Faraday depth gradients.
L-band observations would reveal polarisation at larger heights above the midplane (closer to the observer), a region where the magnetic field configuration is able to create an observable Faraday depth gradient.
However, \cite{2015ApJ...800...92M} found no coincidence between $\HI$ holes and Faraday depth gradients in M\,51 observed in L-band, so it remains to be seen if additional Faraday depth gradients can be detected in NGC\,628
allowing a more comprehensive analysis of the field configuration in the disk-halo interface.

\subsection{Parker loops without helical fields?}
\label{subsec:loops}

Faraday depth is generated by the magnetic field along the line of sight which is the vertical field in a galaxy with a small inclination like NGC\,628. We observed a striking periodic pattern of alternating Faraday depths (see Sect.~\ref{subsec:parkerloops}) from positive to negative Faraday depths (Fig.~\ref{fig:FDprofile}) along one of the spiral arms. The wavelength of this pattern corresponds to 3.7 $\pm$ 0.1\,kpc.
This kind of feature can be described as a magnetic loop extending out of the galaxy's disk into its halo.
Similar loops have been observed in IC\,342 \citep{2015A&A...578A..93B} with a pattern wavelengths of about 4.4\,kpc.

A regular pattern of Faraday depths could be a signature of large-scale helical fields around the spiral arm if the pitch angles of the field (Fig.~\ref{fig:pitchanglemap}), delineating the field component in the galaxy plane, reveal a periodicity with a similar separation. The weak anti-correlation between pitch angle and Faraday depth (Fig.~\ref{fig:FDpitchanglecorr}) shows that the regular magnetic fields observed in NGC\,628 are hardly of helical structure and hence are different from those observed in M\,31 and IC\,342. We conclude that the observed field deviations occur mostly in the vertical direction, in contrast to the models by \cite{2002A&A...386..347H}.

NGC\,628 and IC\,342 have similar Faraday depths, so that the vertical magnetic field strengths (and the electron densities) are similar.
IC\,342 has a star-formation rate per surface area of about 0.016\,M$_{\odot}$ yr$^{-1}$ kpc$^{-2}$ in comparison to 0.0026\,M$_{\odot}$ yr$^{-1}$ kpc$^{-2}$ in NGC\,628 \citep{Calzetti2010}.
A higher star-formation rate should result in a greater outflow speed into the halo. Additionally, the star-formation rate across the disk in NGC\,628 steadily decreased with time (see Sect.~\ref{separatethermal}), thereby preventing a constant outflow in NGC\,628.

The twisting of Parker loops is caused by two main factors, the rotational velocity of the galaxy and the horizontal component of the velocity vector due to gas flow along the magnetic field lines.
NGC\,628 and IC\,342 have similar rotation curves \citep{1992A&A...253..335K,Sofue1996}, while IC\,342 is expected to have a more continuous and stronger outflow into the halo compared to NGC\,628 (Sect.~\ref{separatethermal}).
This has two consequences. Firstly, the polarised emission from NGC\,628 traces mainly the disk. As the Parker loops are not twisted already near to the galaxy plane, the observed magnetic pitch angle is not affected. Secondly, a large halo like in IC\,342 is more efficient in the inflation of Parker loops \citep{Parker1992} because helical fields extend to heights of several kpc, similar to their pattern wavelengths.

If the regular pattern of Faraday depths is indeed caused by Parker loops, this should affect the dynamics of the ionised gas, which would be able to slide down the magnetic field loops with signatures in spectroscopic observations. A follow-up spectroscopic H$\alpha$ observation along the regular pattern of Faraday depths (especially in regions bright in H$\alpha$, see for example Fig.~\ref{fig:HIIpol}) would be beneficial to determine which effect the observed Parker loops have on the dynamics of the ionised gas.

\subsection{Magnetic pitch angles}
\label{subsec:pitch}

The maintenance of both radial and azimuthal components of the magnetic field, and hence a non-zero pitch angle, is an important prediction of dynamo theory \citep{1996ARA&A..34..155B}, which is confirmed by many observations like in NGC\,628. More crucially, the magnetic pitch angle is always larger than the morphological pitch angle of the spiral arms, as was also found for the galaxies M\,83 \citep{2016A&A...585A..21F} and M\,101 \citep{berkhuijsen2016}. This gives evidence for the action of a large-scale dynamo where the magnetic field is not coupled to the gas flow and obtains a significant radial component. Our NGC\,628 data give further support to this model. The smoothness of the magnetic pitch angle is also consistent with dynamo action because the large-scale field is built up over several galactic rotations and is not strongly affected by local features in the spiral arms.

The dynamo models by \cite{2013A&A...556A.147M,2015A&A...578A..94M} assume that a large-scale ordered field is generated everywhere in the disk, while a small-scale dynamo injects turbulent fields only in the spiral arms. This gives polarisation arms between the gaseous arms at all radii but with pitch angles of the polarisation structures and pitch angles of the magnetic field lines that are significantly smaller than those of the gaseous arms. This is in contrast to the observations discussed in this paper. Applying the wavelet transform technique by \cite{2016A&A...585A..21F} to NGC\,628 will allow a more detailed analysis which is essential for a deeper understanding of the interaction between spiral structure and magnetic fields.

The smoothness of the magnetic pitch angle is also worth discussing. As mentioned previously this smoothness is consistent with dynamo action. \cite{2013A&A...556A.147M} performed a 2-D mean-field dynamo model in the ``no-z'' approximation adding injections of a small-scale magnetic field. These injections are the result of supernova shock fronts situated in the spiral arms and represent the effects of the dynamo at smaller scales. In particular, the model with a lower star formation rate (``model 77''), and hence less injection of turbulent field into the spiral arms, can produce a smooth magnetic pitch angle. While this model produces a broad polarisation arm that almost fills the inter-arm regions, the western arm of NGC\,628 is not broad but shows very filamentary structure with the magnetic arm splitting into two (Fig.~\ref{fig:magneticarms}). \cite{2013A&A...556A.147M} pointed out that the field injection could decrease more slowly and smoothly away from the arms and that could produce a more filamentary structure. NGC\,628 poses the interesting question what may occur if the injection of the small-scale magnetic field in the central region disappears. While this decrease in star formation in the central region has only occurred in the past 100\,Myrs, the timescale is probably not long enough to significantly change the dynamo. This calls for further developments of the dynamo models.

Direct evidence for a dynamo-generated large-scale field in the disk can be derived from the observation of a large-scale pattern of Faraday rotation measures or Faraday depths \citep[e.g.][]{2013pss5.book..641B}. However, this method cannot be applied to NGC\,628 due to its small inclination.

\section{Conclusions}
\label{conclusions}

In this work we have presented new observations of NGC\,628 with the JVLA at S-band (2-4\,GHz) and Effelsberg 100-m telescope at 8.35\,GHz and 2.6\,GHz. The wide bandwidth of the JVLA
has enabled us to image NGC\,628 with unprecedented sensitivity and resolution. The wide bandwidth also allowed us not only to perform RM Synthesis but investigate the Faraday depolarisation across the band.
Significant polarisation was observed across the disk especially strong in the interarm regions, similar to NGC\,6946 and IC\,342. Three main polarisation arms are evident but have different properties, with two arms filling the criterion for a magnetic arm.
Until now NGC\,628 has been relatively unexplored in radio continuum but with its extended $\HI$ disk and lack of active star formation in its central region has produced a wealth of interesting magnetic phenomena. We observe evidence for two drivers of magnetic turbulence in the disk-halo connection of NGC\,628, namely, Parker instabilities and superbubbles.

\begin{figure*}
	\centering
	\includegraphics[width=1.0\textwidth]{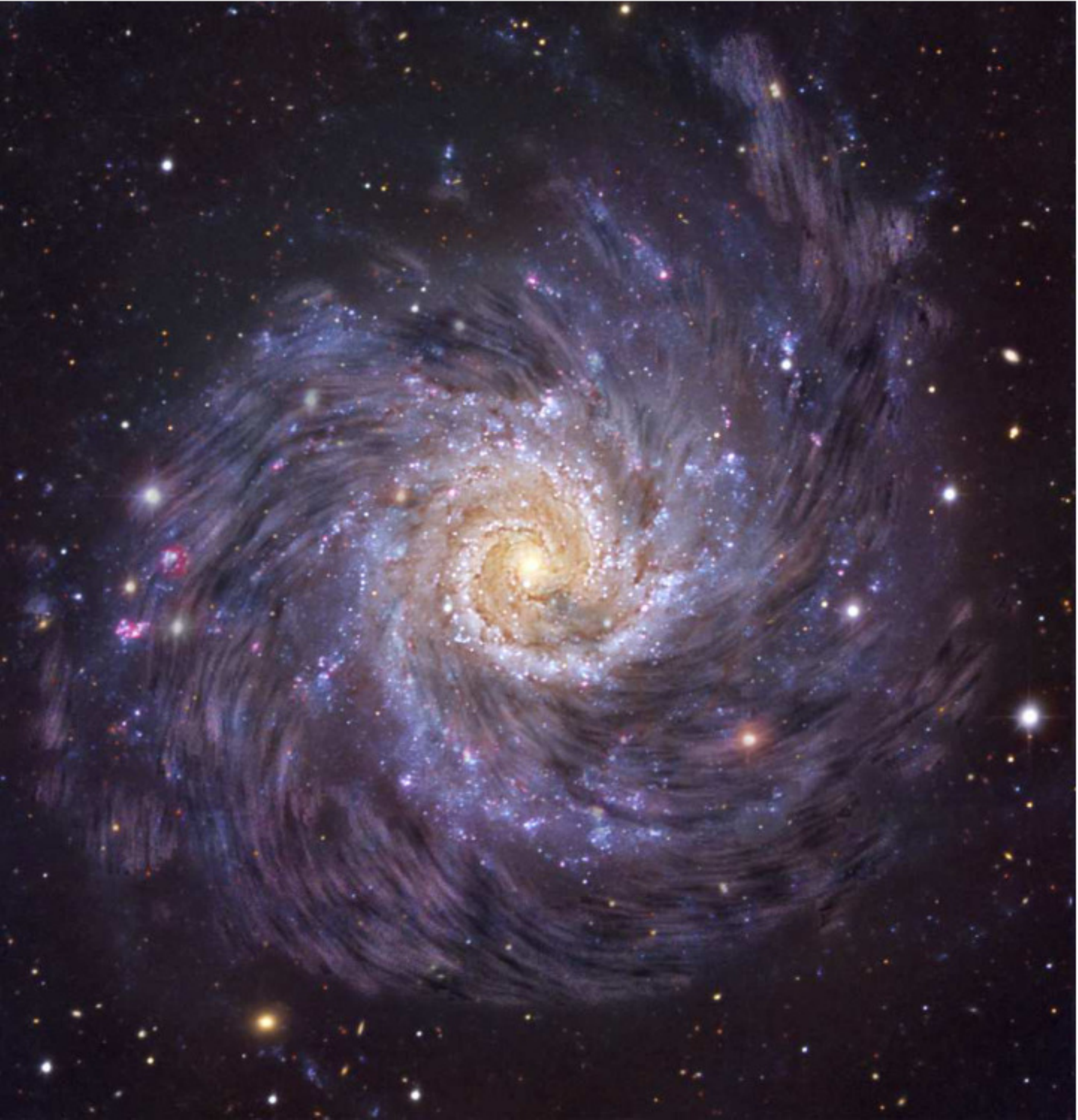}
	\caption{Polarised intensity of NGC\,628 with the magnetic fields illustrated as flow lines using the Line Integral Convolution method \citep{Cabral:IVF}. The background image is an optical image plus H$\alpha$ from the Calar Alto Observatory in Spain, created by Steve Mazlin and Vicent Peris (please see the acknowledgements for the full list of contributors).}
	\label{fig:628StokesPI8arcsecLIC}
\end{figure*}

The main findings of this work can be summarised as follows:

\begin{itemize}
\item From our Effelsberg observations and data from literature we estimate an integrated spectral index for NGC\,628 of $\alpha = -0.79\pm0.06$ which is a normal value for spiral galaxies \citep{1982A&A...116..164G}. The spectrum is also consistent with \cite{Basu2015} that due to the clumpy nature of the ISM the galaxy-integrated spectrum shows no signs of CRE loss mechanisms and remains a power law over much of the radio continuum.
\item A smooth extension to the north of NGC\,628 can be seen in total radio continuum coinciding with the extended $\HI$ disk as well as weak $\HII$ regions located in the outer disk. Additionally, there is evidence that one of the polarisation arms extends far into the extended disk. The polarised arms are seen to traverse along the optical interarm regions, as illustrated as flow lines using the Line Integral Convolution method \citep{Cabral:IVF} in Fig.~\ref{fig:628StokesPI8arcsecLIC}.
\item We observe little radio emission and a high thermal fraction in the central region of the galaxy. The Initial Mass Function seems to be steeper than that in the spiral arms. While the central region hosts enough stars to ionise the gas that emits in H$\alpha$, there are not enough very massive O stars that can become supernovae and produce CREs.
\item We computed magnetic field strengths using the nonthermal emission and find a mean magnetic field strength across the galaxy of 9\,$\mu$G out to a radius of 8\,kpc. In addition, we observe a magnetic field strength of 12\,$\mu$G in the arm regions and 15\,$\mu$G in the largest star forming regions with 7\,$\mu$G in the extended disk beyond a radius of 8\,kpc.
\item We observe a region in the north of the galaxy with intense polarised emission. The degree of field order is higher and the magnetic pitch angle is smaller compared to its immediate surroundings. We find this feature could be consistent with a barrel-shaped expansion taking place preferentially along the direction of the ordered magnetic field. Such a feature is not seen in other galaxies and possibly unique. 
\item We find no clear indications of polarisation originating from the far side of the midplane, as observed in \cite{2010A&A...514A..42B}. Better Faraday depth resolution is required in order to confirm this, e.g. with observations with the JVLA at L-band.
\item In the eastern part of polarisation arm~1 we observe a periodic pattern in Faraday depth with an average pattern wavelength of $3.7\pm0.1$\,kpc which indicates Parker loops. A weak anti-correlation between Faraday depth and magnetic pitch angle suggests that these loops are vertical in nature with little helical twisting.
\item We observe one significant feature in Faraday depth which coincides with an $\HI$ hole. This signifies that the Faraday depth gradient seen in NGC\,6946 by \cite{2012ApJ...754L..35H} is not coincidental. Such phenomena provide
a mechanism for the dynamo process to expel magnetic fields in order to eliminate quenching.
\item We observe a magnetic pitch angle that is systematically larger than the morphological pitch angle which is also seen in M\,83 \citep{2016A&A...585A..21F} and M\,101 \citep{berkhuijsen2016}. This is evidence for the action of a large-scale dynamo where the magnetic field is not coupled to the gas flow and obtains a significant radial component.  The smoothness of the magnetic pitch angle is also consistent with dynamo models, e.g. those by \cite{2013A&A...556A.147M} and \cite{2015A&A...578A..94M}.
\end{itemize}

\begin{acknowledgements}
DDM acknowledges support from ERCStG 307215 (LODESTONE).
This research was performed in the framework of the DFG Research Unit 1254 ``Magnetisation of Interstellar and Intergalactic Media: The Prospects of Low-Frequency Radio Observations''.
Special thanks go to Aritra Basu for his help with this work and to Elly M. Berkhuijsen for carefully reading the paper. We also thank the anonymous referee whose helpful comments improved this paper.
We thank CAHA, Descubre Foundation, DSA, Astronomical Observatory of the University of Valencia (OAUV), Vicent Peris (OAUV), Jos\'e Luis Lamadrid (CEFCA), Jack Harvey (Star Shadow Remote Observatory, SSRO), Steve Mazlin (SSRO), Oriol Lehmkhul, Ivette Rodriguez, and Juan Conejero (PixInsight) for the permission to use their optical image (Fig.~\ref{fig:628StokesPI8arcsecLIC}) of NGC\,628.
\end{acknowledgements}

\bibliographystyle{aa}
\bibliography{ngc628ref}

\end{document}